\documentclass[sigconf,nonacm]{acmart}
\usepackage{enumitem}

\usepackage{tabularx}
\usepackage{url}
\usepackage{booktabs}
\usepackage{multirow}
\usepackage{pifont}
\usepackage{graphicx}
\usepackage{lineno}
\usepackage{makecell} 
\usepackage{adjustbox}

\usepackage[most]{tcolorbox} % 支持多种方框样式

\usepackage{fontawesome5}   
\newcommand{\cmark}{\ding{51}} % ✓
\newcommand{\xmark}{\ding{55}} % ✗
\usepackage{pifont}

\renewcommand{\cmark}{\textcolor{ForestGreen}{\checkmark}}   % 绿色对勾
\renewcommand{\xmark}{\textcolor{red}{\ding{55}}}      % 红色叉号

\usepackage[table,dvipsnames]{xcolor}

\usepackage{ifthen}
\newboolean{include-notes}
\setboolean{include-notes}{true}
\newcommand{\nb}[3]{\ifthenelse{\boolean{include-notes}}{{\colorbox{#2}{\bfseries\sffamily\scriptsize\textcolor{white}{#1}}}{\ \textcolor{#2}{\sf\small\textit{#3}}}}{}}

\usepackage{colortbl}    % needed for \rowcolor/\cellcolor in tables

\usepackage{arydshln}
\usepackage{amsmath, amssymb, amsfonts}
\usepackage{bbold}
\AtBeginDocument{%
  }

\begin{document}

\title{Causal Scaffolding for Physical Reasoning: A Benchmark for Causally-Informed Physical World Understanding in VLMs}


% Use \fnsymbol so any title-page footnote / \thanks renders symbolically.
% Corresponding-author dagger is attached manually below to avoid relying on
% acmart's internal \authornote counter (which is not user-accessible).
\renewcommand{\thefootnote}{\fnsymbol{footnote}}

\author{Tianyi Tang}
\affiliation{%
  \institution{CFAR, IHPC, Agency for Science, Technology and Research (A*STAR)}
  \country{Singapore}}
  \email{Tang_Tianyi@a-star.edu.sg}

\author{Zhuoyi Lin\textsuperscript{$\dagger$}}
\affiliation{%
  \institution{I\textsuperscript{2}R, Agency for Science, Technology and Research (A*STAR)}
  \country{Singapore}}
  \email{lin_zhuoyi@a-star.edu.sg}

\author{Zeyu Feng}
\affiliation{%
  \institution{CFAR, IHPC, Agency for Science, Technology and Research (A*STAR)}
  \country{Singapore}}
  \email{feng_zeyu@a-star.edu.sg}

\author{Tianyi Ma}
\affiliation{%
  \institution{Nanyang Technological University}
  \country{Singapore}}
\affiliation{%
  \institution{CFAR, IHPC, Agency for Science, Technology and Research (A*STAR)}
  \country{Singapore}}
  \email{tianyi008@e.ntu.edu.sg}

\author{Yew-Soon Ong}

\affiliation{%
  \institution{Nanyang Technological University}
  \country{Singapore}}
\affiliation{%
  \institution{CFAR, IHPC, Agency for Science, Technology and Research (A*STAR)}
  \country{Singapore}}
  \email{asysong@ntu.edu.sg}

\author{Ivor Tsang}

\affiliation{%
  \institution{Nanyang Technological University}
  \country{Singapore}}
\affiliation{%
  \institution{CFAR, IHPC, Agency for Science, Technology and Research (A*STAR)}
  \country{Singapore}}
  \email{ivor_tsang@a-star.edu.sg}

\author{Haiyan Yin\textsuperscript{$\dagger$}}

\affiliation{%
  \institution{CFAR, IHPC, Agency for Science, Technology and Research (A*STAR)}
  \country{Singapore}}
  \email{yin_haiyan@a-star.edu.sg}

\renewcommand{\shortauthors}{Tang et al.}

\begin{abstract}
Understanding and reasoning about the physical world is the foundation of intelligent behavior, yet state-of-the-art vision-language models (VLMs) still fail at causal physical reasoning, often producing plausible but incorrect answers. To address this gap, we introduce \textbf{\textsc{CausalPhys}}, a benchmark of over 3,000 carefully curated video- and image-based questions spanning four domains: Perception, Anticipation, Intervention, and Goal Orientation. Each question is paired with an {expert-annotated causal graph} capturing {object–attribute–event} dependencies, enabling interpretable and fine-grained evaluation of causal understanding. Building on this, we formulate a causal-graph-grounded metric that quantitatively measures how well a model’s chain-of-thought reasoning aligns with the correct causal relations, moving beyond answer-only accuracy and enabling systematic diagnosis of VLMs' causal reasoning failures. Using this metric, we conduct a comprehensive analysis of leading VLMs, revealing systematic gaps in capturing causal dependencies and underscoring the need for causality-aware learning. To address these limitations, we further propose \textbf{Causal Rationale-informed Fine-Tuning (CRFT)}, which explicitly aligns VLM reasoning with causal structures. Extensive experiments demonstrate that CRFT substantially enhances both reasoning accuracy and interpretability across multiple model backbones. By unifying dataset curation, causal evaluation, and causality-informed learning, CausalPhys establishes a strong foundation for advancing modern VLMs toward causally grounded physical reasoning. Our code and dataset are available at \href{https://github.com/haorentang/CausalPhys}{\textcolor{blue}{\url{https://github.com/haorentang/CausalPhys}}}.
\end{abstract}

% %% A "teaser" image appears between the author and affiliation
% %% information and the body of the document, and typically spans the


\maketitle

% Manual corresponding-author footnote: mark = fnsymbol(2) = dagger.
% Placed after \maketitle so it appears at the bottom of the title page.
\footnotetext[2]{Corresponding authors.}
% Restore default numeric footnote style for the body.
\renewcommand{\thefootnote}{\arabic{footnote}}
\setcounter{footnote}{0}

\begin{figure*}[!h]
  \centering
  \setlength{\tabcolsep}{3.5pt} % default ~6pt

  % --- 左侧：Pie chart ---
  \begin{minipage}[t]{0.48\textwidth}
    \centering
    \vskip -0.15in  % 如果真的需要再微调
    \includegraphics[width=0.88\linewidth]{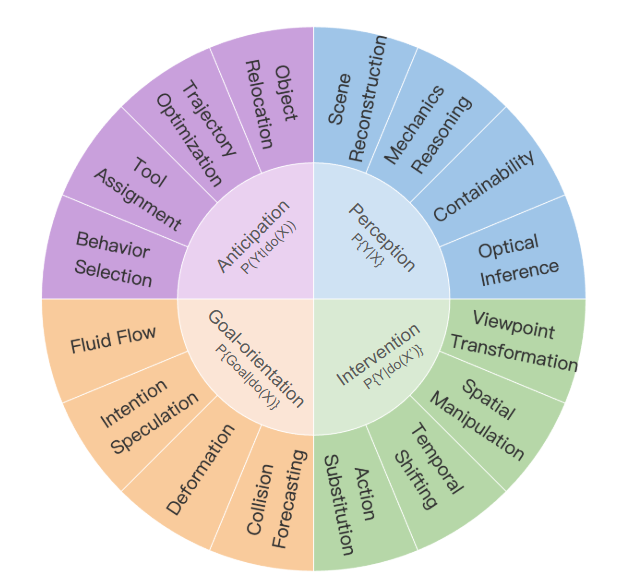}
    \vskip -0.12in
    \captionof{figure}{\textbf{Hierarchical Taxonomy of \textsc{CausalPhys} spanning four categories. }
    Each of the four major categories corresponds to a  \textbf{causal range} \cite{pearl2009causality} ($P\{Y\,|\,X\}$, $P\{Y\,|\,\text{do}(X)\}$, $P\{\text{Goal}\,|\,\text{do}(X)\}$, $P\{Y\,|\,\text{do}(X')\}$). 
    The outer segments enumerate sixteen subcategories that instantiate these primitives.}
    \label{fig:taxonomy}
  \end{minipage}
  \hfill
  % --- 右侧：Table ---
  \begin{minipage}[t]{0.48\textwidth}
  \vspace{0.02in}
    \normalsize
    \centering
    {\fontsize{7.5}{12.2}\selectfont
    \begin{tabular}{l c l c}
    \toprule
    \multicolumn{2}{c}{\textbf{\faEye\, Perception (951)}} &
    \multicolumn{2}{c}{\textbf{\faMagic\, Anticipation (900)}} \\
    \cmidrule(r){1-2} \cmidrule(r){3-4}
    \textbf{Subset} & \textbf{\#Question} &
    \textbf{Subset} & \textbf{\#Question} \\
    \midrule
    Optical Inference     & 252 & Collision Forecasting & 300 \\
    Containability        & 201 & Deformation          & 200 \\
    Scene Reconstruction  & 200 & Fluid Flow           & 200 \\
    Mechanics Reasoning   & 298 & Intention Speculation& 200 \\
    \midrule
    \multicolumn{2}{c}{\textbf{\faCogs \, Intervention (573)}} &
    \multicolumn{2}{c}{\textbf{\faBullseye\, Goal Orientation (638)}} \\
    \cmidrule(r){1-2} \cmidrule(r){3-4}
    \textbf{Subset} & \textbf{\#Question} &
    \textbf{Subset} & \textbf{\#Question} \\
    \midrule
    Spatial Manipulation  & 151 & Object Relocation      & 190 \\
    Action Substitution   &  99 & Tool Assignment        & 100 \\
    Temporal Shifting     & 149 & Behavior Selection     & 229 \\
    Viewpoint Transformation & 174 & Trajectory Optimization & 119 \\
    \bottomrule
    \end{tabular}
    }
    \vspace{0.23in}
    \captionof{table}{\textbf{Statistics of \textsc{CausalPhys}}.  The benchmark comprises 3{,}062 video- and image-based questions spanning 4 causal domains and 16 fine-grained subsets, capturing a broad spectrum of physical reasoning abilities.}
    \label{tab:statis}
  \end{minipage}
\end{figure*}

\section{Introduction}

Understanding and reasoning about the physical world lies at the heart of intelligence, enabling agents to act robustly and adaptively in real-world environments~\citep{srivastava2022behavior, gupta2021embodied}.
Yet today’s vision-language models (VLMs) remain far from human intuition, often struggling with even basic physical interactions.
Robust physical reasoning demands more than visual pattern recognition: agents must infer intrinsic object properties \citep{yi2019clevrer, chen2022comphy}, track spatial and temporal relations among entities \citep{yang2025thinking, wang2024embodiedscan}, interpret evolving physical scenes, and anticipate how interactions unfold to guide planning and avoid costly errors \citep{bear2021physion, dong2025seeing}. 
Humans, by contrast, perform such reasoning effortlessly, guided by an intuitive grasp of physical causality that emerges early in cognitive development \citep{carey2000origin, mccloskey1983intuitive, chow2025physbench}. 
How to equip VLMs with this level of causally grounded understanding remains a central open challenge. Bridging this gap is essential for advancing embodied AI systems that are both reliable and trustworthy.

\begin{figure*}[!t]
    \centering
    \includegraphics[width=0.99\linewidth]{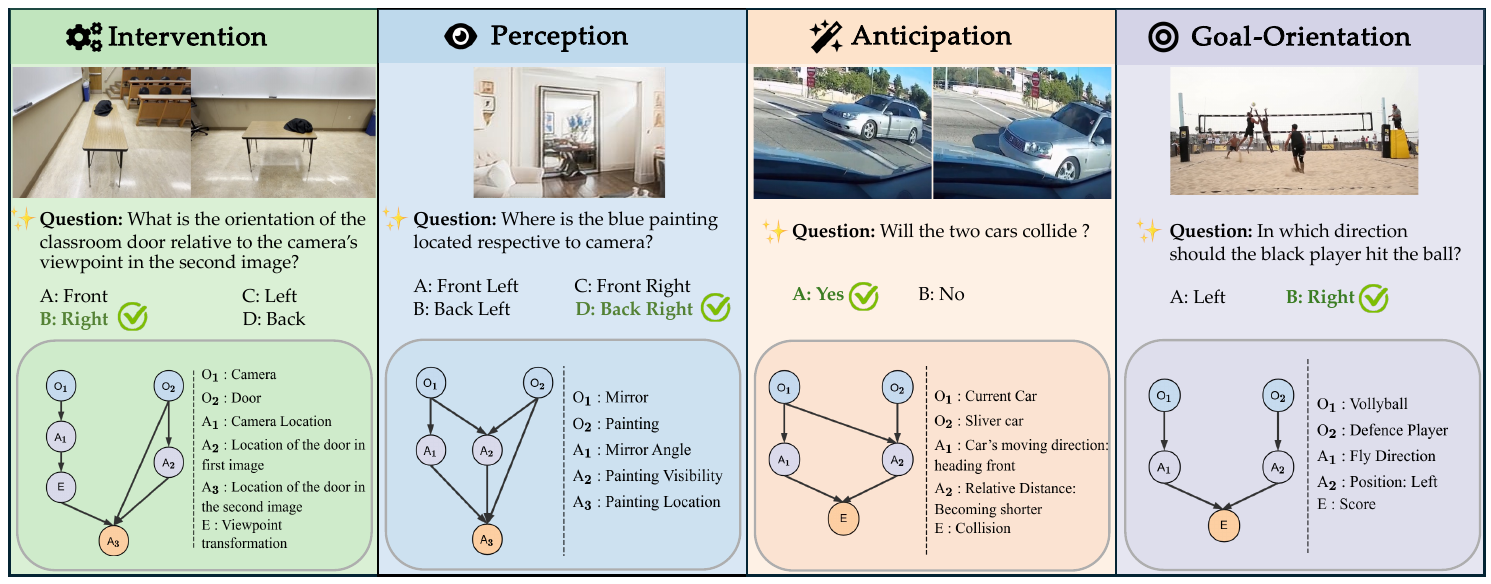} % or .png/.jpg
    \caption{\textbf{Overview of the \textsc{CausalPhys} benchmark.} 
\textsc{CausalPhys} categorizes causally-informed physical understanding across four fundamental domains: (i) {Intervention},  (ii) {Perception}, (iii) {Anticipation}, and (iv) {Goal-Orientation}. Beyond ground-truth question–answer pairs, each question is accompanied by a carefully annotated \textbf{causal directed acyclic graph (DAG)} that captures the underlying \textit{object–attribute–event} dependencies governing the physical dynamics of the scene.}
    \label{fig:overview}
    % \vspace{-0.15in}
\end{figure*}

Recent VLMs excel at multimodal tasks such as visual question answering \citep{antol2015vqa, wu2017visual}, object recognition \citep{carion2020end, radford2021learning}, and image captioning \citep{li2022blip, li2023blip}. Yet extending these successes to dynamic physical reasoning in realistic environments remains an open challenge \citep{bear2021physion,tung2023physion++,chow2025physbench,dong2025seeing}. Relying solely on perception-driven capabilities has proven insufficient for building generalist embodied agents \citep{komanduri2025causalvlbench,foss2025causalvqa,liu2025causal3d,chen2024cello}, often leading to brittle behaviors such as mishandling fragile objects or misjudging grasp affordances. As a concrete example, Fig.~\ref{fig:overview} (Intervention) illustrates that inferring the orientation of a door relative to the camera viewpoint from limited observations is far from trivial. Such reasoning demands sensitivity to latent spatial structures, occluded relationships, and viewpoint transformations that are invisible in isolated images. Ultimately, these cases hinge on anticipating how the world changes under interventions or viewpoint shifts, an ability naturally framed through \textbf{causally informed reasoning}.  
This inferential capacity allows agents to bypass the pitfalls of spurious visual correlations, anchoring their understanding in the structural causal dependencies that dictate physical transitions. Bridging high-dimensional multimodal perception with explicit causal grounding enables a transition from mere observation to intervention-aware reasoning, ultimately fostering a more consistent, generalizable, and mathematically rigorous representation of physical reality.

However, integrating causal reasoning into VLMs remains a fundamental challenge, and we identify three critical gaps.
(1) Current models primarily learn \textit{statistical associations from observational data} rather than underlying causal mechanisms, limiting their ability to reason in dynamic, real-world environments.
(2) Prior benchmarks rarely include \textit{explicit causal annotations}, motivating our construction of a dataset with expert-annotated causal graphs for rigorous evaluation of causal dependencies.
(3) Existing efforts largely \textit{emphasize evaluation rather than training}, leaving open how to effectively instill causal reasoning within multimodal systems.
These challenges call for benchmarks and training paradigms that explicitly foster \textit{causally informed reasoning}, moving VLMs beyond surface correlations toward genuine physical understanding.

To tackle these challenges, we introduce \textbf{\textsc{CausalPhys}}, a comprehensive benchmark of over 3,000 expertly curated video- and image-based questions spanning \textbf{four domains}: {Perception}, {Anticipation}, {Intervention}, and {Goal Orientation}, across \textbf{16 subcategories} (Fig.~\ref{fig:taxonomy} and Table~\ref{tab:statis}). A distinguishing feature of {\textsc{CausalPhys}} is that each question is paired with an \textbf{expert-annotated causal graph} capturing physical interactions and dependencies, enabling {mechanism-level and interpretable evaluation} of VLM reasoning. By coupling visual reasoning with Causal Directed Acyclic Graphs (DAGs), our framework establishes a rigorous foundation for evaluating not just what a model predicts, but why it fails—bridging the gap between surface-level pattern matching and systematic physical world understanding. Utilizing \textsc{CausalPhys}, we identify a systemic decoupling between visual recognition and causal dependency modeling in state-of-the-art VLMs. To address this, we propose \textbf{Causal Rationale-informed Fine-Tuning (CRFT)}, a framework that enforces alignment between a model's latent reasoning and causal graphs. This approach enhances zero-shot generalization and interpretability, bridging the gap between superficial pattern matching and robust physical world understanding.

This work is the first to establish a unified framework connecting \textbf{benchmarking}, \textbf{evaluation}, and \textbf{model improvement} for causal physical reasoning in VLMs.
We aim to deliver actionable insights and close the gap between VLMs and physical world understanding, advancing embodied AI by learning from human reasoning capabilities. 
Overall, this paper makes three key contributions:
\begin{enumerate}
    \item We introduce \textbf{\textsc{CausalPhys}}, the first benchmark coupling physical reasoning tasks with {explicit, expert-annotated causal graphs}, enabling {mechanism-level and interpretable evaluation} beyond surface-level accuracy.
    \item We develop a \textbf{causal-graph-grounded metric} that evaluates whether a model’s reasoning aligns with correct causal dependencies, providing {fine-grained diagnostic insights} beyond answer-only metrics.
    \item We propose \textbf{CRFT}, a causally guided VLM fine-tuning strategy that leverages causal graphs to {enhance VLMs’ accuracy and interpretability} in physical environments.
\end{enumerate}

\section{Related Works}

\paragraph{Physical Benchmarks.}
Early efforts in physical reasoning benchmarks laid the groundwork for evaluating agents’ understanding of basic interactions under simplified conditions~\citep{bear2021physion, tung2023physion++, zhu2023benchmarking}.
Classical benchmarks such as CLEVRER and CoPhy~\citep{yi2019clevrer, chen2022comphy} target elementary visual primitives of spheres, cubes, and rigid-body collisions, thereby focusing on fundamental perceptual and causal concepts.
Subsequent multimodal datasets~\citep{he2024olympiadbench, jiang2024visscience, lu2022learn, hao2025can, zhang2025physreason, azzolini2025cosmos} extended this direction toward commonsense reasoning grounded in linguistic or textual knowledge. While valuable for probing conceptual understanding, these settings often abstract away the perceptual and dynamical complexity inherent to real-world physics.
In parallel, spatial VQA benchmarks~\citep{wang2024embodiedscan, yang2025thinking, li2024proximity, shiri2024empirical} explored geometric relationships and spatial reasoning in 3D environments, marking an important step toward holistic physical scene understanding. 
Recent large-scale benchmarks such as PhysBench~\citep{chow2025physbench} and MVPBench~\citep{dong2025seeing} advance this line by systematically evaluating models’ ability to perceive, anticipate, and describe physical events across diverse settings. 
Both focus on physics-centric understanding through answer-based evaluation, emphasizing prediction accuracy rather than the underlying reasoning process.
\textsc{CausalPhys} complements these efforts by introducing {expert-annotated causal graphs} that capture \textit{object–attribute–event} dependencies, accompanied by {causal-graph-grounded evaluation metrics} for mechanism-level assessment.
We further propose a {causally-informed fine-tuning} strategy that aligns VLM reasoning with causal structure, enabling more consistent and interpretable physical understanding.

\newcommand{\rotcol}[2][-30]{\makecell{#2}}

\begin{table*}[!t]
\centering
\caption{\textbf{Comparison of \textsc{CausalPhys} with existing physical reasoning benchmarks}. While prior datasets are limited by synthetic environments, restricted diversity, or missing causal structure, \textsc{CausalPhys} uniquely integrates \textbf{real-world data}, \textbf{diverse scenes}, and fine-grained \textbf{causal annotations}.}
\label{tab:causalphys_comparison}
\scriptsize

\setlength{\tabcolsep}{5pt}
\renewcommand{\arraystretch}{1.1}

\begin{tabularx}{\textwidth}{l c *{10}{>{\centering\arraybackslash}X}}
\toprule
\multirow{2}{*}{\textbf{Dataset}} &
\multirow{2}{*}{\textbf{Size}} &
\multicolumn{2}{c}{\textbf{Data Type}} &
\multicolumn{2}{c}{\textbf{Data Source}} &
\multicolumn{2}{c}{\textbf{Causal Structure}} &
\multicolumn{3}{c}{\textbf{Causal Node}} \\
\cmidrule(lr){3-4}\cmidrule(lr){5-6}\cmidrule(lr){7-8}\cmidrule(lr){9-11}
& & Image & \makecell{Video} &
\makecell{Real-World\\ Data} & \makecell{Scene\\ Diversity} &
\makecell{Annotation} & \makecell{Flexibility} &
\makecell{Object} & \makecell{Attribute} & \makecell{Event} \\
\midrule
CELLO \citep{chen2024cello} & 14,000+ & \cmark & \xmark & \cmark & \xmark & \cmark & \cmark & \cmark & \xmark & \xmark \\
Causal3D \citep{liu2025causal3d} & - & \cmark & \xmark & \xmark & \xmark & \cmark & \xmark & \xmark & \cmark & \xmark \\
CausalVLBench \citep{komanduri2025causalvlbench} & - & \cmark & \xmark & \xmark & \xmark & \cmark & \xmark & \xmark & \cmark & \xmark \\
PhysBench \citep{chow2025physbench} & 10,000+ & \cmark & \cmark & \cmark & \cmark & \xmark & \xmark & \xmark & \xmark & \xmark \\
Causal VQA \citep{foss2025causalvqa} & 700+ & \xmark & \cmark & \cmark & \cmark & \xmark & \xmark & \xmark & \xmark & \xmark \\
MVP Bench \citep{dong2025seeing} & 1,000+ & \cmark & \xmark & \cmark & \cmark & \xmark & \xmark & \xmark & \xmark & \xmark \\
\midrule
\textbf{\textsc{CausalPhys} (Ours)} & 3,000+ & \cmark & \cmark & \cmark & \cmark & \cmark & \cmark & \cmark & \cmark & \cmark \\
\bottomrule
\end{tabularx}
\end{table*}

\paragraph{Causal Reasoning Datasets.}
While causal reasoning has been extensively studied for LLMs \citep{jin2023cladder, jiralerspong2024efficient, rajendran2024learning,li-etal-2025-multimodal-causal}, equivalent efforts in the VLMs remain comparatively nascent. Early multimodal studies encode causal knowledge at a focused level of granularity. 
For example, CELLO~\citep{chen2024cello} represents nodes as perceptible objects and models local relations such as “object 1 supports object 2,” while other works instantiate structural causal models primarily for interpretability rather than data-driven reasoning~\citep{fu2025unveiling}.
Recent VLM benchmarks, exemplified by CausalVLBench~\citep{komanduri2025causalvlbench} and Causal3D~\citep{liu2025causal3d}, introduce \textit{fixed-structure causal graphs} within synthetic or template-based scenes, where both entities and relations are predefined in the prompts.
These controlled settings are valuable for isolating local causal dependencies but insufficient for probing holistic, multi-object interactions in real-world physical environments.
In contrast, {\textsc{CausalPhys}} generalizes this paradigm through a \textit{flexible-form causal annotation schema} grounded in \textit{expert-annotated DAGs} with rich human insights that capture \textit{object–attribute–event dependencies} across diverse static and dynamic scenes.
It is further accompanied by \textit{causal evaluation criteria} with graph-grounded reasoning metrics and a \textit{causally informed fine-tuning strategy}, enabling more consistent and interpretable causal reasoning in VLMs.

\section{The \textsc{CausalPhys} Benchmark}

We introduce \textbf{\textsc{CausalPhys}}, a large-scale, rigorously curated benchmark for evaluating VLMs on causally-informed physical reasoning. We outline the benchmark design and formalism for causally informed physical reasoning in Sec.~\ref{sec:3.1}, detail the expert-driven annotation workflow in Sec.~\ref{sec:3.2}, and introduce a causal-graph-grounded evaluation framework in Sec.~\ref{sec:3.3} for mechanism-level reasoning.
Comprehensive results in Sec.~\ref{sec:3.4} reveal systematic reasoning gaps in current VLMs and highlight clear directions toward causally grounded model improvement.

% \cite{DBLP:conf/iccv/CuiZZYWW23,DBLP:conf/cvpr/MouraZZ25,DBLP:conf/iccv/GoyalKMMWKHFYMH17,foss2025causalvqa,DBLP:journals/pami/GraumanWBCCFGHJKLLMNRRR25,DBLP:journals/corr/abs-2506-21458,Damen2021PAMI,degusseme2025datasetbenchmarkroboticcloth,kantine_domotic_pouringCoffee_expert,villekuosmanen_agilex_clean_pour_water,villekuosmanen_agilex_pour_water_cup_full} , and every single instance is paired with its own causal graph (Fig.~\ref{fig:overview}). 

\subsection{Benchmark Overview} \label{sec:3.1}
\textsc{CausalPhys} comprises over 3,000 expert-curated image- and video-based questions (Fig.~\ref{tab:statis}) drawn from 11 public datasets, covering diverse physical scenarios ranging from collisions to fluid dynamics (see Appendix ~\ref{ap:a.1}).
Each benchmark instance includes a visual scene, a question, a ground-truth answer, and an expert-annotated causal graph capturing object–attribute–event dependencies. 
An illustrative example is shown in Fig.~\ref{fig:overview}.

\paragraph{Taxonomy.} 
We organize \textsc{CausalPhys} around the levels of causal understanding defined by\textbf{ Pearl’s causal ladder}~\cite{pearl2009causality}: \textit{association}, \textit{intervention}, and \textit{counterfactuals}.
This hierarchy guides \textbf{four task categories}: {Perception}, {Anticipation}, {Intervention}, and {Goal-Orientation} (Fig.~\ref{fig:taxonomy}).
\textit{Perception} corresponds to the association rung, testing whether models recognize objects and physical attributes.
\textit{Anticipation} and \textit{Intervention} probe the interventional level, assessing a model’s ability to predict or reason under explicit manipulations.
\textit{Goal-Orientation} aligns with counterfactual reasoning, inferring actions that would achieve a desired outcome. 
This causal grounding offers a principled framework for assessing how far VLMs can ascend the causal hierarchy of physical reasoning. Each of the four categories is further divided into four subcategories targeting specific \textit{physical mechanisms}, \textit{geometry}, \textit{dynamics}, \textit{interaction}, and \textit{transformation}, providing a fine-grained and systematic evaluation of VLMs across diverse aspects of the physical world.
 
% \paragraph{Structured Task Instances.} Grounding the task taxonomy in the causal hierarchy provides CausalPhys with a principled lens to reveal not only which causal levels VLMs can handle, but also where their reasoning breaks down. Each benchmark instance is represented as a structured tuple $\mathcal{I} = (X, Q, Y^{*}, \mathcal{G})$, where $X$ denotes the visual input (image or video), $Q$ is the multiple-choice question, $Y^{*}$ is the ground-truth answer, and $\mathcal{G}$ is the instance-specific causal graph encoding the underlying physical mechanism. Each question provides two to four answer options with exactly one correct choice, ensuring clean and unambiguous evaluation across models. Critically, pairing each question with a causal graph allows CausalPhys to assess not only whether a model selects the right answer, but also whether its reasoning aligns with the causal structure of the scene.

\begin{figure*}[t]
    \centering
    \includegraphics[width=1\linewidth]{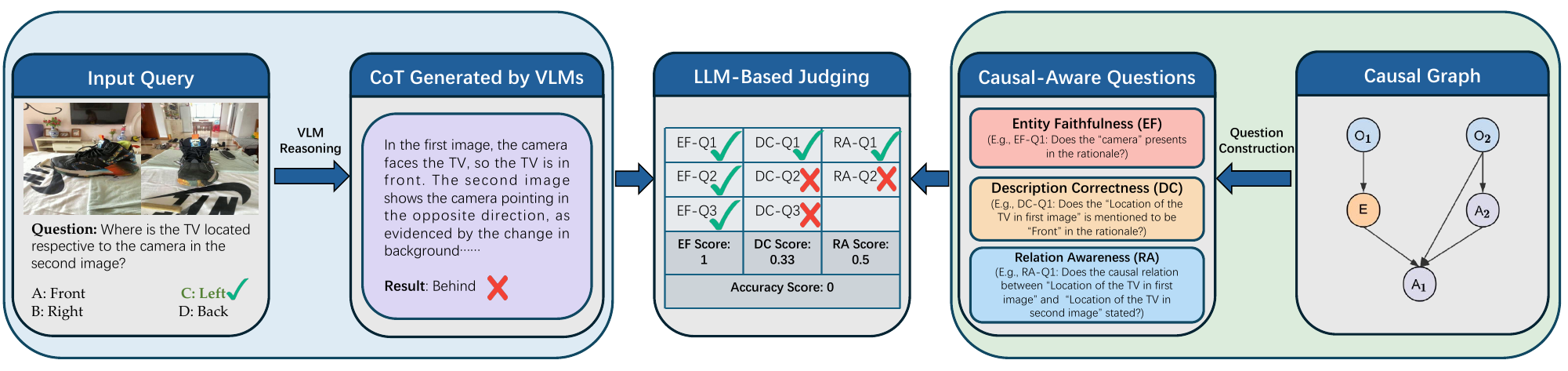} % or .png/.jpg
    \caption{\textbf{Evaluation pipeline for causal-graph-grounded metrics.} 
Given a visual input and a corresponding query, the tested VLM first produces a rationale-styled CoT and a final answer.
A judge LLM then evaluates the reasoning outputs using three causal-aware metrics: Entity Faithfulness (EF), Description Correctness (DC), and Relation Awareness (RA), and computes the final Accuracy (ACC).}
    \label{fig:metric}
   % \vspace{-0.15in}
\end{figure*}

\paragraph{Causal Graph Representation (typed DAG)}
Each question in \textbf{\textsc{CausalPhys}} is paired with a structured causal graph that exposes the underlying physical mechanism, allowing models to be evaluated on how they reason, not just whether they are correct.
Formally, each graph is a directed acyclic graph $\mathcal{G} = (\mathcal{V}, \mathcal{E})$, where nodes $\mathcal{V}$ represent causal variables and edges $\mathcal{E}$ denote directed dependencies.
To capture the heterogeneous components of physical reasoning, every node $v \in \mathcal{V}$ is defined as a typed tuple $(v_{\tau}, v_{n}, v_{d})$: $v_{\tau}$ specifies the semantic type, $v_{n}$ gives the name of entity, and the optional $v_{d}$ stores descriptive details.
Nodes are typed into three semantic classes: \textbf{Objects} ($\mathbf{O}$), representing physical entities (e.g., door, ball); \textbf{Attributes} ($\mathbf{A}$), representing their states (e.g., position, distance, velocity); and \textbf{Events} ($\mathbf{E}$), representing transformations or interactions (e.g., rotating, colliding, pouring).
This typed DAG formalism provides a unified representation of the causal dynamics that link what exists, how it changes, and why, forming the backbone of mechanism-level reasoning in {\textsc{CausalPhys}}.
% \paragraph{Typed Node Structure.}
% To capture the heterogeneous components involved in physical reasoning, each node $v \in \mathcal{V}$ is represented as a typed tuple $(v_{\tau}, v_{n}, v_{d})$. Here, $v_{\tau}$ specifies the semantic type, $v_{n}$ provides a canonical name, and the optional field $v_{d}$ stores fine-grained descriptive information. This representation allows the graph to jointly encode physical entities, their states, and the events that modify them. The type label $v_{\tau}$ assigns each node to exactly one of three semantic categories: \textbf{Objects} ($\mathbf{O}$), representing persistent physical entities (e.g., door, ball, container); \textbf{Attributes} ($\mathbf{A}$), representing state variables associated with objects (e.g., position, distance, velocity, angle); and \textbf{Events} ($\mathbf{E}$), representing physical transformations or interactions (e.g., rotating, colliding, pouring).

\paragraph{Causal Dependencies.}
Directed edges in the causal graph express how physical quantities and events influence one another. These edges cover a broad range of dependencies, including attributes describing objects, events acting on objects, events modifying attributes, and cascaded event-event interactions where one transformation triggers another. Formally, for any nodes $u, v \in \mathcal{V}$,
\[
(u, v) \in \mathcal{E}
\quad \Longleftrightarrow \quad
u \text{ is a direct cause of } v.
\]
This design yields a fine-grained and typed dependency structure that comprehensively represents the causal mechanisms underlying each physical scenario, enabling evaluation not only of \textit{what} a model predicts but also \textit{how} it arrives at that prediction.

\subsection{Data Collection Workflow} \label{sec:3.2}

We constructed \textsc{CausalPhys} through a rigorously designed and fully auditable workflow to ensure \textit{causal correctness, annotation reliability, and reproducibility} (see Appendix~\ref{ap:a.4} for full documentation).
The benchmark's visual foundations are derived from established datasets (Appendix~\ref{ap:a.2}), upon which ten STEM-background experts, including researchers in physics and robotics, synthesized all questions and causal graphs. Each annotator completed a rigorous calibration phase (Appendix~\ref{ap:a.3}) focusing on our formal causal-graph schema and physical reasoning taxonomy, ensuring high inter-annotator consistency and precise grounding of visual events into causal structures. We continuously monitored inter-annotator agreement across graph annotations, with all conflicts adjudicated by senior annotators. Comprehensive statistics on annotator profiles and agreement metrics are provided in Appendix~\ref{ap:a.1}. 

Our workflow proceeds in five tightly controlled stages:
(a) \textit{Data Acquisition}: instances sourced from 11 public datasets with complete provenance tracking;
(b) \textit{Question Formulation}: annotators design physically grounded questions paired with verified answers;
(c) \textit{Data Processing}: media are standardized and aligned with their corresponding annotations;
(d) \textit{Causal Graph Construction}: each instance is encoded from a Mermaid draft into a typed JSON causal graph with validators;
(e) \textit{Quality Assurance}: all items undergo double annotation and adjudication, removing samples with insufficient visual evidence, inconsistent annotations, or textual biases.

To ensure transparency and community extensibility, we release the entire pipeline end-to-end, including dataset selection scripts, annotation guidelines, Mermaid DAG templates, and JSON-schema validators (Appendix ~\ref{ap:a}). This level of documentation makes \textsc{CausalPhys} one of the few multimodal reasoning benchmarks that is \textit{highly transparent, auditable, and reproducible by design}.

\subsection{Evaluation Metrics} \label{sec:3.3}

Our proposed evaluation metrics move beyond traditional `answer-only' evaluation toward a mechanism-level diagnostic paradigm. To this end, we formulate a novel
\textbf{causal-graph-grounded evaluation framework} (Fig.~\ref{fig:metric}).  
Given a visual input $X$ and a query $Q$, the evaluated model produces a rationale
$R$ and a final answer $Y$ to be jointly evaluated:
\begin{equation}  
(R, Y) = \text{VLM}(X, Q).
\end{equation}  

\paragraph{LLM-based Judging.} 
% Since $R$ is natural-language text, determining whether it mentions a specific entity cannot rely on exact string matching, as semantically equivalent phrases may appear in different surface forms. We therefore employ a \emph{judge LLM} (e.g., GPT-4o~\cite{hurst2024gpt}) to provide binary (True/False) decisions for each evaluation criterion. Formally, we define a scoring function
The rationale evaluation compares the model’s generated explanation $R$, expressed in natural language, against the ground-truth causal graph.
We employ a dedicated \textit{judge LLM} (e.g., GPT-4o~\cite{hurst2024gpt}) to make binary (True/False) decisions for each evaluation item:
\[
\mathcal{M}(z, R) \in \{0,1\},
\]
where $\mathcal{M}(z, R)=1$ if the judge confirms that the target element $z$ (an entity, relation, or description) is correctly reflected in $R$.

% In our evaluation design, the judge LLM is tasked only with verifying concrete, fact-based conditions derived from the ground-truth causal graph. These checks involve determining whether the rationale correctly mentions specific entities, relations, or physical descriptions, tasks for which modern LLMs are known to perform reliably. Appendix~3.1 reports a human cross-validation study showing strong agreement between the LLM judge and human annotators, supporting the robustness of this evaluation.

The judge operates under strictly fact-based criteria derived from the ground-truth causal graph, verifying the presence and correctness of relevant entities, attributes, and causal dependencies.
This design ensures the evaluation remains both semantically aware and objectively grounded, scaling reliably across natural-language rationales.
\paragraph{Reliability of the LLM-as-Judge.}
We validate the judge against a trained human annotator and a second frontier model, and confirm its robustness to paraphrasing; the full agreement study (Cohen's $\kappa$ for LLM--Human, Cross-LLM, and paraphrase stability) is reported in Appendix~\ref{ap:reliability}.

\newcommand{\rot}[2][-25]{\rotatebox[origin=c]{#1}{#2}}
\begin{table*}[t]
\centering
\setlength{\tabcolsep}{4pt}
\renewcommand{\arraystretch}{1.15}
\Large
\resizebox{\textwidth}{!}{%
\begin{tabular}{lccccccc:cccc}
\hline
\multirow{2}{*}{\textbf{Model}}& \multicolumn{7}{c}{\textbf{Open-Source Models}} & \multicolumn{4}{c}{\textbf{Closed-Source Models}} \\
\cline{2-6} \cline{7-12}
 & \rot{\normalsize \makecell{InternVL3\cite{zhu2025internvl3exploringadvancedtraining}}} & \rot{\normalsize \makecell{Qwen3-VL \cite{yang2025qwen3technicalreport}}}
 & \rot{\normalsize \makecell{Qwen2.5-VL \cite{qwen2025qwen25technicalreport}}} & \rot{\normalsize \makecell{Qwen2-VL \cite{yang2024qwen2technicalreport}}} & \rot{\normalsize \makecell{Llama \cite{grattafiori2024llama3herdmodels}}} & \rot{\normalsize \makecell{Phi-4-Multimodal \cite{abdin2024phi4technicalreport}}}& \rot{\normalsize \makecell{Mistral-Small-3.2 \cite{mistral_small_3_2025} }}& \rot{\normalsize \makecell{GPT-4o \cite{openai2024gpt4technicalreport}}} & \rot{\normalsize \makecell{GPT-4o-mini\cite{openai2024gpt4technicalreport}}} & \rot{\normalsize \makecell{Gemini-2.5-Flash \cite{comanici2025gemini25pushingfrontier}}} & \rot{\normalsize \makecell{Claude-Sonnet-4 \cite{anthropic_claude4_2025}}}\\
\hline
\textbf{Size} & 78B & 32B & 3B & 7B & 11B & 5.6B & 24B & - & - & - & -\\
\hline
\multicolumn{12}{l}{\cellcolor{yellow!10}\faMagic\,\textbf{Anticipation}} \\
\midrule

Accuracy (ACC) $\uparrow$   & 0.5800& 0.5189& 0.2944& 0.5222& 0.3333& 0.5533& 0.4100& \textbf{0.6011}& \underline{0.5911}& 0.5822 & 0.5322\\
Entity Faithfulness (EF) $\uparrow$   & \textbf{0.6211}& 0.5910& 0.2700& 0.5100& 0.5290& 0.5570& 0.4926& \underline{0.5935}& 0.5706& 0.5820 & 0.5798\\
Relation Awareness (RA) $\uparrow$ & \underline{0.2338}& 0.2088& 0.0797& 0.1710& 0.1736& 0.1719& 0.1808& \textbf{0.2346}& 0.2021& 0.2061 & 0.2238\\
Description Correctness (DC) $\uparrow$ & \textbf{0.4243}& 0.3428& 0.1217& 0.2586& 0.2789& 0.3012& 0.2559& \underline{0.3979}& 0.3303& 0.3714 & 0.3481\\

\rowcolor{yellow!10}\multicolumn{12}{l}{\faEye\, \textbf{Perception}} \\\midrule
Accuracy (ACC) $\uparrow$    & \textbf{0.6257}& 0.5689& 0.4490& 0.5205& 0.3985& 0.5573& 0.4826& 0.5889& \underline{0.5983}& 0.5920 & 0.5868\\
Entity Faithfulness (EF) $\uparrow$  & \textbf{0.7873}& \underline{0.7822}& 0.7112& 0.6562& 0.6965& 0.7141& 0.7221& 0.7738& 0.7687& 0.7411 & 0.7349\\
Relation Awareness (RA) $\uparrow$ & \textbf{0.3976}& \underline{0.3884}& 0.2692& 0.2325& 0.2490& 0.2822& 0.3680& 0.3407& 0.3027& 0.3092 & 0.3488\\
Description Correctness (DC) $\uparrow$ & \textbf{0.4664}& 0.4049& 0.3483& 0.3133& 0.3457& 0.3501& 0.3479& \underline{0.4621}& 0.4014& 0.4574 & 0.4218\\

\rowcolor{yellow!10}\multicolumn{12}{l}{\faCogs \,\textbf{Intervention}} \\\midrule
Accuracy (ACC) $\uparrow$    & \textbf{0.5707}& 0.4799& 0.3246& 0.4764& 0.3211& 0.4852& 0.4852& \textbf{0.5707}& 0.5131& 0.5567 & \underline{0.5672}\\
Entity Faithfulness (EF) $\uparrow$  & 0.6547& 0.5941& 0.3954& 0.5563& 0.5044& 0.5501& 0.6240& \underline{0.6592}& 0.6295& 0.6451 & \textbf{0.6941}\\
Relation Awareness (RA) $\uparrow$ & \underline{0.2858}& 0.2666& 0.1493& 0.2003& 0.1762& 0.1925& 0.2483& 0.2762& 0.2496& 0.2464 & \textbf{0.3076}\\
Description Correctness (DC) $\uparrow$ & \textbf{0.5985}& 0.5516& 0.2991& 0.5661& 0.3781& 0.4659& 0.5562& 0.5742& \underline{0.5873}& 0.5827 & 0.5278\\

\rowcolor{yellow!10} \multicolumn{12}{l}{\faBullseye \, \textbf{Goal-Orientation}} \\ \midrule
Accuracy (ACC) $\uparrow$   & \underline{0.5799}& 0.5172& 0.3103& 0.4906& 0.3009& 0.4483& 0.4796& \textbf{0.5878}& 0.5157& 0.5439 & 0.4702\\
Entity Faithfulness (EF) $\uparrow$  & \textbf{0.7064}& \underline{0.6978}& 0.5372& 0.6207& 0.5762& 0.6440& 0.6858& 0.6784& 0.6764& 0.6614 & 0.6470\\
Relation Awareness (RA) $\uparrow$ & \underline{0.2855}& \textbf{0.2856}& 0.1819& 0.1931& 0.2078& 0.2036& 0.2410& 0.2376& 0.2238& 0.1966 & 0.2238\\
Description Correctness (DC) $\uparrow$ & \textbf{0.4339}& 0.3564& 0.1341& 0.2475& 0.2264& 0.3153& \underline{0.3817}& 0.3447& 0.3250& 0.3385 & 0.3439\\

\rowcolor{teal!10} \multicolumn{12}{l}{\faChartBar \, \textbf{Average}} \\\midrule
 Accuracy (ACC) $\uparrow$  & \textbf{0.5924}& 0.5268& 0.3514& 0.5065& 0.3445& 0.5199& 0.4611& \underline{0.5888}& 0.5630& 0.5725 & 0.5428\\
 Entity Faithfulness (EF) $\uparrow$  & \textbf{0.6968}& \underline{0.6795}& 0.4862& 0.5871& 0.5862& 0.6226& 0.6287& \underline{0.6795}& 0.6652& 0.6597 & 0.6634\\
 Relation Awareness (RA) $\uparrow$ & \textbf{0.3052}& 0.2914& 0.1729& 0.2002& 0.2046& 0.2166& 0.2641& 0.2760& 0.2468& 0.2437 & \underline{0.2783}\\
 Description Correctness (DC) $\uparrow$ & \textbf{0.4720}& 0.4040& 0.2278& 0.3308& 0.3073& 0.3501& 0.3669& \underline{0.4397}& 0.3994& 0.4308 & 0.4037\\

\hline
\end{tabular}%
}
\caption{\textbf{Benchmark evaluation results on \textsc{CausalPhys}.} We report performance of state-of-the-art open- and closed-source VLMs across four domains (Anticipation, Perception, Intervention, and Goal Orientation). Metrics include \textbf{Accuracy (ACC)}, \textbf{Entity Faithfulness (EF)}, \textbf{Relation Awareness (RA)}, and \textbf{Description Correctness (DC)}. Results reveal that while models achieve moderate accuracy and entity-level consistency, they struggle with relation-level reasoning (RA), indicating persistent gaps in capturing causal dependencies. These systematic weaknesses underscore the need for causally-informed approaches such as our proposed CRFT.}
\label{table3}
\end{table*} 

\begin{figure*}[t]
    \centering
    \includegraphics[width=0.78\linewidth]{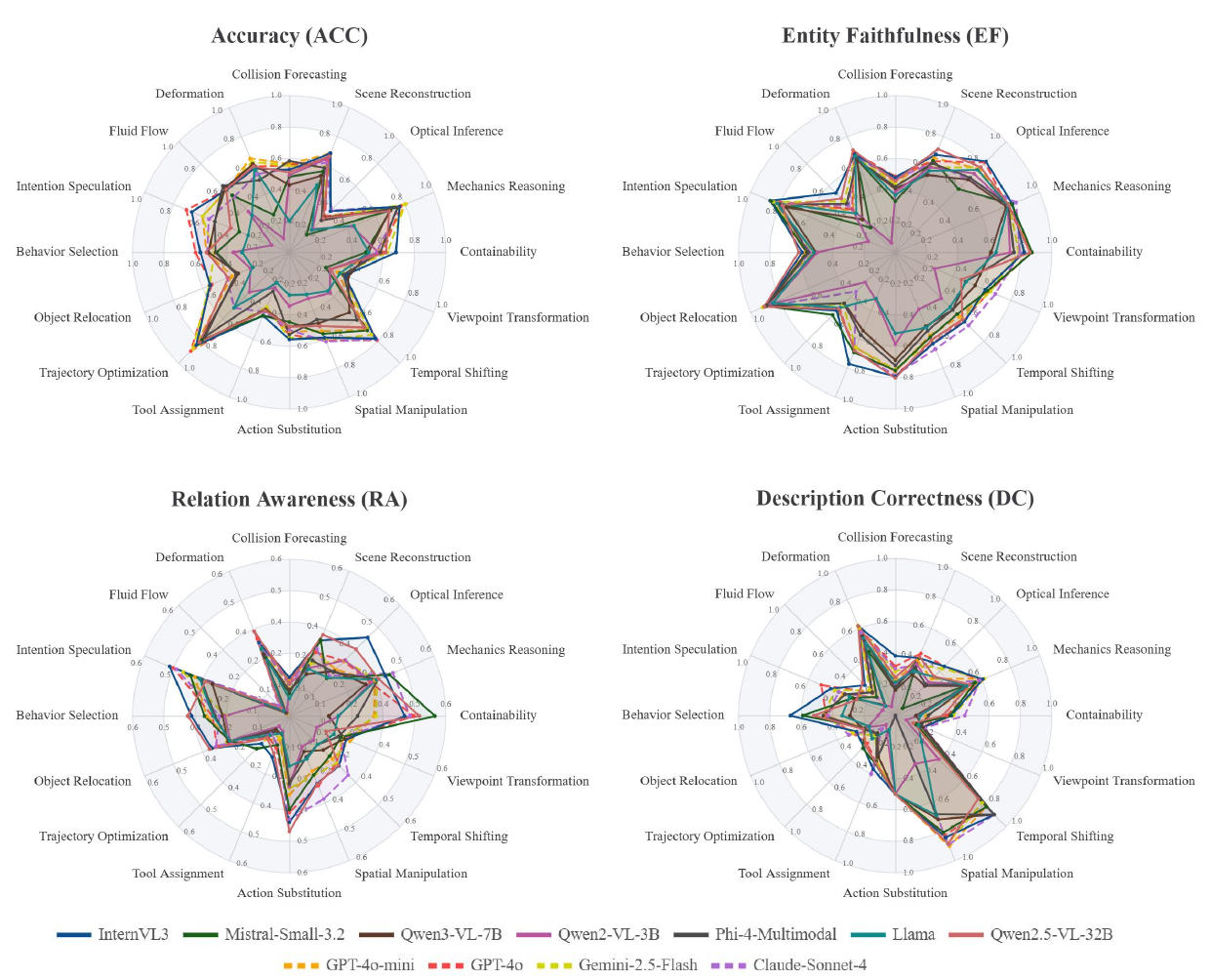} % or .png/.jpg
    \caption{\textbf{Radar plots comparing 11 tested VLMs over 16 \textsc{CausalPhys} subcategories.} Evaluation spans four causal-graph-grounded metrics: Accuracy (ACC), Relation Awareness (RA), Description Correctness (DC), and Entity Faithfulness (EF).}
    \label{fig:causalphys_overview}
    \vskip -0.1in
\end{figure*}

\paragraph{Metric Definitions.}
Beyond answer accuracy, \textsc{CausalPhys}  measures whether a model’s reasoning is causally aligned with the underlying physical mechanisms. Given the ground-truth answer $Y^{*}$ and causal graph $\mathcal{G}=(\mathcal{V},\mathcal{E})$, we define four complementary metrics that jointly capture correctness, causal grounding, and descriptive coherence:

\begin{enumerate}[leftmargin=1.5em]

    \item \textbf{Accuracy (ACC):} measures whether the predicted answer matches the ground truth.
    \begin{equation}
        \text{ACC} = \mathbb{1}\{\, Y = Y^{*} \,\}.
    \end{equation}
    \item \textbf{Entity Faithfulness (EF):} assesses whether the rationale explicitly mentions the entities in the causal graph.
    \begin{equation}
        \text{EF} = \frac{1}{|\mathcal{V}|}
          \sum_{v \in \mathcal{V}}
          \, \mathcal{M}(v_{n},R).
    \end{equation}

    \item \textbf{Relation Awareness (RA):} tests whether the rationale captures the directed causal dependencies specified in the graph.  
    \begin{equation}
        \text{RA}
        = \frac{1}{|\mathcal{E}|}
          \sum_{(u, v) \in \mathcal{E}}
         \mathcal{M}((u_n,v_n),R).
        \end{equation}
    
    \item \textbf{Description Correctness (DC):} evaluates whether the rationale describes each attribute or event consistently with its ground-truth annotation $v_{d}$.
\begin{equation}
    \text{DC}
    = \frac{1}{|A \cup E|}
      \sum_{v \in (A \cup E)}
       \mathcal{M}(v_{d}, R).
\end{equation}

\end{enumerate}
% Grounding evaluation in explicit causal graphs enables \textbf{fine-grained diagnostics of physical reasoning}, revealing not only whether VLMs are correct but also how their reasoning aligns with the true causal structure of the physical world, and where it diverges.
These metrics provide a {multi-level diagnostic of causal reasoning}, jointly capturing outcome accuracy, structural grounding, and descriptive fidelity.
% This unified evaluation framework bridges \textit{ causal graph structures} with \textit{natural-language rationales}, enabling principled and fine-grained assessment of how VLMs reason about physical world.

% \subsection{Benchmarking VLMs on Physical World Understanding}\label{sec:3.4}
\subsection{Benchmark Results of VLMs on \textsc{CausalPhys}}\label{sec:3.4}
We evaluate a broad suite of VLMs on \textsc{CausalPhys}, revealing three key patterns that characterize their strengths and limitations in causal physical reasoning.

\paragraph{(I) Understanding physical relations remains a fundamental bottleneck for VLMs.}
While modern Vision--Language Models (VLMs) demonstrate proficiency in purely perceptual tasks, their performance \textbf{precipitously declines} when reasoning necessitates an understanding of physical relations (Fig.~\ref{fig:causalphys_overview}, Table~\ref{table3}). Relation-intensive subsets, specifically viewpoint transformation and optical inference, pose the greatest challenge; most models perform \textbf{near-chance} (below 40\% or $\approx0.3$), exposing a systemic inability to integrate spatial geometry with causal dependencies. Apparent successes in tasks like trajectory prediction are often artifacts of \textbf{spurious pattern recognition} (e.g., object presence) rather than genuine causal synthesis. Consequently, the robust encoding of spatial and dynamical relations remains a critical open challenge.

\paragraph{(II) Performance parity between open-source and proprietary systems.}
A salient finding from \textsc{CausalPhys} is that open-source models perform \textbf{on par} with proprietary systems across nearly all categories (e.g., InternVL3 vs.~GPT-4o), diverging from established trends on general multimodal benchmarks. This parity suggests that proprietary scale and private datasets are \textbf{insufficient} to bridge the gap in physical reasoning. Instead, \textbf{model capacity} emerges as the primary determinant of success: smaller variants (e.g., Qwen2-VL 3B) consistently underperform relative to their larger counterparts (7B, 32B). This pattern underscores that while scaling is essential, current paradigms have yet to translate increased parameter counts into superior causal generalization.

\paragraph{(III) The persistent gap between entity recognition and relational reasoning.}
Across all categories, we identify a consistent \textbf{Entity Faithfulness (EF)-Relation Awareness (RA) gap}: models reliably identify constituent objects and attributes (high EF, $\approx 0.7$) but fail to synthesize them into coherent causal structures (low RA, $\approx 0.2$-$0.3$). This disparity highlights a \textbf{structural decoupling} in current VLMs: while the vision encoder successfully extracts individual semantic tokens, the multimodal fusion layers fail to compose these entities into a valid physical world model. This failure to bridge the gap from ``what'' to ``how'' represents a critical bottleneck for deploying VLMs in dynamic, real-world environments.

\paragraph{Statistical reliability.}
A bootstrap sensitivity analysis (10{,}000 resamples) confirms that these patterns are not artifacts of sample composition: overall accuracy and the model ranking remain stable across resamples within a narrow 95\% CI, indicating that \textsc{CausalPhys} yields statistically reliable comparisons. Full numbers and per-model statistics are reported in Appendix~\ref{ap:bootstrap}.

\section{From Answers to Reasons:  Causal Rationale Fine-Tuning }
The empirical evidence from our benchmark (Sec.~\ref{sec:3.4}) converges on a key conclusion: optimal VLM performance is achieved when models generate not only the conclusion, but also the causal structure. In light of this observation, we propose \textbf{Causal Rationale Fine-Tuning (CRFT)}, which acts as a structural regularizer for the model's reasoning manifold. By aligning latent reasoning chains with grounded causal structures, CRFT effectively mitigates the `Guesser' behavior prevalent in standard SFT, fostering zero-shot generalization and ensuring that accuracy is a byproduct of robust causal grounding rather than shallow heuristic shortcuts.
% Our benchmark analysis (Sec.~\ref{sec:2.4}) reveals a consistent pattern: 
% VLMs perform better when they articulate not only the final answer, 
% but also the causal relations that explain it. 
% Motivated by this insight, we propose \textbf{Causal Rationale Fine-Tuning (CRFT)}, 
% a training paradigm that explicitly grounds rationales in causal graphs, 
% teaching VLMs to reason through mechanisms rather than surface correlations.
\paragraph{Gold Rationale Construction.} 
Given an instance $(X,Q,Y^{*},\mathcal{G})$ as defined in Section~\ref{sec:3.1},
we generate \textbf{gold causal rationales} $R_{gold}$ utilizing a teacher LLM (e.g., GPT-4o~\cite{openai2024gpt4technicalreport}). 
Each rationale is required to (i) explicitly reference nodes and edges in $\mathcal{G}$, 
(ii) trace intermediate causal implications, and (iii) conclude with $Y^{*}$. This ensures that rationales are \textit{faithful to the causal graph}, providing structured supervision beyond free-form text.

\paragraph{Training Objective.} For training, we concatenate the gold-rationale and the ground-truth answer into a single sequence $(R_{gold},Y^{*})$ and fine-tune the target VLM $\pi_\theta$ to maximize its likelihood under a weighted supervision:  
{\normalsize
\begin{align}
      \mathcal{L}_{\mathrm{CRFT}}(\theta) \!
= \!
-\mathbb{E}\,\!\Big[
&\lambda_r \!\!\sum_{t\in \mathrm{idx}(R_{gold})} \!\!\log \pi_\theta(s_t|X,Q,s_{<t})  \notag \\
+ &\lambda_y \!\!\sum_{t\in \mathrm{idx}(Y^{*})} \!\!\log \pi_\theta(s_t|X,Q,s_{<t})
\Big],
\label{eq:crft}  
\end{align}
}
where $\lambda_r$ and $\lambda_y$ balance rationale and answer supervision. 
By anchoring fine-tuning to causal rationales, CRFT drives VLMs to \textbf{internalize causal mechanisms} instead of merely memorizing surface correlations.
% By anchoring fine-tuning to causal rationales, CRFT compels VLMs to \textbf{internalize causal pathways} rather than memorize surface correlations. 
The model is guided not just to predict the correct answer, but to trace \textit{why} the answer follows, aligning its reasoning with the ground-truth causal graph. 
This shift transforms evaluation into learning: it produces predictions that are more accurate, reasoning that is more interpretable, and models that are more reliable for physical decision-making. 
This specific alignment is critically dependent on the \textsc{CausalPhys}, where every instance comes with explicit causal structure and gold rationale, making CRFT both principled and practically feasible.  

We implement CRFT using a strategic 90/10 split of the \textsc{CausalPhys} corpus, allocating the majority for causal alignment while reserving a distinct subset for rigorous out-of-distribution evaluation. This configuration is empirically validated to be highly efficient, demonstrating that even a moderate scale of expert-guided rationales can effectively regularize the VLMs' reasoning manifold and catalyze the emergence of robust causal grounding.

\begin{figure*}[t]
    \centering
    \includegraphics[width=1\linewidth]{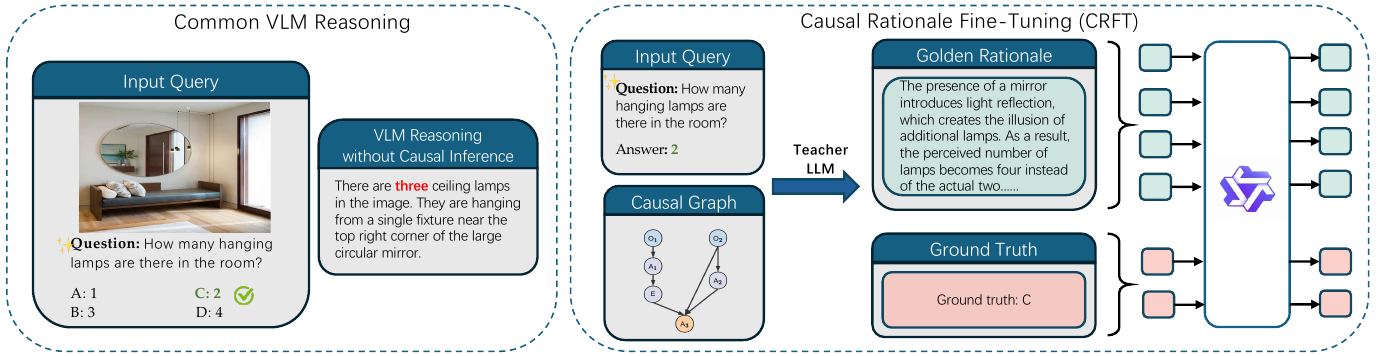} % or .png/.jpg
    \caption{\textbf{Workflow of Causal Rationale Fine-Tuning (CRFT).}
\textbf{Left}: Baseline VLM reasoning fails when causal dependencies are not modeled. \textbf{Right}: CRFT uses causal graphs and teacher-generated gold rationales to guide VLMs toward physically grounded and causally coherent reasoning, jointly supervising both the rationale and the final answer.}

    \label{fig:SFT}
    \vspace{0.1in}
\end{figure*}\label{metric}

\begin{table}[t]
\centering
\caption{
Comparison of Qwen2-VL-7B variants on \textsc{CausalPhys}. CoT-SFT is a chain-of-thought SFT baseline with rationale length matched to CRFT, isolating the effect of \emph{causal structure} from that of supervision richness.}
\label{tab:qwen2vl_results}
\renewcommand{\arraystretch}{1.05}
\resizebox{\linewidth}{!}{
\begin{tabular}{lcccc}
\toprule
\textbf{Methods} & \textbf{ACC $\uparrow$} & \textbf{EF $\uparrow$} & \textbf{RA $\uparrow$} & \textbf{DC $\uparrow$} \\
\midrule
Qwen2-VL-7B                  & 0.5349 & \textbf{0.5978} & 0.2130 & 0.2905 \\
Qwen2-VL-7B SFT (answer-only)& 0.6762 & 0.3247 & 0.0911 & 0.2645 \\
Qwen2-VL-7B CoT-SFT          & 0.6570 & 0.4280 & 0.2080 & 0.2160 \\
\midrule
Qwen2-VL-7B CRFT             & \textbf{0.7066} & 0.5969 & \textbf{0.2554} & \textbf{0.3493} \\
\bottomrule
\end{tabular}
}
\end{table}

\paragraph{Experiment results}For the experiments, we split the dataset into training and testing sets, applied uniformly across all subcategories. As shown in Table~\ref{tab:qwen2vl_results}, the SFT (answer-only fine-tuning) results reveal a critical limitation: although SFT achieves a satisfactory Accuracy (ACC) score, its performance on Entity Faithfulness (EF), Description Correctness (DC), and especially Relation Awareness (RA) drops dramatically.
% observed from the fine-tuning results, we can conclude that although SFT (answer-only fine-tuning) achieves a satisfactory Accuracy (ACC) score, its performance on Entity Faithfulness (EF), Description Correctness (DC), and especially Relation Awareness (RA) drops dramatically. 
This suggests that answer-only supervision encourages the model to optimize for surface-level prediction accuracy, but at the cost of its ability to capture and reflect the underlying causal reasoning process. In other words, SFT fine-tuning tends to make the model behave like a “guesser,” prioritizing concluding final answers based on shallow experience rather than over structured, interpretable reasoning chains.

To verify that the gains stem from \emph{causal structure} rather than longer supervision, we compare against a CoT-SFT baseline with rationale length matched to CRFT but free-form rather than graph-grounded. CoT-SFT trails CRFT on every causal metric (EF $0.4280$ vs.\ $0.5969$, RA $0.2080$ vs.\ $0.2554$), and its accuracy ($0.6570$) even dips below answer-only SFT ($0.6762$). This indicates that unstructured rationales inject noise without coherent causal grounding, and that the structured, graph-anchored signal in CRFT, not rationale length alone, is the key driver of the improvement.

In contrast, the proposed CRFT explicitly integrates causal relations into the learning strategy. The results demonstrate that CRFT not only preserves competitive accuracy but also substantially improves EF, DC, and RA scores compared to the original model, the answer-only SFT variant, and the CoT-SFT baseline. This indicates that CRFT effectively encourages the model to ground its answers in a more faithful and structured causal rationale, aligning outputs more closely with human-like reasoning.

\paragraph{Cross-benchmark generalization.}
To test whether CRFT learns transferable causal reasoning rather than overfitting to \textsc{CausalPhys}, we evaluate the CRFT-trained model on \textbf{PhysBench}~\cite{chow2025physbench}, an independent physical-reasoning benchmark, without any further tuning. As shown in Table~\ref{tab:physbench}, CRFT improves Qwen2-VL-7B by \textbf{+4.0\% overall} (58.3$\rightarrow$62.3), with consistent gains on Dynamics (+3.8) and Relationships (+8.1) and no degradation elsewhere. The largest improvement on relational reasoning mirrors the EF--RA gap identified in our benchmark, indicating that the causal grounding instilled by CRFT transfers across benchmarks rather than memorizing \textsc{CausalPhys}-specific patterns.

\begin{table}[t]
\centering
\caption{\textbf{Cross-benchmark transfer on PhysBench.} The CRFT-trained model is evaluated on PhysBench with no further tuning, improving overall accuracy by +4.0\%.}
\label{tab:physbench}
\renewcommand{\arraystretch}{1.05}
\resizebox{\linewidth}{!}{
\begin{tabular}{lccccc}
\toprule
\textbf{Model} & \textbf{Dynamics} & \textbf{Relationships} & \textbf{Property} & \textbf{Scene} & \textbf{Overall} \\
\midrule
Qwen2-VL-7B (Vanilla) & 50.6 & 73.0 & \textbf{61.7} & 54.1 & 58.3 \\
\quad + CRFT          & \textbf{54.4} & \textbf{81.1} & \textbf{61.7} & \textbf{56.8} & \textbf{62.3} \\
\bottomrule
\end{tabular}
}
\end{table}

% \paragraph{Results of CRFT} \haiyan{hurry hurry}

\section{Conclusion}
%In this paper, we present CausalPhys, a comprehensive benchmark that assesses VLMs' physical reasoning capability in real-world environments.  By incorporating explicit ground-truth causal graph annotations, CausalPhys enables more precise assessment of VLMs' physical understanding and reasoning capabilities in real-world environments.  Our systematic evaluation of existing VLMs on CausalPhys identifies substantial limitations in their physical reasoning capabilities.  Consequently, we further propose CausalRationale, a causally inspired rationale fine-tuning strategy that scaffolds VLM reasoning via teacher–student training. 
In this paper, we introduced \textsc{CausalPhys}, a benchmark that evaluates physical reasoning through expert-annotated causal graphs providing explicit mechanism-level supervision. It enables systematic and interpretable evaluation of VLM reasoning across perception, anticipation, intervention, and goal-oriented physical tasks. Empirical results show that even state-of-the-art VLMs degrade sharply when causal consistency is required, revealing a persistent gap between visual correlation and causally coherent physical reasoning. To bridge this gap, we propose {Causal Rationale Fine-Tuning (CRFT)}, a fine-tuning paradigm that explicitly aligns VLM reasoning with causal structures.

Looking forward, \textsc{CausalPhys} establishes a foundation for studying causal generalization across stochastic dynamics, multi-agent interactions, and embodied environments. As immediate next steps, we are extending this line of work toward AI assurance and data security and privacy, aiming for causally coherent, human-aligned AI systems that are trustworthy in real-world deployment.

\section{Ethical Considerations}
\label{ap:ethics}

We have carefully considered the ethical implications of constructing and releasing \textsc{CausalPhys}, and evaluation pipeline to ensure that the benchmark adheres to widely accepted standards of research integrity and social responsibility.

\begin{itemize}[leftmargin=12pt]
    \item \textbf{Informed Consent and Human Subjects.} All activities comply with the \textbf{ACM Publications Policy on Research Involving Human Participants}. We obtained \textbf{explicit informed consent} from all annotators, who were briefed on research goals and their right to withdraw at any stage. The protocol followed institutional ethical guidelines.

    \item \textbf{Annotator Welfare.} Annotators were compensated at fair-market rates. We implemented a pre-screening pipeline to ensure no exposure to harmful or offensive content, prioritizing participant psychological safety.

    \item \textbf{Data Privacy and Licensing.} Source visual data (Appendix~\ref{ap:a.2}) strictly adhere to original licensing terms, excluding any content with redistribution restrictions. We performed rigorous manual audits to remove personal identifiers, ensuring no biometric or sensitive metadata is released.

    \item \textbf{Responsible Usage.} The dataset is restricted to non-commercial research. We conducted bias audits to ensure content neutrality and provide documentation to prevent misuse in sensitive automated decision-making contexts.
\end{itemize}

\section{Acknowledgments}
This research is supported by the National Research Foundation, Singapore under its AI Singapore Programme (AISG Award No: AISG-NMLP-2024-003), the National Research Foundation, Singapore and Infocomm Media Development Authority under its Trust Tech Funding Initiative, and the National Research Foundation, Singapore under its Smart Nation \& Digital Government Translational R\&D Grant (TRANS) (Award No.\ TRANS2023-TGC03). Any opinions, findings and conclusions or recommendations expressed in this material are those of the author(s) and do not reflect the views of the National Research Foundation, Singapore, the Agency for Science, Technology and Research, or the Infocomm Media Development
Authority.

We would also like to thank Hongtian Cai, Jianghan Zhu, Chenfei Liu, Jiageng Wang, Jing Wang, Huiying Zheng, Jiali Yu, and Xuanlin Zhang for their dedicated efforts in data collection, data annotation and figure preparation, which were essential to the completion of this work.

\bibliographystyle{ACM-Reference-Format}
\bibliography{sample-base}


\begin{thebibliography}{62}

%%% the Web address.

\ifx \showCODEN    \undefined \def \showCODEN     #1{\unskip}     \fi
\ifx \showISBNx    \undefined \def \showISBNx     #1{\unskip}     \fi
\ifx \showISBNxiii \undefined \def \showISBNxiii  #1{\unskip}     \fi
\ifx \showISSN     \undefined \def \showISSN      #1{\unskip}     \fi
\ifx \showLCCN     \undefined \def \showLCCN      #1{\unskip}     \fi
\ifx \shownote     \undefined \def \shownote      #1{#1}          \fi
\ifx \showarticletitle \undefined \def \showarticletitle #1{#1}   \fi
\ifx \showURL      \undefined \def \showURL       {\relax}        \fi
% invisible to TeX
\providecommand\bibfield[2]{#2}
\providecommand\bibinfo[2]{#2}
\providecommand\natexlab[1]{#1}
\providecommand\showeprint[2][]{arXiv:#2}

\bibitem[Abdin et~al\mbox{.}(2024)]%
        {abdin2024phi4technicalreport}
\bibfield{author}{\bibinfo{person}{Marah Abdin}, \bibinfo{person}{Jyoti Aneja}, \bibinfo{person}{Harkirat Behl}, {et~al\mbox{.}}} \bibinfo{year}{2024}\natexlab{}.
\newblock \bibinfo{title}{Phi-4 Technical Report}.
\newblock
\showeprint[arxiv]{2412.08905}~[cs.CL]
\urldef\tempurl%
\url{https://arxiv.org/abs/2412.08905}
\showURL{%
\tempurl}


\bibitem[Anthropic(2025)]%
        {anthropic_claude4_2025}
\bibfield{author}{\bibinfo{person}{Anthropic}.} \bibinfo{year}{2025}\natexlab{}.
\newblock \bibinfo{title}{Introducing Claude 4}.
\newblock \bibinfo{howpublished}{\url{https://www.anthropic.com/news/claude-4}}.
\newblock
\newblock
\shownote{Accessed: 2025-09-25}.


\bibitem[Antol et~al\mbox{.}(2015)]%
        {antol2015vqa}
\bibfield{author}{\bibinfo{person}{Stanislaw Antol}, \bibinfo{person}{Aishwarya Agrawal}, \bibinfo{person}{Jiasen Lu}, \bibinfo{person}{Margaret Mitchell}, \bibinfo{person}{Dhruv Batra}, \bibinfo{person}{C~Lawrence Zitnick}, {and} \bibinfo{person}{Devi Parikh}.} \bibinfo{year}{2015}\natexlab{}.
\newblock \showarticletitle{Vqa: Visual question answering}. In \bibinfo{booktitle}{\emph{Proceedings of the IEEE international conference on computer vision}}. \bibinfo{pages}{2425--2433}.
\newblock


\bibitem[Ayman({[n.\,d.]})]%
        {ayman_cup_dataset}
\bibfield{author}{\bibinfo{person}{Samuel Ayman}.} \bibinfo{year}{[n.\,d.]}\natexlab{}.
\newblock \bibinfo{title}{{Cup Dataset [Kaggle]}}.
\newblock
\newblock
\shownote{Accessed: 2025-11-21}.


\bibitem[Azzolini et~al\mbox{.}(2025)]%
        {azzolini2025cosmos}
\bibfield{author}{\bibinfo{person}{Alisson Azzolini}, \bibinfo{person}{Junjie Bai}, \bibinfo{person}{Hannah Brandon}, \bibinfo{person}{Jiaxin Cao}, \bibinfo{person}{Prithvijit Chattopadhyay}, \bibinfo{person}{Huayu Chen}, \bibinfo{person}{Jinju Chu}, \bibinfo{person}{Yin Cui}, \bibinfo{person}{Jenna Diamond}, \bibinfo{person}{Yifan Ding}, {et~al\mbox{.}}} \bibinfo{year}{2025}\natexlab{}.
\newblock \showarticletitle{Cosmos-reason1: From physical common sense to embodied reasoning}.
\newblock \bibinfo{journal}{\emph{arXiv preprint arXiv:2503.15558}} (\bibinfo{year}{2025}).
\newblock


\bibitem[Bear et~al\mbox{.}(2021)]%
        {bear2021physion}
\bibfield{author}{\bibinfo{person}{Daniel~M Bear}, \bibinfo{person}{Elias Wang}, \bibinfo{person}{Damian Mrowca}, \bibinfo{person}{Felix~J Binder}, \bibinfo{person}{Hsiao-Yu~Fish Tung}, \bibinfo{person}{RT Pramod}, \bibinfo{person}{Cameron Holdaway}, \bibinfo{person}{Sirui Tao}, \bibinfo{person}{Kevin Smith}, \bibinfo{person}{Fan-Yun Sun}, {et~al\mbox{.}}} \bibinfo{year}{2021}\natexlab{}.
\newblock \showarticletitle{Physion: Evaluating physical prediction from vision in humans and machines}.
\newblock \bibinfo{journal}{\emph{arXiv preprint arXiv:2106.08261}} (\bibinfo{year}{2021}).
\newblock


\bibitem[Carey(2000)]%
        {carey2000origin}
\bibfield{author}{\bibinfo{person}{Susan Carey}.} \bibinfo{year}{2000}\natexlab{}.
\newblock \showarticletitle{The origin of concepts}.
\newblock \bibinfo{journal}{\emph{Journal of Cognition and Development}} \bibinfo{volume}{1}, \bibinfo{number}{1} (\bibinfo{year}{2000}), \bibinfo{pages}{37--41}.
\newblock


\bibitem[Carion et~al\mbox{.}(2020)]%
        {carion2020end}
\bibfield{author}{\bibinfo{person}{Nicolas Carion}, \bibinfo{person}{Francisco Massa}, \bibinfo{person}{Gabriel Synnaeve}, \bibinfo{person}{Nicolas Usunier}, \bibinfo{person}{Alexander Kirillov}, {and} \bibinfo{person}{Sergey Zagoruyko}.} \bibinfo{year}{2020}\natexlab{}.
\newblock \showarticletitle{End-to-end object detection with transformers}. In \bibinfo{booktitle}{\emph{European conference on computer vision}}. Springer, \bibinfo{pages}{213--229}.
\newblock


\bibitem[Chen et~al\mbox{.}(2024)]%
        {chen2024cello}
\bibfield{author}{\bibinfo{person}{Meiqi Chen}, \bibinfo{person}{Bo Peng}, \bibinfo{person}{Yan Zhang}, {and} \bibinfo{person}{Chaochao Lu}.} \bibinfo{year}{2024}\natexlab{}.
\newblock \showarticletitle{Cello: Causal evaluation of large vision-language models}.
\newblock \bibinfo{journal}{\emph{arXiv preprint arXiv:2406.19131}} (\bibinfo{year}{2024}).
\newblock


\bibitem[Chen et~al\mbox{.}(2022)]%
        {chen2022comphy}
\bibfield{author}{\bibinfo{person}{Zhenfang Chen}, \bibinfo{person}{Kexin Yi}, \bibinfo{person}{Yunzhu Li}, \bibinfo{person}{Mingyu Ding}, \bibinfo{person}{Antonio Torralba}, \bibinfo{person}{Joshua~B Tenenbaum}, {and} \bibinfo{person}{Chuang Gan}.} \bibinfo{year}{2022}\natexlab{}.
\newblock \showarticletitle{Comphy: Compositional physical reasoning of objects and events from videos}.
\newblock \bibinfo{journal}{\emph{arXiv preprint arXiv:2205.01089}} (\bibinfo{year}{2022}).
\newblock


\bibitem[Chow et~al\mbox{.}(2025)]%
        {chow2025physbench}
\bibfield{author}{\bibinfo{person}{Wei Chow}, \bibinfo{person}{Jiageng Mao}, \bibinfo{person}{Boyi Li}, \bibinfo{person}{Daniel Seita}, \bibinfo{person}{Vitor Guizilini}, {and} \bibinfo{person}{Yue Wang}.} \bibinfo{year}{2025}\natexlab{}.
\newblock \showarticletitle{Physbench: Benchmarking and enhancing vision-language models for physical world understanding}.
\newblock \bibinfo{journal}{\emph{arXiv preprint arXiv:2501.16411}} (\bibinfo{year}{2025}).
\newblock


\bibitem[Chung et~al\mbox{.}(2024)]%
        {DBLP:journals/jmlr/ChungHLZTFL00BW24}
\bibfield{author}{\bibinfo{person}{Hyung~Won Chung}, \bibinfo{person}{Le Hou}, \bibinfo{person}{Shayne Longpre}, \bibinfo{person}{Barret Zoph}, \bibinfo{person}{Yi Tay}, \bibinfo{person}{William Fedus}, \bibinfo{person}{Yunxuan Li}, \bibinfo{person}{Xuezhi Wang}, \bibinfo{person}{Mostafa Dehghani}, \bibinfo{person}{Siddhartha Brahma}, \bibinfo{person}{Albert Webson}, \bibinfo{person}{Shixiang~Shane Gu}, \bibinfo{person}{Zhuyun Dai}, \bibinfo{person}{Mirac Suzgun}, \bibinfo{person}{Xinyun Chen}, \bibinfo{person}{Aakanksha Chowdhery}, \bibinfo{person}{Alex Castro{-}Ros}, \bibinfo{person}{Marie Pellat}, \bibinfo{person}{Kevin Robinson}, \bibinfo{person}{Dasha Valter}, \bibinfo{person}{Sharan Narang}, \bibinfo{person}{Gaurav Mishra}, \bibinfo{person}{Adams Yu}, \bibinfo{person}{Vincent~Y. Zhao}, \bibinfo{person}{Yanping Huang}, \bibinfo{person}{Andrew~M. Dai}, \bibinfo{person}{Hongkun Yu}, \bibinfo{person}{Slav Petrov}, \bibinfo{person}{Ed~H. Chi}, \bibinfo{person}{Jeff Dean}, \bibinfo{person}{Jacob Devlin},
  \bibinfo{person}{Adam Roberts}, \bibinfo{person}{Denny Zhou}, \bibinfo{person}{Quoc~V. Le}, {and} \bibinfo{person}{Jason Wei}.} \bibinfo{year}{2024}\natexlab{}.
\newblock \showarticletitle{Scaling Instruction-Finetuned Language Models}.
\newblock \bibinfo{journal}{\emph{J. Mach. Learn. Res.}}  \bibinfo{volume}{25} (\bibinfo{year}{2024}), \bibinfo{pages}{70:1--70:53}.
\newblock
\urldef\tempurl%
\url{https://jmlr.org/papers/v25/23-0870.html}
\showURL{%
\tempurl}


\bibitem[Comanici et~al\mbox{.}(2025)]%
        {comanici2025gemini25pushingfrontier}
\bibfield{author}{\bibinfo{person}{Gheorghe Comanici}, \bibinfo{person}{Eric Bieber}, \bibinfo{person}{Mike Schaekermann}, {et~al\mbox{.}}} \bibinfo{year}{2025}\natexlab{}.
\newblock \bibinfo{title}{Gemini 2.5: Pushing the Frontier with Advanced Reasoning, Multimodality, Long Context, and Next Generation Agentic Capabilities}.
\newblock
\showeprint[arxiv]{2507.06261}~[cs.CL]
\urldef\tempurl%
\url{https://arxiv.org/abs/2507.06261}
\showURL{%
\tempurl}


\bibitem[Cui et~al\mbox{.}(2023)]%
        {DBLP:conf/iccv/CuiZZYWW23}
\bibfield{author}{\bibinfo{person}{Yutao Cui}, \bibinfo{person}{Chenkai Zeng}, \bibinfo{person}{Xiaoyu Zhao}, \bibinfo{person}{Yichun Yang}, \bibinfo{person}{Gangshan Wu}, {and} \bibinfo{person}{Limin Wang}.} \bibinfo{year}{2023}\natexlab{}.
\newblock \showarticletitle{SportsMOT: {A} Large Multi-Object Tracking Dataset in Multiple Sports Scenes}. In \bibinfo{booktitle}{\emph{{IEEE/CVF} International Conference on Computer Vision, {ICCV} 2023, Paris, France, October 1-6, 2023}}. \bibinfo{publisher}{{IEEE}}, \bibinfo{pages}{9887--9897}.
\newblock
\href{https://doi.org/10.1109/ICCV51070.2023.00910}{doi:\nolinkurl{10.1109/ICCV51070.2023.00910}}


\bibitem[Damen et~al\mbox{.}(2021)]%
        {Damen2021PAMI}
\bibfield{author}{\bibinfo{person}{Dima Damen}, \bibinfo{person}{Hazel Doughty}, \bibinfo{person}{Giovanni~Maria Farinella}, \bibinfo{person}{Sanja Fidler}, \bibinfo{person}{Antonino Furnari}, \bibinfo{person}{Evangelos Kazakos}, \bibinfo{person}{Davide Moltisanti}, \bibinfo{person}{Jonathan Munro}, \bibinfo{person}{Toby Perrett}, \bibinfo{person}{Will Price}, {and} \bibinfo{person}{Michael Wray}.} \bibinfo{year}{2021}\natexlab{}.
\newblock \showarticletitle{The EPIC-KITCHENS Dataset: Collection, Challenges and Baselines}.
\newblock \bibinfo{journal}{\emph{IEEE Transactions on Pattern Analysis and Machine Intelligence (TPAMI)}} \bibinfo{volume}{43}, \bibinfo{number}{11} (\bibinfo{year}{2021}), \bibinfo{pages}{4125--4141}.
\newblock
\href{https://doi.org/10.1109/TPAMI.2020.2991965}{doi:\nolinkurl{10.1109/TPAMI.2020.2991965}}


\bibitem[Dong et~al\mbox{.}(2025)]%
        {dong2025seeing}
\bibfield{author}{\bibinfo{person}{Zhuobai Dong}, \bibinfo{person}{Junchao Yi}, \bibinfo{person}{Ziyuan Zheng}, \bibinfo{person}{Haochen Han}, \bibinfo{person}{Xiangxi Zheng}, \bibinfo{person}{Alex~Jinpeng Wang}, \bibinfo{person}{Fangming Liu}, {and} \bibinfo{person}{Linjie Li}.} \bibinfo{year}{2025}\natexlab{}.
\newblock \showarticletitle{Seeing is Not Reasoning: MVPBench for Graph-based Evaluation of Multi-path Visual Physical CoT}.
\newblock \bibinfo{journal}{\emph{arXiv preprint arXiv:2505.24182}} (\bibinfo{year}{2025}).
\newblock


\bibitem[Foss et~al\mbox{.}(2025)]%
        {foss2025causalvqa}
\bibfield{author}{\bibinfo{person}{Aaron Foss}, \bibinfo{person}{Chloe Evans}, \bibinfo{person}{Sasha Mitts}, \bibinfo{person}{Koustuv Sinha}, \bibinfo{person}{Ammar Rizvi}, {and} \bibinfo{person}{Justine~T Kao}.} \bibinfo{year}{2025}\natexlab{}.
\newblock \showarticletitle{CausalVQA: A Physically Grounded Causal Reasoning Benchmark for Video Models}.
\newblock \bibinfo{journal}{\emph{arXiv preprint arXiv:2506.09943}} (\bibinfo{year}{2025}).
\newblock


\bibitem[Fu et~al\mbox{.}(2025)]%
        {fu2025unveiling}
\bibfield{author}{\bibinfo{person}{Jiarun Fu}, \bibinfo{person}{Lizhong Ding}, \bibinfo{person}{Hao Li}, \bibinfo{person}{Pengqi Li}, \bibinfo{person}{Qiuning Wei}, {and} \bibinfo{person}{Xu Chen}.} \bibinfo{year}{2025}\natexlab{}.
\newblock \showarticletitle{Unveiling and causalizing cot: A causal pespective}.
\newblock \bibinfo{journal}{\emph{arXiv preprint arXiv:2502.18239}} (\bibinfo{year}{2025}).
\newblock


\bibitem[Goyal et~al\mbox{.}(2017)]%
        {DBLP:conf/iccv/GoyalKMMWKHFYMH17}
\bibfield{author}{\bibinfo{person}{Raghav Goyal}, \bibinfo{person}{Samira~Ebrahimi Kahou}, \bibinfo{person}{Vincent Michalski}, \bibinfo{person}{Joanna Materzynska}, \bibinfo{person}{Susanne Westphal}, \bibinfo{person}{Heuna Kim}, \bibinfo{person}{Valentin Haenel}, \bibinfo{person}{Ingo Fr{\"{u}}nd}, \bibinfo{person}{Peter Yianilos}, \bibinfo{person}{Moritz Mueller{-}Freitag}, \bibinfo{person}{Florian Hoppe}, \bibinfo{person}{Christian Thurau}, \bibinfo{person}{Ingo Bax}, {and} \bibinfo{person}{Roland Memisevic}.} \bibinfo{year}{2017}\natexlab{}.
\newblock \showarticletitle{The "Something Something" Video Database for Learning and Evaluating Visual Common Sense}. In \bibinfo{booktitle}{\emph{{IEEE} International Conference on Computer Vision, {ICCV} 2017, Venice, Italy, October 22-29, 2017}}. \bibinfo{publisher}{{IEEE} Computer Society}, \bibinfo{pages}{5843--5851}.
\newblock
\href{https://doi.org/10.1109/ICCV.2017.622}{doi:\nolinkurl{10.1109/ICCV.2017.622}}


\bibitem[Grattafiori et~al\mbox{.}(2024)]%
        {grattafiori2024llama3herdmodels}
\bibfield{author}{\bibinfo{person}{Aaron Grattafiori}, \bibinfo{person}{Abhimanyu Dubey}, \bibinfo{person}{Abhinav Jauhri}, {et~al\mbox{.}}} \bibinfo{year}{2024}\natexlab{}.
\newblock \bibinfo{title}{The Llama 3 Herd of Models}.
\newblock
\showeprint[arxiv]{2407.21783}~[cs.AI]
\urldef\tempurl%
\url{https://arxiv.org/abs/2407.21783}
\showURL{%
\tempurl}


\bibitem[Grauman et~al\mbox{.}(2025)]%
        {DBLP:journals/pami/GraumanWBCCFGHJKLLMNRRR25}
\bibfield{author}{\bibinfo{person}{Kristen Grauman}, \bibinfo{person}{Andrew Westbury}, \bibinfo{person}{Eugene Byrne}, \bibinfo{person}{Vincent Cartillier}, {and} \bibinfo{person}{al Zachary Chavis~et}.} \bibinfo{year}{2025}\natexlab{}.
\newblock \showarticletitle{Ego4D: Around the World in 3,600 Hours of Egocentric Video}.
\newblock \bibinfo{journal}{\emph{{IEEE} Trans. Pattern Anal. Mach. Intell.}} \bibinfo{volume}{47}, \bibinfo{number}{11} (\bibinfo{year}{2025}), \bibinfo{pages}{9468--9509}.
\newblock
\href{https://doi.org/10.1109/TPAMI.2024.3381075}{doi:\nolinkurl{10.1109/TPAMI.2024.3381075}}


\bibitem[Gupta et~al\mbox{.}(2021)]%
        {gupta2021embodied}
\bibfield{author}{\bibinfo{person}{Agrim Gupta}, \bibinfo{person}{Silvio Savarese}, \bibinfo{person}{Surya Ganguli}, {and} \bibinfo{person}{Li Fei-Fei}.} \bibinfo{year}{2021}\natexlab{}.
\newblock \showarticletitle{Embodied intelligence via learning and evolution}.
\newblock \bibinfo{journal}{\emph{Nature communications}} \bibinfo{volume}{12}, \bibinfo{number}{1} (\bibinfo{year}{2021}), \bibinfo{pages}{5721}.
\newblock


\bibitem[Gusseme et~al\mbox{.}(2025)]%
        {degusseme2025datasetbenchmarkroboticcloth}
\bibfield{author}{\bibinfo{person}{Victor-Louis~De Gusseme}, \bibinfo{person}{Thomas Lips}, \bibinfo{person}{Remko Proesmans}, \bibinfo{person}{Julius Hietala}, \bibinfo{person}{Giwan Lee}, \bibinfo{person}{Jiyoung Choi}, \bibinfo{person}{Jeongil Choi}, \bibinfo{person}{Geon Kim}, {and} \bibinfo{person}{al Phayuth Yonrith~et}.} \bibinfo{year}{2025}\natexlab{}.
\newblock \bibinfo{title}{A Dataset and Benchmark for Robotic Cloth Unfolding Grasp Selection: The ICRA 2024 Cloth Competition}.
\newblock
\showeprint[arxiv]{2508.16749}~[cs.RO]
\urldef\tempurl%
\url{https://arxiv.org/abs/2508.16749}
\showURL{%
\tempurl}


\bibitem[Hao et~al\mbox{.}(2025)]%
        {hao2025can}
\bibfield{author}{\bibinfo{person}{Yunzhuo Hao}, \bibinfo{person}{Jiawei Gu}, \bibinfo{person}{Huichen~Will Wang}, \bibinfo{person}{Linjie Li}, \bibinfo{person}{Zhengyuan Yang}, \bibinfo{person}{Lijuan Wang}, {and} \bibinfo{person}{Yu Cheng}.} \bibinfo{year}{2025}\natexlab{}.
\newblock \showarticletitle{Can mllms reason in multimodality? emma: An enhanced multimodal reasoning benchmark}.
\newblock \bibinfo{journal}{\emph{arXiv preprint arXiv:2501.05444}} (\bibinfo{year}{2025}).
\newblock


\bibitem[He et~al\mbox{.}(2024)]%
        {he2024olympiadbench}
\bibfield{author}{\bibinfo{person}{Chaoqun He}, \bibinfo{person}{Renjie Luo}, \bibinfo{person}{Yuzhuo Bai}, \bibinfo{person}{Shengding Hu}, \bibinfo{person}{Zhen~Leng Thai}, \bibinfo{person}{Junhao Shen}, \bibinfo{person}{Jinyi Hu}, \bibinfo{person}{Xu Han}, \bibinfo{person}{Yujie Huang}, \bibinfo{person}{Yuxiang Zhang}, {et~al\mbox{.}}} \bibinfo{year}{2024}\natexlab{}.
\newblock \showarticletitle{Olympiadbench: A challenging benchmark for promoting agi with olympiad-level bilingual multimodal scientific problems}.
\newblock \bibinfo{journal}{\emph{arXiv preprint arXiv:2402.14008}} (\bibinfo{year}{2024}).
\newblock


\bibitem[Hurst et~al\mbox{.}(2024)]%
        {hurst2024gpt}
\bibfield{author}{\bibinfo{person}{Aaron Hurst}, \bibinfo{person}{Adam Lerer}, \bibinfo{person}{Adam~P Goucher}, \bibinfo{person}{Adam Perelman}, \bibinfo{person}{Aditya Ramesh}, \bibinfo{person}{Aidan Clark}, \bibinfo{person}{AJ Ostrow}, \bibinfo{person}{Akila Welihinda}, \bibinfo{person}{Alan Hayes}, \bibinfo{person}{Alec Radford}, {et~al\mbox{.}}} \bibinfo{year}{2024}\natexlab{}.
\newblock \showarticletitle{Gpt-4o system card}.
\newblock \bibinfo{journal}{\emph{arXiv preprint arXiv:2410.21276}} (\bibinfo{year}{2024}).
\newblock


\bibitem[Jiang et~al\mbox{.}(2025)]%
        {DBLP:journals/corr/abs-2502-09621}
\bibfield{author}{\bibinfo{person}{Dongzhi Jiang}, \bibinfo{person}{Renrui Zhang}, \bibinfo{person}{Ziyu Guo}, \bibinfo{person}{Yanwei Li}, \bibinfo{person}{Yu Qi}, \bibinfo{person}{Xinyan Chen}, \bibinfo{person}{Liuhui Wang}, \bibinfo{person}{Jianhan Jin}, \bibinfo{person}{Claire Guo}, \bibinfo{person}{Shen Yan}, \bibinfo{person}{Bo Zhang}, \bibinfo{person}{Chaoyou Fu}, \bibinfo{person}{Peng Gao}, {and} \bibinfo{person}{Hongsheng Li}.} \bibinfo{year}{2025}\natexlab{}.
\newblock \showarticletitle{MME-CoT: Benchmarking Chain-of-Thought in Large Multimodal Models for Reasoning Quality, Robustness, and Efficiency}.
\newblock \bibinfo{journal}{\emph{CoRR}}  \bibinfo{volume}{abs/2502.09621} (\bibinfo{year}{2025}).
\newblock
\showeprint[arXiv]{2502.09621}
\href{https://doi.org/10.48550/ARXIV.2502.09621}{doi:\nolinkurl{10.48550/ARXIV.2502.09621}}


\bibitem[Jiang et~al\mbox{.}(2024)]%
        {jiang2024visscience}
\bibfield{author}{\bibinfo{person}{Zhihuan Jiang}, \bibinfo{person}{Zhen Yang}, \bibinfo{person}{Jinhao Chen}, \bibinfo{person}{Zhengxiao Du}, \bibinfo{person}{Weihan Wang}, \bibinfo{person}{Bin Xu}, {and} \bibinfo{person}{Jie Tang}.} \bibinfo{year}{2024}\natexlab{}.
\newblock \showarticletitle{Visscience: An extensive benchmark for evaluating k12 educational multi-modal scientific reasoning}.
\newblock \bibinfo{journal}{\emph{arXiv preprint arXiv:2409.13730}} (\bibinfo{year}{2024}).
\newblock


\bibitem[Jin et~al\mbox{.}(2023)]%
        {jin2023cladder}
\bibfield{author}{\bibinfo{person}{Zhijing Jin}, \bibinfo{person}{Yuen Chen}, \bibinfo{person}{Felix Leeb}, \bibinfo{person}{Luigi Gresele}, \bibinfo{person}{Ojasv Kamal}, \bibinfo{person}{Zhiheng Lyu}, \bibinfo{person}{Kevin Blin}, \bibinfo{person}{Fernando Gonzalez~Adauto}, \bibinfo{person}{Max Kleiman-Weiner}, \bibinfo{person}{Mrinmaya Sachan}, {et~al\mbox{.}}} \bibinfo{year}{2023}\natexlab{}.
\newblock \showarticletitle{Cladder: Assessing causal reasoning in language models}.
\newblock \bibinfo{journal}{\emph{Advances in Neural Information Processing Systems}}  \bibinfo{volume}{36} (\bibinfo{year}{2023}), \bibinfo{pages}{31038--31065}.
\newblock


\bibitem[Jiralerspong et~al\mbox{.}(2024)]%
        {jiralerspong2024efficient}
\bibfield{author}{\bibinfo{person}{Thomas Jiralerspong}, \bibinfo{person}{Xiaoyin Chen}, \bibinfo{person}{Yash More}, \bibinfo{person}{Vedant Shah}, {and} \bibinfo{person}{Yoshua Bengio}.} \bibinfo{year}{2024}\natexlab{}.
\newblock \showarticletitle{Efficient causal graph discovery using large language models}.
\newblock \bibinfo{journal}{\emph{arXiv preprint arXiv:2402.01207}} (\bibinfo{year}{2024}).
\newblock


\bibitem[Kantine({[n.\,d.]})]%
        {kantine_domotic_pouringCoffee_expert}
\bibfield{author}{\bibinfo{person}{Kantine}.} \bibinfo{year}{[n.\,d.]}\natexlab{}.
\newblock \bibinfo{title}{{DOMOTIC PouringCoffee Expert Dataset [Hugging Face]}}.
\newblock
\newblock
\shownote{Accessed: 2025-11-20}.


\bibitem[Komanduri et~al\mbox{.}(2025)]%
        {komanduri2025causalvlbench}
\bibfield{author}{\bibinfo{person}{Aneesh Komanduri}, \bibinfo{person}{Karuna Bhaila}, {and} \bibinfo{person}{Xintao Wu}.} \bibinfo{year}{2025}\natexlab{}.
\newblock \showarticletitle{CausalVLBench: Benchmarking Visual Causal Reasoning in Large Vision-Language Models}.
\newblock \bibinfo{journal}{\emph{arXiv preprint arXiv:2506.11034}} (\bibinfo{year}{2025}).
\newblock


\bibitem[Kuosmanen({[n.\,d.]})]%
        {villekuosmanen_agilex_clean_pour_water}
\bibfield{author}{\bibinfo{person}{Ville Kuosmanen}.} \bibinfo{year}{[n.\,d.]}\natexlab{}.
\newblock \bibinfo{title}{{AGILEX Clean Pour Water Dataset [Hugging Face]}}.
\newblock
\newblock
\shownote{Accessed: 2025-11-20}.


\bibitem[Li et~al\mbox{.}(2023)]%
        {li2023blip}
\bibfield{author}{\bibinfo{person}{Junnan Li}, \bibinfo{person}{Dongxu Li}, \bibinfo{person}{Silvio Savarese}, {and} \bibinfo{person}{Steven Hoi}.} \bibinfo{year}{2023}\natexlab{}.
\newblock \showarticletitle{Blip-2: Bootstrapping language-image pre-training with frozen image encoders and large language models}. In \bibinfo{booktitle}{\emph{International conference on machine learning}}. PMLR, \bibinfo{pages}{19730--19742}.
\newblock


\bibitem[Li et~al\mbox{.}(2022)]%
        {li2022blip}
\bibfield{author}{\bibinfo{person}{Junnan Li}, \bibinfo{person}{Dongxu Li}, \bibinfo{person}{Caiming Xiong}, {and} \bibinfo{person}{Steven Hoi}.} \bibinfo{year}{2022}\natexlab{}.
\newblock \showarticletitle{Blip: Bootstrapping language-image pre-training for unified vision-language understanding and generation}. In \bibinfo{booktitle}{\emph{International conference on machine learning}}. PMLR, \bibinfo{pages}{12888--12900}.
\newblock


\bibitem[Li et~al\mbox{.}(2024)]%
        {li2024proximity}
\bibfield{author}{\bibinfo{person}{Jianing Li}, \bibinfo{person}{Xi Nan}, \bibinfo{person}{Ming Lu}, \bibinfo{person}{Li Du}, {and} \bibinfo{person}{Shanghang Zhang}.} \bibinfo{year}{2024}\natexlab{}.
\newblock \showarticletitle{Proximity qa: Unleashing the power of multi-modal large language models for spatial proximity analysis}.
\newblock \bibinfo{journal}{\emph{arXiv preprint arXiv:2401.17862}} (\bibinfo{year}{2024}).
\newblock


\bibitem[Li et~al\mbox{.}(2025)]%
        {li-etal-2025-multimodal-causal}
\bibfield{author}{\bibinfo{person}{Zhiyuan Li}, \bibinfo{person}{Heng Wang}, \bibinfo{person}{Dongnan Liu}, \bibinfo{person}{Chaoyi Zhang}, \bibinfo{person}{Ao Ma}, \bibinfo{person}{Jieting Long}, {and} \bibinfo{person}{Weidong Cai}.} \bibinfo{year}{2025}\natexlab{}.
\newblock \showarticletitle{Multimodal Causal Reasoning Benchmark: Challenging Multimodal Large Language Models to Discern Causal Links Across Modalities}. In \bibinfo{booktitle}{\emph{Findings of the Association for Computational Linguistics: ACL 2025}}. \bibinfo{publisher}{Association for Computational Linguistics}, \bibinfo{address}{Vienna, Austria}, \bibinfo{pages}{5509--5533}.
\newblock
\showISBNx{979-8-89176-256-5}
\href{https://doi.org/10.18653/v1/2025.findings-acl.288}{doi:\nolinkurl{10.18653/v1/2025.findings-acl.288}}


\bibitem[Liu et~al\mbox{.}(2025)]%
        {liu2025causal3d}
\bibfield{author}{\bibinfo{person}{Disheng Liu}, \bibinfo{person}{Yiran Qiao}, \bibinfo{person}{Wuche Liu}, \bibinfo{person}{Yiren Lu}, \bibinfo{person}{Yunlai Zhou}, \bibinfo{person}{Tuo Liang}, \bibinfo{person}{Yu Yin}, {and} \bibinfo{person}{Jing Ma}.} \bibinfo{year}{2025}\natexlab{}.
\newblock \showarticletitle{Causal3d: A comprehensive benchmark for causal learning from visual data}.
\newblock \bibinfo{journal}{\emph{arXiv preprint arXiv:2503.04852}} (\bibinfo{year}{2025}).
\newblock


\bibitem[Lu et~al\mbox{.}(2022)]%
        {lu2022learn}
\bibfield{author}{\bibinfo{person}{Pan Lu}, \bibinfo{person}{Swaroop Mishra}, \bibinfo{person}{Tanglin Xia}, \bibinfo{person}{Liang Qiu}, \bibinfo{person}{Kai-Wei Chang}, \bibinfo{person}{Song-Chun Zhu}, \bibinfo{person}{Oyvind Tafjord}, \bibinfo{person}{Peter Clark}, {and} \bibinfo{person}{Ashwin Kalyan}.} \bibinfo{year}{2022}\natexlab{}.
\newblock \showarticletitle{Learn to explain: Multimodal reasoning via thought chains for science question answering}.
\newblock \bibinfo{journal}{\emph{Advances in Neural Information Processing Systems}}  \bibinfo{volume}{35} (\bibinfo{year}{2022}), \bibinfo{pages}{2507--2521}.
\newblock


\bibitem[McCloskey et~al\mbox{.}(1983)]%
        {mccloskey1983intuitive}
\bibfield{author}{\bibinfo{person}{Michael McCloskey}, \bibinfo{person}{Allyson Washburn}, {and} \bibinfo{person}{Linda Felch}.} \bibinfo{year}{1983}\natexlab{}.
\newblock \showarticletitle{Intuitive physics: the straight-down belief and its origin.}
\newblock \bibinfo{journal}{\emph{Journal of Experimental Psychology: Learning, Memory, and Cognition}} \bibinfo{volume}{9}, \bibinfo{number}{4} (\bibinfo{year}{1983}), \bibinfo{pages}{636}.
\newblock


\bibitem[Moura et~al\mbox{.}(2025)]%
        {DBLP:conf/cvpr/MouraZZ25}
\bibfield{author}{\bibinfo{person}{Daniel~C. Moura}, \bibinfo{person}{Shizhan Zhu}, {and} \bibinfo{person}{Orly Zvitia}.} \bibinfo{year}{2025}\natexlab{}.
\newblock \showarticletitle{Nexar Dashcam Collision Prediction Dataset and Challenge}. In \bibinfo{booktitle}{\emph{{IEEE/CVF} Conference on Computer Vision and Pattern Recognition Workshops, {CVPR} Workshops 2025, Nashville, TN, USA, June 11-15, 2025}}. \bibinfo{publisher}{Computer Vision Foundation / {IEEE}}, \bibinfo{pages}{2583--2591}.
\newblock
\urldef\tempurl%
\url{https://openaccess.thecvf.com/content/CVPR2025W/WAD/html/Moura\_Nexar\_Dashcam\_Collision\_Prediction\_Dataset\_and\_Challenge\_CVPRW\_2025\_paper.html}
\showURL{%
\tempurl}


\bibitem[OpenAI et~al\mbox{.}(2024)]%
        {openai2024gpt4technicalreport}
\bibfield{author}{\bibinfo{person}{OpenAI}, \bibinfo{person}{Josh Achiam}, \bibinfo{person}{Steven Adler}, \bibinfo{person}{Sandhini Agarwal}, \bibinfo{person}{Lama Ahmad}, \bibinfo{person}{Ilge Akkaya}, \bibinfo{person}{Florencia~Leoni Aleman}, {and} \bibinfo{person}{et al}.} \bibinfo{year}{2024}\natexlab{}.
\newblock \bibinfo{title}{GPT-4 Technical Report}.
\newblock
\showeprint[arxiv]{2303.08774}~[cs.CL]
\urldef\tempurl%
\url{https://arxiv.org/abs/2303.08774}
\showURL{%
\tempurl}


\bibitem[Pearl(2009)]%
        {pearl2009causality}
\bibfield{author}{\bibinfo{person}{Judea Pearl}.} \bibinfo{year}{2009}\natexlab{}.
\newblock \bibinfo{booktitle}{\emph{Causality}}.
\newblock \bibinfo{publisher}{Cambridge university press}.
\newblock


\bibitem[Qwen et~al\mbox{.}(2025)]%
        {qwen2025qwen25technicalreport}
\bibfield{author}{\bibinfo{person}{Qwen}, \bibinfo{person}{:}, \bibinfo{person}{An Yang}, \bibinfo{person}{Baosong Yang}, \bibinfo{person}{Beichen Zhang}, {et~al\mbox{.}}} \bibinfo{year}{2025}\natexlab{}.
\newblock \bibinfo{title}{Qwen2.5 Technical Report}.
\newblock
\showeprint[arxiv]{2412.15115}~[cs.CL]
\urldef\tempurl%
\url{https://arxiv.org/abs/2412.15115}
\showURL{%
\tempurl}


\bibitem[Radford et~al\mbox{.}(2021)]%
        {radford2021learning}
\bibfield{author}{\bibinfo{person}{Alec Radford}, \bibinfo{person}{Jong~Wook Kim}, \bibinfo{person}{Chris Hallacy}, \bibinfo{person}{Aditya Ramesh}, \bibinfo{person}{Gabriel Goh}, \bibinfo{person}{Sandhini Agarwal}, \bibinfo{person}{Girish Sastry}, \bibinfo{person}{Amanda Askell}, \bibinfo{person}{Pamela Mishkin}, \bibinfo{person}{Jack Clark}, {et~al\mbox{.}}} \bibinfo{year}{2021}\natexlab{}.
\newblock \showarticletitle{Learning transferable visual models from natural language supervision}. In \bibinfo{booktitle}{\emph{International conference on machine learning}}. PmLR, \bibinfo{pages}{8748--8763}.
\newblock


\bibitem[Rajendran et~al\mbox{.}(2024)]%
        {rajendran2024learning}
\bibfield{author}{\bibinfo{person}{Goutham Rajendran}, \bibinfo{person}{Simon Buchholz}, \bibinfo{person}{Bryon Aragam}, \bibinfo{person}{Bernhard Sch{\"o}lkopf}, {and} \bibinfo{person}{Pradeep Ravikumar}.} \bibinfo{year}{2024}\natexlab{}.
\newblock \showarticletitle{Learning interpretable concepts: Unifying causal representation learning and foundation models}.
\newblock \bibinfo{journal}{\emph{arXiv preprint arXiv:2402.09236}} (\bibinfo{year}{2024}).
\newblock


\bibitem[Shiri et~al\mbox{.}(2024)]%
        {shiri2024empirical}
\bibfield{author}{\bibinfo{person}{Fatemeh Shiri}, \bibinfo{person}{Xiao-Yu Guo}, \bibinfo{person}{Mona~Golestan Far}, \bibinfo{person}{Xin Yu}, \bibinfo{person}{Gholamreza Haffari}, {and} \bibinfo{person}{Yuan-Fang Li}.} \bibinfo{year}{2024}\natexlab{}.
\newblock \showarticletitle{An empirical analysis on spatial reasoning capabilities of large multimodal models}.
\newblock \bibinfo{journal}{\emph{arXiv preprint arXiv:2411.06048}} (\bibinfo{year}{2024}).
\newblock


\bibitem[Srivastava et~al\mbox{.}(2022)]%
        {srivastava2022behavior}
\bibfield{author}{\bibinfo{person}{Sanjana Srivastava}, \bibinfo{person}{Chengshu Li}, \bibinfo{person}{Michael Lingelbach}, \bibinfo{person}{Roberto Mart{\'\i}n-Mart{\'\i}n}, \bibinfo{person}{Fei Xia}, \bibinfo{person}{Kent~Elliott Vainio}, \bibinfo{person}{Zheng Lian}, \bibinfo{person}{Cem Gokmen}, \bibinfo{person}{Shyamal Buch}, \bibinfo{person}{Karen Liu}, {et~al\mbox{.}}} \bibinfo{year}{2022}\natexlab{}.
\newblock \showarticletitle{Behavior: Benchmark for everyday household activities in virtual, interactive, and ecological environments}. In \bibinfo{booktitle}{\emph{Conference on robot learning}}. PMLR, \bibinfo{pages}{477--490}.
\newblock


\bibitem[Team(2025)]%
        {mistral_small_3_2025}
\bibfield{author}{\bibinfo{person}{Mistral~AI Team}.} \bibinfo{year}{2025}\natexlab{}.
\newblock \bibinfo{title}{Mistral Small 3: Apache 2.0, 81\% MMLU, 150 tokens/s}.
\newblock \bibinfo{howpublished}{\url{https://mistral.ai/news/mistral-small-3}}.
\newblock
\newblock
\shownote{Accessed: 2025-09-25}.


\bibitem[Tung et~al\mbox{.}(2023)]%
        {tung2023physion++}
\bibfield{author}{\bibinfo{person}{Hsiao-Yu Tung}, \bibinfo{person}{Mingyu Ding}, \bibinfo{person}{Zhenfang Chen}, \bibinfo{person}{Daniel Bear}, \bibinfo{person}{Chuang Gan}, \bibinfo{person}{Josh Tenenbaum}, \bibinfo{person}{Dan Yamins}, \bibinfo{person}{Judith Fan}, {and} \bibinfo{person}{Kevin Smith}.} \bibinfo{year}{2023}\natexlab{}.
\newblock \showarticletitle{Physion++: Evaluating physical scene understanding that requires online inference of different physical properties}.
\newblock \bibinfo{journal}{\emph{Advances in Neural Information Processing Systems}}  \bibinfo{volume}{36} (\bibinfo{year}{2023}), \bibinfo{pages}{67048--67068}.
\newblock


\bibitem[Van~Engelenburg et~al\mbox{.}(2024)]%
        {van2024msd}
\bibfield{author}{\bibinfo{person}{Casper Van~Engelenburg}, \bibinfo{person}{Fatemeh Mostafavi}, \bibinfo{person}{Emanuel Kuhn}, \bibinfo{person}{Yuntae Jeon}, \bibinfo{person}{Michael Franzen}, \bibinfo{person}{Matthias Standfest}, \bibinfo{person}{Jan van Gemert}, {and} \bibinfo{person}{Seyran Khademi}.} \bibinfo{year}{2024}\natexlab{}.
\newblock \showarticletitle{MSD: A Benchmark Dataset for Floor Plan Generation of Building Complexes}. In \bibinfo{booktitle}{\emph{European Conference on Computer Vision}}. Springer, \bibinfo{pages}{60--75}.
\newblock


\bibitem[Wang et~al\mbox{.}(2024)]%
        {wang2024embodiedscan}
\bibfield{author}{\bibinfo{person}{Tai Wang}, \bibinfo{person}{Xiaohan Mao}, \bibinfo{person}{Chenming Zhu}, \bibinfo{person}{Runsen Xu}, \bibinfo{person}{Ruiyuan Lyu}, \bibinfo{person}{Peisen Li}, \bibinfo{person}{Xiao Chen}, \bibinfo{person}{Wenwei Zhang}, \bibinfo{person}{Kai Chen}, \bibinfo{person}{Tianfan Xue}, {et~al\mbox{.}}} \bibinfo{year}{2024}\natexlab{}.
\newblock \showarticletitle{Embodiedscan: A holistic multi-modal 3d perception suite towards embodied ai}. In \bibinfo{booktitle}{\emph{Proceedings of the IEEE/CVF Conference on Computer Vision and Pattern Recognition}}. \bibinfo{pages}{19757--19767}.
\newblock


\bibitem[Wu et~al\mbox{.}(2017)]%
        {wu2017visual}
\bibfield{author}{\bibinfo{person}{Qi Wu}, \bibinfo{person}{Damien Teney}, \bibinfo{person}{Peng Wang}, \bibinfo{person}{Chunhua Shen}, \bibinfo{person}{Anthony Dick}, {and} \bibinfo{person}{Anton Van Den~Hengel}.} \bibinfo{year}{2017}\natexlab{}.
\newblock \showarticletitle{Visual question answering: A survey of methods and datasets}.
\newblock \bibinfo{journal}{\emph{Computer Vision and Image Understanding}}  \bibinfo{volume}{163} (\bibinfo{year}{2017}), \bibinfo{pages}{21--40}.
\newblock


\bibitem[Yang et~al\mbox{.}(2025a)]%
        {yang2025qwen3technicalreport}
\bibfield{author}{\bibinfo{person}{An Yang}, \bibinfo{person}{Anfeng Li}, \bibinfo{person}{Baosong Yang}, {et~al\mbox{.}}} \bibinfo{year}{2025}\natexlab{a}.
\newblock \bibinfo{title}{Qwen3 Technical Report}.
\newblock
\showeprint[arxiv]{2505.09388}~[cs.CL]
\urldef\tempurl%
\url{https://arxiv.org/abs/2505.09388}
\showURL{%
\tempurl}


\bibitem[Yang et~al\mbox{.}(2024)]%
        {yang2024qwen2technicalreport}
\bibfield{author}{\bibinfo{person}{An Yang}, \bibinfo{person}{Baosong Yang}, \bibinfo{person}{Binyuan Hui}, {et~al\mbox{.}}} \bibinfo{year}{2024}\natexlab{}.
\newblock \bibinfo{title}{Qwen2 Technical Report}.
\newblock
\showeprint[arxiv]{2407.10671}~[cs.CL]
\urldef\tempurl%
\url{https://arxiv.org/abs/2407.10671}
\showURL{%
\tempurl}


\bibitem[Yang et~al\mbox{.}(2025b)]%
        {yang2025thinking}
\bibfield{author}{\bibinfo{person}{Jihan Yang}, \bibinfo{person}{Shusheng Yang}, \bibinfo{person}{Anjali~W Gupta}, \bibinfo{person}{Rilyn Han}, \bibinfo{person}{Li Fei-Fei}, {and} \bibinfo{person}{Saining Xie}.} \bibinfo{year}{2025}\natexlab{b}.
\newblock \showarticletitle{Thinking in space: How multimodal large language models see, remember, and recall spaces}. In \bibinfo{booktitle}{\emph{Proceedings of the Computer Vision and Pattern Recognition Conference}}. \bibinfo{pages}{10632--10643}.
\newblock


\bibitem[Yi et~al\mbox{.}(2019)]%
        {yi2019clevrer}
\bibfield{author}{\bibinfo{person}{Kexin Yi}, \bibinfo{person}{Chuang Gan}, \bibinfo{person}{Yunzhu Li}, \bibinfo{person}{Pushmeet Kohli}, \bibinfo{person}{Jiajun Wu}, \bibinfo{person}{Antonio Torralba}, {and} \bibinfo{person}{Joshua~B Tenenbaum}.} \bibinfo{year}{2019}\natexlab{}.
\newblock \showarticletitle{Clevrer: Collision events for video representation and reasoning}.
\newblock \bibinfo{journal}{\emph{arXiv preprint arXiv:1910.01442}} (\bibinfo{year}{2019}).
\newblock


\bibitem[Yin et~al\mbox{.}(2025)]%
        {DBLP:journals/corr/abs-2506-21458}
\bibfield{author}{\bibinfo{person}{Baiqiao Yin}, \bibinfo{person}{Qineng Wang}, \bibinfo{person}{Pingyue Zhang}, \bibinfo{person}{Jianshu Zhang}, \bibinfo{person}{Kangrui Wang}, \bibinfo{person}{Zihan Wang}, \bibinfo{person}{Jieyu Zhang}, \bibinfo{person}{Keshigeyan Chandrasegaran}, \bibinfo{person}{Han Liu}, \bibinfo{person}{Ranjay Krishna}, \bibinfo{person}{Saining Xie}, \bibinfo{person}{Manling Li}, \bibinfo{person}{Jiajun Wu}, {and} \bibinfo{person}{Li Fei{-}Fei}.} \bibinfo{year}{2025}\natexlab{}.
\newblock \showarticletitle{Spatial Mental Modeling from Limited Views}.
\newblock \bibinfo{journal}{\emph{CoRR}}  \bibinfo{volume}{abs/2506.21458} (\bibinfo{year}{2025}).
\newblock
\showeprint[arXiv]{2506.21458}
\href{https://doi.org/10.48550/ARXIV.2506.21458}{doi:\nolinkurl{10.48550/ARXIV.2506.21458}}


\bibitem[Zhang et~al\mbox{.}(2025)]%
        {zhang2025physreason}
\bibfield{author}{\bibinfo{person}{Xinyu Zhang}, \bibinfo{person}{Yuxuan Dong}, \bibinfo{person}{Yanrui Wu}, \bibinfo{person}{Jiaxing Huang}, \bibinfo{person}{Chengyou Jia}, \bibinfo{person}{Basura Fernando}, \bibinfo{person}{Mike~Zheng Shou}, \bibinfo{person}{Lingling Zhang}, {and} \bibinfo{person}{Jun Liu}.} \bibinfo{year}{2025}\natexlab{}.
\newblock \showarticletitle{Physreason: A comprehensive benchmark towards physics-based reasoning}.
\newblock \bibinfo{journal}{\emph{arXiv preprint arXiv:2502.12054}} (\bibinfo{year}{2025}).
\newblock


\bibitem[Zhu et~al\mbox{.}(2025b)]%
        {zhu2025internvl3exploringadvancedtraining}
\bibfield{author}{\bibinfo{person}{Jinguo Zhu}, \bibinfo{person}{Weiyun Wang}, \bibinfo{person}{Zhe Chen}, {et~al\mbox{.}}} \bibinfo{year}{2025}\natexlab{b}.
\newblock \bibinfo{title}{InternVL3: Exploring Advanced Training and Test-Time Recipes for Open-Source Multimodal Models}.
\newblock
\showeprint[arxiv]{2504.10479}~[cs.CV]
\urldef\tempurl%
\url{https://arxiv.org/abs/2504.10479}
\showURL{%
\tempurl}


\bibitem[Zhu et~al\mbox{.}(2025a)]%
        {zhu2025intokenrationalityoptimizationaccurate}
\bibfield{author}{\bibinfo{person}{Mingye Zhu}, \bibinfo{person}{Yi Liu}, \bibinfo{person}{Zheren Fu}, \bibinfo{person}{Quan Wang}, {and} \bibinfo{person}{Yongdong Zhang}.} \bibinfo{year}{2025}\natexlab{a}.
\newblock \bibinfo{title}{In-Token Rationality Optimization: Towards Accurate and Concise LLM Reasoning via Self-Feedback}.
\newblock
\showeprint[arxiv]{2511.09865}~[cs.CL]
\urldef\tempurl%
\url{https://arxiv.org/abs/2511.09865}
\showURL{%
\tempurl}


\bibitem[Zhu et~al\mbox{.}(2023)]%
        {zhu2023benchmarking}
\bibfield{author}{\bibinfo{person}{Mingwei Zhu}, \bibinfo{person}{Leigang Sha}, \bibinfo{person}{Yu Shu}, \bibinfo{person}{Kangjia Zhao}, \bibinfo{person}{Tiancheng Zhao}, {and} \bibinfo{person}{Jianwei Yin}.} \bibinfo{year}{2023}\natexlab{}.
\newblock \showarticletitle{Benchmarking sequential visual input reasoning and prediction in multimodal large language models}.
\newblock \bibinfo{journal}{\emph{arXiv preprint arXiv:2310.13473}} (\bibinfo{year}{2023}).
\newblock


\end{thebibliography}

\appendix

\clearpage
\section*{Supplementary Material}
\section{Data Collection and Annotation Guidelines} \label{ap:a}

Our dataset comprises over 3{,}000 instances covering a wide range of real-world scenarios, including household activities, sports, traffic, and cooking, making the annotation process both diverse and non-trivial. To ensure annotation quality and domain robustness, we recruited ten well-trained STEM graduate students as annotators, with each annotator responsible for roughly 300 instances spanning two assigned subcategories.

The annotation pipeline is structured into three stages:
\begin{enumerate}
\item \textbf{Data Collection}: sampling and temporally clipping raw videos from multiple publicly available datasets.
\item \textbf{Question Construction}: designing causal-grounded questions aligned with the four categories: Perception, Anticipation, Intervention, and Goal-Orientation.
\item \textbf{Data Annotation}: creating causal graphs, writing rationales, and producing final labels formatted in \texttt{JSON}.
\end{enumerate}

\subsection{Data Collection}\label{ap:a.1}

We curate data from a broad range of publicly available video and image datasets, including EPIC-Kitchens \cite{Damen2021PAMI}, SportsMOT \cite{DBLP:conf/iccv/CuiZZYWW23}, Something-Something \cite{DBLP:conf/iccv/GoyalKMMWKHFYMH17}, Ego4D \cite{DBLP:journals/pami/GraumanWBCCFGHJKLLMNRRR25}, CausalVQA \cite{foss2025causalvqa}, MSD \cite{van2024msd}, Pouring Water \cite{kantine_domotic_pouringCoffee_expert,villekuosmanen_agilex_clean_pour_water}, Nexar Collision Prediction \cite{DBLP:conf/cvpr/MouraZZ25}, MindCube \cite{DBLP:journals/corr/abs-2506-21458}, Cups \cite{ayman_cup_dataset}, and Robotic Clothes \cite{degusseme2025datasetbenchmarkroboticcloth}.  
These sources were deliberately chosen because they contain rich temporal interactions, object contacts, force-transfer patterns, and human–object manipulation dynamics, which are elements that naturally encode causal cues essential for constructing our four reasoning categories.  
Based on assigned subcategories, annotators select source materials that adequately support causal-grounded question creation. All instances are then manually reviewed and annotated following our unified guidelines. Table~\ref{table:dataset_mapping} summarizes the mapping between visual sources  and subcategories.

\paragraph{Data formats.}  
Our dataset includes three modalities: \emph{images}, \emph{image sequences}, and \emph{videos}.  
Images may originate from image-centric datasets, individual frames extracted from videos, or frames sampled from image sequences.  
Image sequences are either sourced directly from sequence-oriented datasets or constructed by sampling up to 8 evenly spaced frames from a video.  
Video instances are curated as short clips capped at 5 seconds, a choice that balances annotation effort while providing sufficient temporal context for identifying causal dependencies such as state transitions, collisions, occlusions, or tool-use dynamics.

\paragraph{Image/Video resolution.}  
Most raw footage appears in high-resolution formats (e.g., 1920\,$\times$\,1080). To ensure consistent visual appearance and simplify downstream processing, we uniformly downsample all images and videos to 442\,$\times$\,442.  
This normalization reduces cross-dataset variability, preserves essential spatial cues for causal interpretation, and maintains compatibility with GPU memory constraints during model training.

\paragraph{Scene and causal pattern diversity.}  
The assembled data collectively covers a wide spectrum of causal patterns, including force propagation, action–reaction events, viewpoint changes, occlusion–disocclusion cycles, object deformation, everyday manipulation tasks, and safety-critical scenarios (e.g., vehicle collisions).  
Such diversity ensures that the dataset captures both short-term physical dynamics and higher-level causal structures, providing a comprehensive basis for constructing causal graphs and evaluating process-level reasoning in VLMs.

% req: \usepackage{booktabs,tabularx,makecell,xcolor}
% ------- preamble 需要 -------
% \usepackage{booktabs,tabularx,array,graphicx,xcolor,makecell}

% in preamble:
% \usepackage{booktabs,tabularx,array,graphicx,xcolor,makecell}

% preamble macros assumed already defined:
% \usepackage{booktabs,tabularx,array,graphicx,xcolor,makecell}
% \newcommand{\cmark}{\textcolor{green}{\checkmark}}
% \newcolumntype{Y}{>{\centering\arraybackslash}X}
\newcolumntype{Y}{>{\centering\arraybackslash}X}
\renewcommand{\rotcol}[2][-30]{%
  \rotatebox[origin=c]{#1}{\tiny\begin{tabular}{@{}c@{}}#2\end{tabular}}%
}

% preamble macros assumed already defined:
% \usepackage{booktabs,tabularx,array,graphicx,xcolor,makecell}
% \newcommand{\cmark}{\textcolor{green}{\checkmark}}
% \newcolumntype{Y}{>{\centering\arraybackslash}X}
% \newcommand{\rotcol}[2][-30]{\rotatebox[origin=c]{#1}{\tiny\begin{tabular}{@{}c@{}}#2\end{tabular}}}

\begin{table*}[h]
\centering
\caption{\textbf{Mapping between publicly available visual sources and the subcategories of CausalPhys.} Only the raw visual inputs (video frames or images) originate from these datasets while the questions are self-designed.  Checkmark (\cmark) indicates coverage.}
\label{tab:datasets_subcategories_transposed}
\setlength{\tabcolsep}{3pt}
\renewcommand{\arraystretch}{1.15}
\scriptsize

% NOTE: increased to 11 dataset columns
\begin{tabularx}{\textwidth}{l*{11}{Y}}
\toprule
\textbf{Subcategory} &
\rotcol{\scriptsize Ego4D\cite{DBLP:journals/pami/GraumanWBCCFGHJKLLMNRRR25}} &
\rotcol{\scriptsize Causal\\\scriptsize VQA\cite{foss2025causalvqa}} &
\rotcol{\scriptsize Epic\\\scriptsize Kitchen\cite{Damen2021PAMI}} &
\rotcol{\scriptsize MSD\cite{van2024msd}} &
\rotcol{\scriptsize Pouring \\ \scriptsize Water \cite{kantine_domotic_pouringCoffee_expert,villekuosmanen_agilex_clean_pour_water},} &
\rotcol{\scriptsize Sports\\\scriptsize MOT\cite{DBLP:conf/iccv/CuiZZYWW23}} &
\rotcol{\scriptsize Nexar\\\scriptsize Collision\\\scriptsize Prediction\cite{DBLP:conf/cvpr/MouraZZ25}} &
\rotcol{\scriptsize MindCube\cite{DBLP:journals/corr/abs-2506-21458}} &
\rotcol{\scriptsize Robotic \\ \scriptsize Clothes\cite{degusseme2025datasetbenchmarkroboticcloth}} &
\rotcol{\scriptsize Something\\\scriptsize Something\cite{DBLP:conf/iccv/GoyalKMMWKHFYMH17}} &
\rotcol{\scriptsize Cups\cite{ayman_cup_dataset}} \\
% type row (tiny, unrotated)
\midrule
\textbf{Type} &
\scriptsize Video &
\scriptsize Video &
\scriptsize Video &
\scriptsize Image &
\scriptsize Seq.\ Images &
\scriptsize Video &
\scriptsize Video &
\scriptsize Seq.\ Images &
\scriptsize Video &
\scriptsize Video &
\scriptsize Image \\
\midrule
\multicolumn{12}{l}{\textbf{Perception}} \\
Scene Reconstruction      &   &   &   &   &   &   &   &   &   & \cmark &   \\
Mechanics Reasoning       & \cmark &   & \cmark &   &   & \cmark &   &   &   & \cmark &   \\
Containability            &   &   &   &   &   &   &   &   &   &   & \cmark \\
Optical Inference         &   &   &   & \cmark &   &   &   &   &   &   &   \\
\midrule
\multicolumn{12}{l}{\textbf{Intervention}} \\
Viewpoint Transformation  &   &   &   &   &   &   &   & \cmark &   &   &   \\
Spatial Manipulation      & \cmark & \cmark & \cmark &   &   & \cmark &   &   &   &   &   \\
Temporal Shifting         & \cmark & \cmark & \cmark &   &   &   &   &   &   &   &   \\
Action Substitution       &   &   &   &   &   & \cmark &   &   &   &   &   \\
\midrule
\multicolumn{12}{l}{\textbf{Goal-orientation}} \\
Object Relocation         & \cmark & \cmark & \cmark &   &   &   &   &   &   &   &   \\
Trajectory Optimization   & \cmark & \cmark & \cmark &   &   & \cmark &   &   &   &   &   \\
Tool Assignment           &   & \cmark & \cmark &   &   &   &   &   &   &   &   \\
Behavior Selection        & \cmark & \cmark & \cmark &   &   & \cmark &   &   &   &   &   \\
\midrule
\multicolumn{12}{l}{\textbf{Anticipation}} \\
Collision Forecasting     &   &   &   &   &   &   & \cmark &   &   &   &   \\
Deformation               &   &   &   &   &   &   &   &   & \cmark &   &   \\
Fluid Flow                &   &   &   &   & \cmark &   &   &   &   &   &   \\
Intention Speculation     & \cmark & \cmark & \cmark &   &   &   &   &   &   &   &   \\
\bottomrule
\label{table:dataset_mapping}
\end{tabularx}
\end{table*}

\subsection{Question Creation}\label{ap:a.2}

All CausalPhys questions, answer options, and causal graphs were \emph{entirely authored} by our annotators and do not reuse, rephrase, or re-annotate any existing benchmark questions. The only reused component is the \emph{raw visual frames}, which serve solely as the visual input.

We categorize causal-grounded questions into four major types: \textbf{Perception}, \textbf{Anticipation}, \textbf{Intervention}, and \textbf{Goal-Orientation}, each targeting a different level of the causal reasoning hierarchy.

\paragraph{Perception.}  
Perception questions assess factual understanding that requires reasoning beyond surface-level recognition. Typical templates include ``Can~\dots~be~\dots?'', ``What is~\dots?'', or ``Where is~\dots?''.  
Examples include \textit{containability} (``Can the cups be nested?''), which requires structural reasoning about object geometry, and spatial queries such as ``Where is the chair relative to the camera?’’ that may require analyzing reflections or occlusions.

\paragraph{Anticipation.}  
Anticipation questions require predicting future outcomes based on current observations. Each question includes an explicit \textit{evidence cue} from which the prediction must be derived. Common templates include ``Will~\dots?'', ``What will~\dots?'', or ``Where will~\dots?''.  
Examples include \textit{fluid flow} (``Where will the liquid flow?'') and \textit{collision prediction} (``Will the car collide?'').

\paragraph{Intervention.}  
Intervention questions introduce a hypothetical action or modification, corresponding to a causal \textit{do-operator}, and evaluate the effect under this manipulated scenario. They are typically phrased as ``If~\dots, what will~\dots?''.  
Examples include \textit{viewpoint transformation} (``If the viewpoint changes, where will the door appear?'') and \textit{action substitution} (``If the defense player could jump one meter higher, would he block the ball?'').

\paragraph{Goal-Orientation.}  
Goal-Orientation questions specify a target objective and ask for the optimal action required to achieve it. The reasoning must incorporate constraints, affordances, and physical feasibility in the scene. Templates include ``To achieve~\dots, what should~\dots?'' or ``If we want to~\dots, what should~\dots?''.  
Examples include \textit{tool assignment} (``If we want to disassemble the toy car, which tool should we use?'') and \textit{trajectory selection} (``If we want to score a goal, which direction should we shoot?'').

\paragraph{Specific Instructions}  
Annotators follow a standardized set of instructions to ensure consistency and clarity:
\begin{itemize}
    \item Each question must be grounded in an image, a sequence of images, or a video clip.
    \item All questions must be written in clear, grammatically correct English.
    \item Questions must be unambiguous and answerable exclusively via the provided multiple-choice options.
    \item Annotators must submit: (i) the question, (ii) the answer options, (iii) the correct answer, and (iv) the causal graph.
    \item All fields must strictly follow the required formatting specifications to maintain structural consistency.
\end{itemize}
\label{annotation}

\paragraph{Review Process}  
To ensure correctness, fairness, and robustness, all annotations undergo a three-stage review pipeline:
\begin{enumerate}
    \item \textit{Self-check}: initial verification by the annotator.
    \item \textit{Independent peer review}: cross-review by another annotator to detect ambiguity or logical inconsistencies.
    \item \textit{Textual/visual bias check}: each question is validated to ensure it cannot be answered using only the text or only the visual input.
\end{enumerate}

This multi-layered process ensures high-quality, causally grounded questions and promotes consistency across the benchmark.

\subsection{Annotation Structure}\label{ap:a.3}

\begin{figure*}[h]
    \centering
    \includegraphics[width=0.6\textwidth]{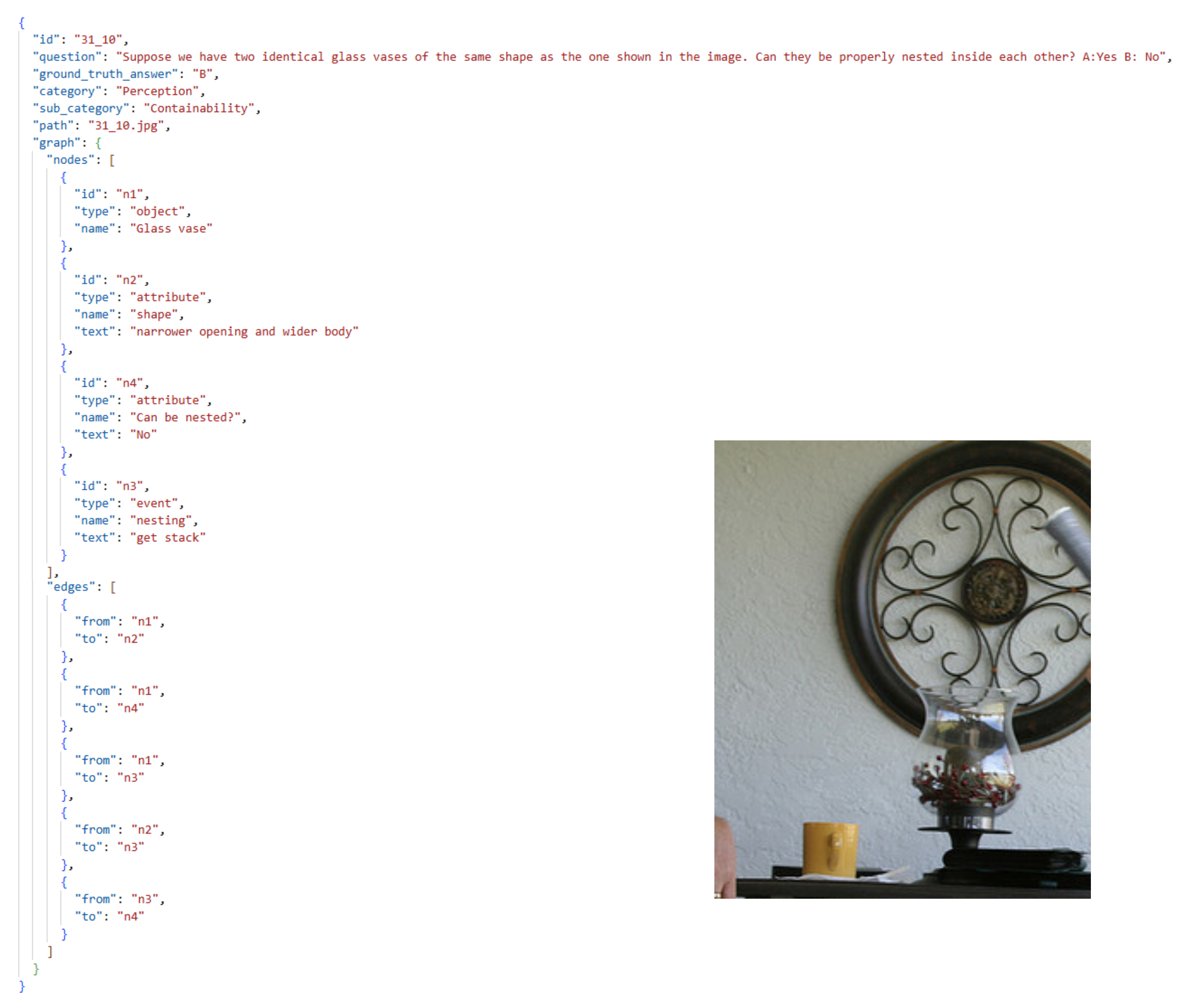}
    \caption{\textbf{An illustrative example of our JSON-based causal annotation format.} Each instance includes a question, visual input, ground-truth answer, and a structured causal graph. This format ensures that each component is easy to access, verify, and integrate into downstream evaluation pipelines.}
    \label{fig:annotaion json}
\end{figure*}

\paragraph{Annotation example}  
As illustrated in Figure~\ref{fig:annotaion json}, each annotation instance is stored in JSON format and contains the following keys:  

\begin{itemize}
    \item \texttt{"id"}: The identifier of the current data instance.  
    \item \texttt{"question"}: The question created for this instance.  
    \item \texttt{"ground\_truth\_answer"}: The ground-truth answer, denoted by one of \texttt{A|B|C|D}.  
    \item \texttt{"path"}: The path(s) to the visual data. This field supports three types of input: a single image path, a single video path, or a list of image paths representing an image sequence.  
    \item \texttt{"category"}: One of the four main categories.  
    \item \texttt{"sub\_category"}: One of the sixteen subcategories.  
    \item \texttt{"graph"}: The annotated ground-truth causal graph, consisting of \texttt{nodes} (objects, attributes, or events) and \texttt{edges} (relations between nodes).  
\end{itemize}

\paragraph{Causal Graph Annotation}

Although the final annotations are stored in \texttt{JSON}, directly constructing causal graphs in raw \texttt{JSON} is cumbersome and error-prone.  
To streamline the annotation process, we employ \texttt{Mermaid} as an intermediate graph-editing interface. Mermaid provides real-time visualization, intuitive graph manipulation, and significantly reduces structural annotation errors. 
 
For example:  
If the causal entity corresponds to the object \textit{vase}, the node is annotated as \texttt{O:Vase}. If the entity corresponds to the attribute: the shape of the vase is a narrower opening and wider body, the node is annotated as \texttt{A: shape: narrower opening and wider body}.  

\begin{figure*}[h]
    \centering
    \includegraphics[width=0.6\textwidth]{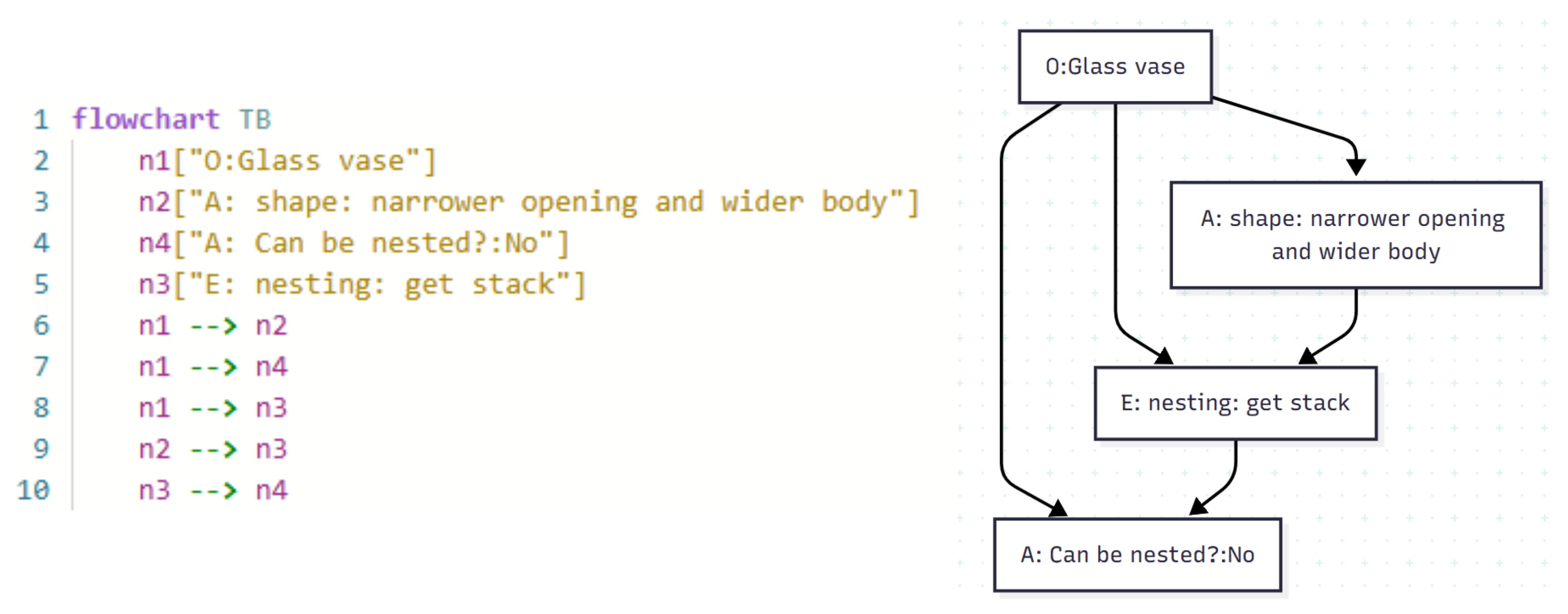}
    \caption{\textbf{Mermaid graph annotation of a causal graph}. It supports real-time visualization and easy editing, enabling annotators to interactively refine node–edge structures.This format is also highly interpretable and can be seamlessly converted to our JSON causal annotation schema.}
    \label{fig:mmd}
\end{figure*}

The annotated causal graph is saved in \texttt{.mmd} format as shown in Figure~\ref{fig:mmd} and then automatically converted into the standardized \texttt{JSON} format.  

\subsection{Annotation Platform}\label{ap:a.4}
To facilitate large-scale data labeling, we designed a dedicated annotation platform that automatically iterates through the video and image dataset, presenting each instance sequentially for annotation. This system significantly reduces manual effort by streamlining the workflow and ensuring that annotators can focus entirely on content creation rather than file handling.  

\begin{figure*}[h]
    \centering
    \fbox{\includegraphics[height=0.9\textheight]{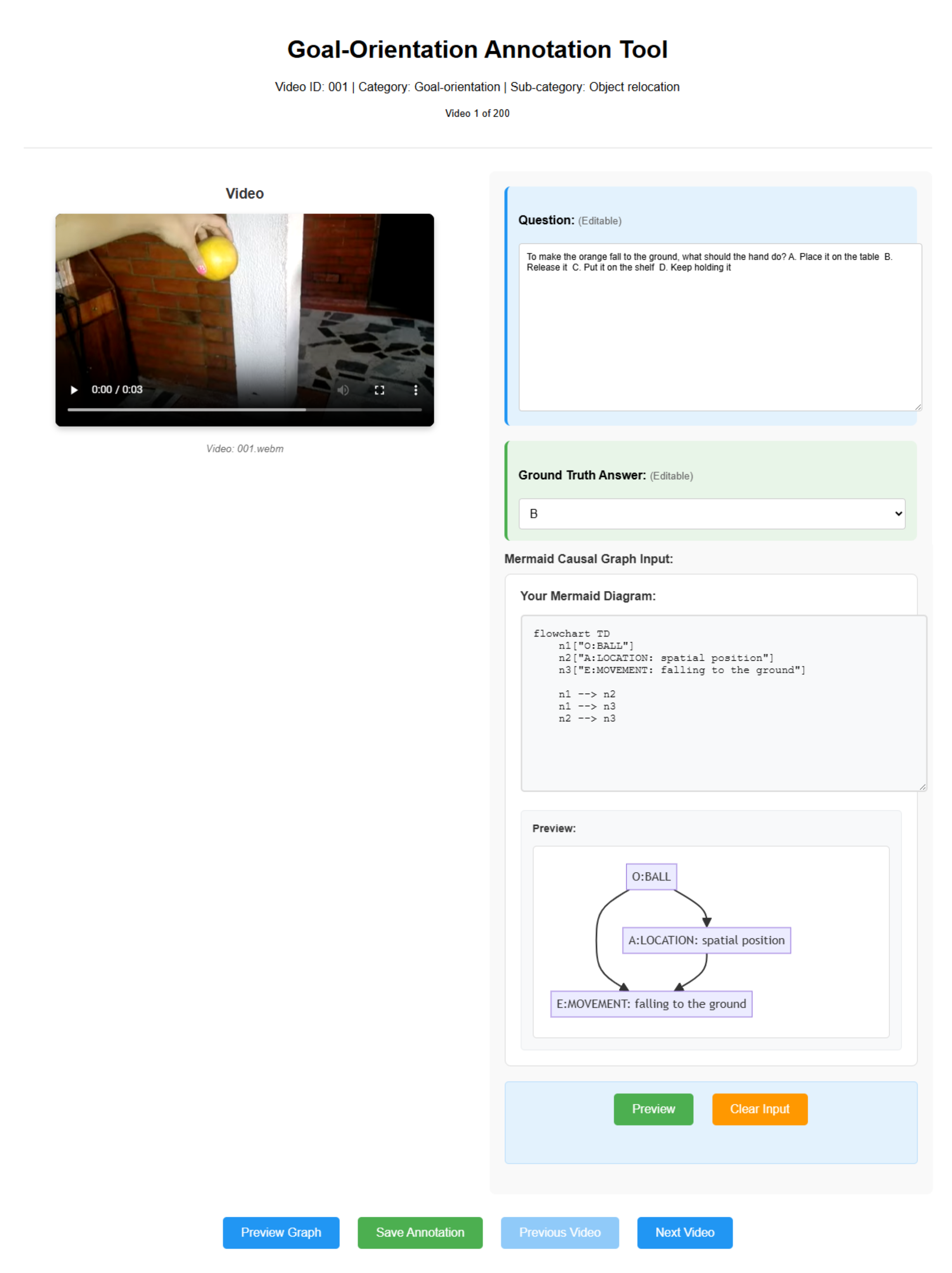}}
    \caption{\textbf{Designed GUI-based annotation platform}. It integrates visual input display, question design, causal graph editing, and rendered graph preview, providing an easily accessible and well-visualized interface that greatly reduces annotation effort. }
    \label{fig:gui}
\end{figure*}

For each data instance, the platform provides an interactive graphical user interface (GUI), as illustrated in Figure~\ref{fig:gui}. On the left-hand side of the interface, the video (or image) corresponding to the current instance is displayed, allowing annotators to carefully observe the scene. On the right-hand side, several editable blocks guide the annotation process. Specifically, annotators can:  
\begin{itemize}
    \item \textbf{Formulate a Question.} Each instance requires a question relevant to the visual content. The question is editable in a designated text block, enabling annotators to phrase it in a multiple-choice format that reflects the underlying causal or goal-oriented reasoning of the scenario.  

    \item \textbf{Define Answer Options.} Alongside the question, annotators specify multiple-choice options (e.g., actions that could be taken in the video). These options allow for structured evaluation of models on causal reasoning and decision-making.  

    \item \textbf{Select the Ground-Truth Answer.} From the defined options, annotators must identify the correct choice, which is recorded as the ground-truth label. This ensures that the dataset captures unambiguous supervisory signals for training and evaluation.  

    \item \textbf{Construct the Causal Graph.} To go beyond question-answer annotation, our platform integrates a causal graph editor using Mermaid syntax. Annotators can input nodes (representing objects, attributes, or events) and edges (capturing causal dependencies). The tool automatically renders a visual preview of the graph, allowing annotators to validate the structure before saving. This ensures that every instance is accompanied not only by a question-answer pair but also by a structured causal representation.  
\end{itemize}

GUI provides utility functions such as previewing the rendered causal graph, clearing inputs, saving the current annotation, or moving to the previous/next video in the dataset. Together, these functions make the annotation process more efficient, standardized, and less error-prone.  

Overall, this annotation platform supports a unified pipeline for curating complex datasets that combine natural video/image observations with structured causal annotations, thus laying the foundation for high-quality benchmarks in causal reasoning tasks.

% \newpage

\begin{table}[t]
\centering
\small
\setlength{\tabcolsep}{4pt}
\begin{tabular}{lcccc}
\toprule
 \textbf{Reliability Check}  & \textbf{EF} & \textbf{RA} & \textbf{DC} & \textbf{Overall} \\
\midrule
\textbf{LLM--Human Agreement ($\kappa$)} & 0.86 & 0.83 & 0.76 & 0.82 \\
\textbf{Cross-LLM Consistency ($\kappa$)} & 0.89 & 0.88 & 0.78 & 0.84 \\
\textbf{Paraphrase Stability ($\kappa$)}  & 0.92 & 0.86 & 0.96 & 0.90 \\
\bottomrule
\end{tabular}
\caption{\textbf{Reliability evaluation of the LLM-as-Judge} on 320 samples (20 per subcategory). \textbf{LLM--Human} reports Cohen's $\kappa$ between GPT-5.4 and a human annotator; \textbf{Cross-LLM} reports $\kappa$ between GPT-5.4 and Claude Sonnet~4.6; \textbf{Paraphrase Stability} reports consistency under semantically equivalent rewrites.}
\label{tab:judge_reliability}
\end{table}

\section{Additional Experiments and Analyses}\label{ap:addexp}

\subsection{LLM-as-Judge Reliability Validation} \label{ap:reliability}
Prior work shows that LLM-based evaluators provide semantically informed, objectively grounded judgments \cite{DBLP:journals/corr/abs-2502-09621}. We validate our setup on \textbf{320 instances}. While the main paper uses GPT-4o as the judge, here we re-judge with two stronger, more recent SOTA models (GPT-5.4 and Claude Sonnet~4.6) and assess three complementary axes of reliability (Table~\ref{tab:judge_reliability}). Both agreement settings exceed $0.80$ (almost-perfect), and LLM--Human agreement is comparable to Cross-LLM, confirming that the judge is human-comparable, reproducible, and not sensitive to the choice of judge model.

\paragraph{Human Cross-Validation.}
Comparing GPT-5.4 against a trained human annotator (blinded to the model outputs) on the EF/RA/DC questions yields a strong overall agreement of Cohen's $\kappa = 0.82$ (EF $0.86$, RA $0.83$, DC $0.76$), indicating human-comparable reliability when evaluating causal-graph--grounded rationales.

\paragraph{Cross-LLM Consistency.}
Re-judging the same instances with Claude Sonnet~4.6 under identical prompts gives $\kappa = 0.84$ overall (EF $0.89$, RA $0.88$, DC $0.78$). Because this matches the LLM--Human level, the evaluation is stable across frontier judges and not an artifact of the specific model.

\paragraph{Paraphrase Stability.}
Rewriting each question into three semantically equivalent paraphrases and counting a verdict as consistent only when all paraphrases agree, the judge achieves a $0.90$ consistency rate, showing strong invariance to surface wording.

\subsection{Teacher-Invariant Causal Structure}\label{ap:teacher}
To verify that CRFT learns the underlying causal structure rather than imitating the style of a specific teacher LLM, we re-generate \emph{all} training rationales with an independent teacher (Claude Sonnet~4) under the identical graph-anchored prompt and re-train Qwen2-VL-7B with the same configuration. At the rationale level the two teachers differ substantially in surface form (ROUGE-L $= 0.38$) yet convey near-identical causal content (BERTScore $= 0.91$), indicating that the supervision is graph-anchored rather than style-anchored. Consequently, the two resulting models perform comparably across all metrics (Table~\ref{tab:teacher_swap}; ACC $0.707$ vs.\ $0.713$, RA $0.255$ vs.\ $0.267$), confirming that CRFT captures \textbf{teacher-invariant causal structure} rather than teacher-specific stylistic priors.

\begin{table}[t]
\centering

\renewcommand{\arraystretch}{1.05}
\resizebox{\linewidth}{!}{
\begin{tabular}{lcccc}
\toprule
\textbf{Teacher} & \textbf{ACC $\uparrow$} & \textbf{EF $\uparrow$} & \textbf{RA $\uparrow$} & \textbf{DC $\uparrow$} \\
\midrule
CRFT (GPT-4o)          & 0.7066 & 0.5969 & 0.2554 & 0.3493 \\
CRFT (Claude Sonnet~4) & 0.7133 & 0.6133 & 0.2667 & 0.3333 \\
\bottomrule
\end{tabular}
}
\caption{\textbf{Teacher-swap analysis.} Re-generating all gold rationales with an independent teacher (Claude Sonnet~4) yields comparable CRFT performance, confirming that CRFT learns teacher-invariant causal structure. Rationale-level similarity between teachers: ROUGE-L $=0.38$, BERTScore $=0.91$.}
\label{tab:teacher_swap}
\end{table}

\subsection{Bootstrap Sensitivity Analysis}\label{ap:bootstrap}
To confirm the results are not artifacts of sample composition, we resample the evaluation set with replacement \textbf{10{,}000} times and recompute the metrics, reporting 95\% CIs as the $2.5$/$97.5$ percentiles. Over the full benchmark (\textbf{N$\,=\,$3{,}062}), overall accuracy is \textbf{58.9\%} (bootstrap std \textbf{0.88\%}, 95\% CI \textbf{[57.2\%, 60.6\%]}). The narrow interval and the stable model ranking across resamples indicate that \textsc{CausalPhys} yields statistically reliable comparisons; per-model statistics are released with the artifacts.

\newpage

\clearpage
\newpage
\clearpage

\section{Prompts}\label{ap:b}

This section describes the prompt configurations used across different LLMs and VLMs under various evaluation settings. We present the prompts for: (i) VQA with rationale, (ii) VQA answer-only, (iii) gold rationale generation, and (iv) rationale correctness evaluation.

\subsection{VQA Prompt with Rationale}\label{ap:b.1}

The VLM reasoning prompt instructs the model to produce a step-by-step rationale prior to answering the question. The agent is asked to carefully analyze the input image or image sequence, articulate its reasoning process, and then provide the final answer.  
This prompt is used for both VLM evaluation and for training and evaluating CRFT, as it explicitly elicits interpretable intermediate reasoning.

\begin{tcolorbox}[
  enhanced,
  breakable,
  colback=green!5,
  colframe=green!40!black,
  boxrule=0pt,
  arc=2mm,
  left=2mm,
  right=2mm,
  top=1mm,
  bottom=1mm,
  title=VQA Prompt with Rationale,
  fontupper=\small
]

You are a precise Vision--Language QA assistant. \\

\#\# Goals \\
- Read the user's question and (if provided) a SEQUENCE of images in the given order \\
- Provide a one-sentence rationale and your answer \\

\#\# Sequence Handling \\
- If multiple images are provided, treat them as an ordered sequence (e.g., frames of a video) \\
- Consider temporal consistency and cross-frame cues when reasoning \\

\#\# Conservative Reasoning \\
- Rely only on information available in the images and the question \\
- Be explicit and concise; avoid speculation \\

\#\# Hard Format Constraints (must obey exactly) \\
Output MUST include: \\
1. Generate a clear, step-by-step rationale (max 8 sentences) wrapped in \verb|<rationale>...</rationale>| \\
2. Your answer must be in EXACTLY ONE CAPITAL LETTER: A, B, C, or D wrapped in \verb|<result>...</result>|
\end{tcolorbox}

\subsection{Gold Rationale Generation} \label{ap:b.3}

Based on the ground-truth causal graph, the teacher LLM is required to generate a gold rationale that reflects the reasoning implied by the nodes and edges. The agent is provided with the question, causal graph,visual input and ground-truth answer. The rationale must be written in natural language as a reasoning process leading to the final answer, rather than as a description of the causal graph itself. This gold rationale will be used for further CRFT supervision.

\begin{tcolorbox}[
  enhanced,
  breakable,
  colback=blue!5,
  colframe=blue!30,
  boxrule=0pt,
  arc=2mm,
  left=2mm,
  right=2mm,
  top=1mm,
  bottom=1mm,
  title=Gold Rationale Generation Prompt,
  fontupper=\small
]
You are a reasoning assistant that analyzes visual scenarios and provides step-by-step reasoning. \\

\#\# Input format \\
You will receive: \\
- A question about the visual scenario \\
- A image or a sequence of images \\
- A ground truth answer (A, B, C, or D) \\
- Supporting information about objects, their properties, and relationships \\

\#\# Task \\
Generate a clear, step-by-step rationale that answers the question in natural language. \\

\#\# Requirements \\
1. Write an objective, answer-focused rationale in natural language \\
2. Treat the supporting information as reference only (do not describe it) \\
3. Write ONE coherent paragraph (max 8 sentences) that flows naturally \\
4. Include relevant elements from the reference only when needed for reasoning (do not enumerate them) \\
5. Follow the correct logical order: causes must appear before their effects \\
6. If an element has a description, state it clearly and exactly as provided \\
7. Use natural, everyday language (avoid terms like "entity", "relation", "graph", "structure") \\
8. Ensure proper grammar and spelling \\
9. Make the explanation easy to understand and self-contained \\
10. Present the reasoning as a logical analysis of the situation \\

\#\# Output format \\
- Single paragraph only \\
- No bullet points, lists, or special formatting \\
- Plain English text \\
- Complete explanation that follows the logical reasoning sequence \\

\#\# Important \\
The supporting information (entities, descriptions, relations) is for reference only. Do NOT describe or list it. \\
Use it implicitly to justify the answer. Focus on explaining why the answer is correct in plain language.
\end{tcolorbox}

\subsection{Rationale Correctness Judgement Prompt}\label{ap:b.4}

Based on the defined rationale evaluation metrics, three categories of questions will be constructed. The evaluator LLM will be given a rationale generated by the VLM along with a sequence of true/false questions. Its task is to evaluate the rationale by answering each question and output the results as a list in YAML format. This prompt is used in rationale evaluation.

\begin{tcolorbox}[
  enhanced,
  breakable,
  colback=purple!5,
  colframe=purple!50!black,
  boxrule=0pt,
  arc=2mm,
  left=2mm,
  right=2mm,
  top=1mm,
  bottom=1mm,
  title=LLM as a judge causal relationship prompt,
  fontupper=\small
]
You are a meticulous evaluator. Read the problem and the model's rationale, then answer a list of True/False questions strictly based on that rationale. Do not use outside knowledge or the image. If the rationale is ambiguous or does not state the fact, answer False. \\[6pt]

Answer using ONLY the specified YAML schema. Do not add extra commentary. \\[6pt]

INPUT \\[4pt]
- problem: The multiple-choice question with options \\
- rationale: The model's rationale paragraph(s) \\
- questions: A list of True/False questions. Each item has: \\
\quad - id: opaque identifier (string) \\
\quad - text: the T/F question \\[6pt]

JUDGING PRINCIPLES \\[4pt]
- True only if the rationale explicitly supports the statement with clear mention or an unambiguous entailment. \\
- False if absent, unclear, contradicted, or only weakly implied. \\
- Allow synonyms/coreference (e.g., ``kicker'' for ``Fighter''), but do not infer beyond text. \\
- For causal relation questions, require a clear causal/influence expression (e.g., X causes/leads to/affects Y; Y depends on X). Mere co-occurrence is insufficient. \\[6pt]

OUTPUT FORMAT (YAML) \\[4pt]
answers: \\
\quad - id: \verb|<string>| \\
\quad ~~answer: true$|$false \\[6pt]

EXAMPLE \\[4pt]
problem: "Which direction should he kick to hit the target? A. Left B. Right" \\
rationale: "The pad is to the left of the kicker; therefore he should kick left to hit it." \\
questions: \\
~~- id: "0" \\
~~~~text: "Is object 'pad' mentioned in the rationale?" \\
~~- id: "1" \\
~~~~text: "Is the causal relation between 'Kick direction' and 'Pad Location' correctly expressed?" \\[6pt]

answers: \\
~~- id: "0" \\
~~~~answer: true \\
~~- id: "1" \\
~~~~answer: true
\end{tcolorbox}

\section{Experiment Details}\label{ap:c}

\subsection{Metric Questions Construction} \label{ap:c.1}

In Section~3.2 of the main paper, we introduced three LLM-as-Judge evaluation metrics, Entity Faithfulness (EF), Relation Awareness (RA), and Description Correctness (DC). To reduce the degrees of freedom in free-form model outputs and ensure stable judgement, we convert each metric into a set of canonical True/False verification questions.

For any entity represented as
\[
v = (\langle \text{type}\rangle,\ \langle \text{name}\rangle,\ \langle \text{description}\rangle),
\]
and for any entity pair $(v_1, v_2)$, the corresponding questions are instantiated as follows:
\begin{itemize}
    \item \textbf{EF}: ``Does the $\langle\text{type}_1\rangle\ \langle\text{name}_1\rangle$ appear in the rationale?''
    \item \textbf{DC}: ``Is the $\langle\text{type}_1\rangle\ \langle\text{name}_1\rangle$ described as `$\langle\text{description}_1\rangle$' in the rationale?''
    \item \textbf{RA}: ``Is a direct causal relation between $\langle\text{type}_1\rangle\ \langle\text{name}_1\rangle$ and $\langle\text{type}_2\rangle\ \langle\text{name}_2\rangle$ explicitly stated in the rationale?''
\end{itemize}

% \noindent\textbf{Example.}  
% Given an event entity: 
% \[
% v = (\text{event},\ \text{viewpoint transformation},\ \text{counterclockwise }90^\circ),
% \]
% the instantiated questions become:
% \begin{itemize}
%     \item \textbf{EF}: ``Does the viewpoint transformation appear in the rationale?''
%     \item \textbf{DC}: ``Is the viewpoint transformation described as counterclockwise $90^\circ$ in the rationale?''
% \end{itemize}

% This template construction yields a controlled binary judgment space, ensuring that EF, RA, and DC can be evaluated reliably and consistently using LLM-as-Judge.

% LLM-as-Judge reliability validation has been moved to
% Appendix~\ref{ap:reliability} (Additional Experiments and Analyses).

\subsection{CRFT Details}\label{ap:c.3}

\paragraph{Graph-Anchored Rationale Construction.}
Each instance is paired with a human-annotated causal graph
$G=(V,E)$ containing object, attribute, and event nodes.
We prompt GPT-4o to convert this graph into a textual rationale that faithfully reflects its structure.
GPT-4o does \emph{not} generate the graph itself; it only verbalizes the human-provided graph, making the supervision graph-anchored rather than model-anchored.
We further verify that the resulting rationales neither introduce information absent from the graph nor directly recite the graph.

\begin{figure}[!t]
    \centering
    \includegraphics[width=1\linewidth]{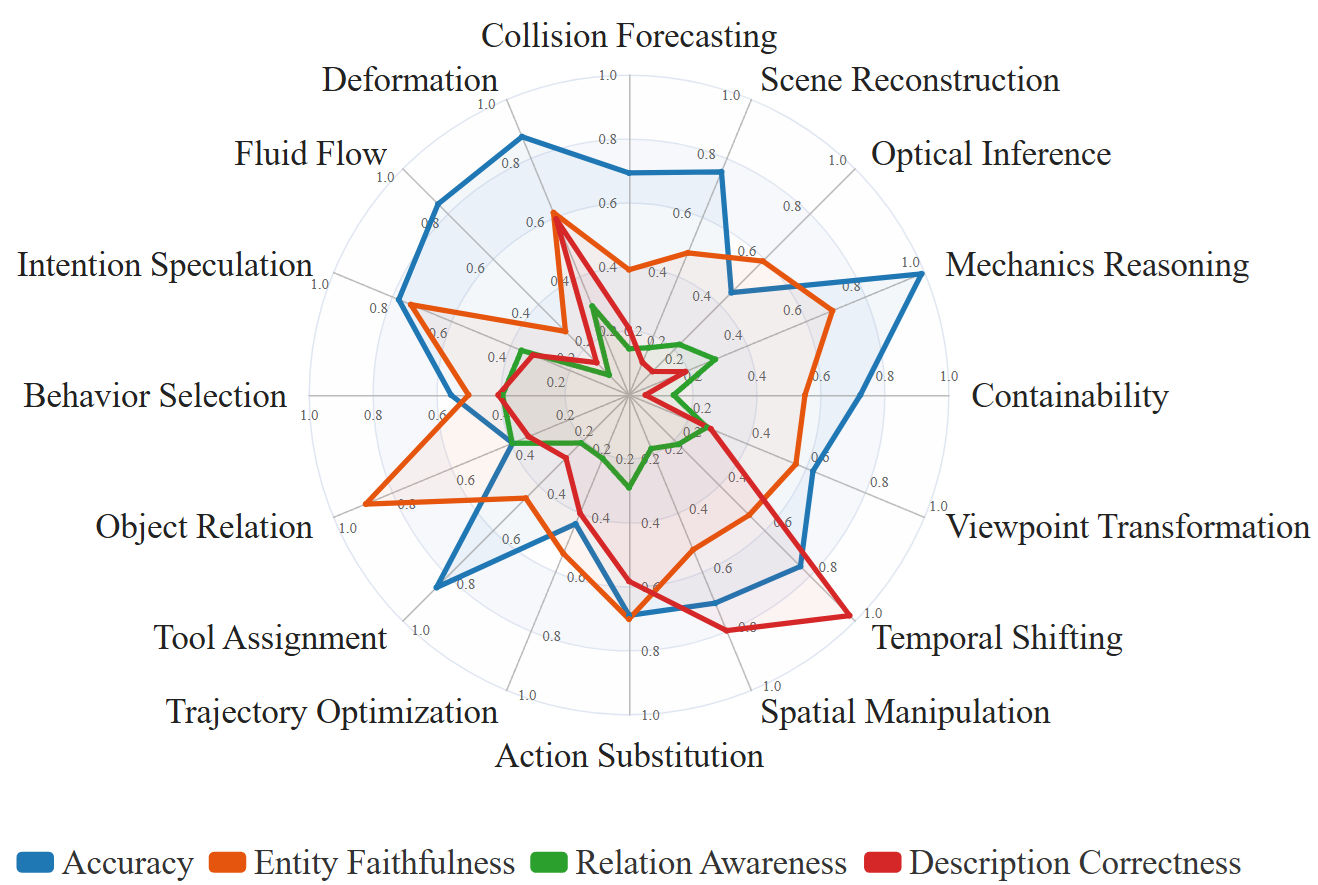}
    \caption{\textbf{Detailed evaluation of CRFT.} CRFT shows promising improvements across multiple reasoning categories, with especially strong gains in Anticipation, suggesting that causal-rationale fine-tuning is particularly effective for forward physical reasoning.}
    \label{fig:crft_pie}
\end{figure}

\paragraph{Training Objective.}
As described in Section~4 of the main paper, CRFT minimizes a 
mixed objective consisting of (i) the rationale-level 
cross-entropy loss $\mathcal{L}_{\text{rat}}$ and 
(ii) the answer-level loss $\mathcal{L}_{\text{ans}}$. 
Following prior work showing that combining intermediate 
rationale supervision with final-answer supervision yields a denser and 
more stable learning signal \cite{DBLP:journals/jmlr/ChungHLZTFL00BW24},
we optimize both components:
\[
\mathcal{L} = \lambda_{\text{rat}} \mathcal{L}_{\text{rat}}
           + \lambda_{\text{ans}} \mathcal{L}_{\text{ans}}.
\]
This mixed supervision is particularly useful in our setting, where MCQ answers are often only one token and answer-only optimization can be unstable.
A small sweep over
$\lambda_{\text{rat}} \in {0.1, 0.2, 0.5}$ and
$\lambda_{\text{ans}} \in {0.5, 1, 2}$
showed that $\lambda_{\text{rat}}=0.2$ and $\lambda_{\text{ans}}=1$ yield the most stable optimization, consistent with prior findings on balanced rationale--answer weighting \cite{zhu2025intokenrationalityoptimizationaccurate}.

\paragraph{Training Configuration.}
We use a 9:1 split, resulting in 2,765 training instances and 297 test instances.
Qwen-VL-7B is fine-tuned with LoRA on a single NVIDIA L40S 40GB GPU for 6 epochs, taking approximately 8 hours.
We use batch size 8, learning rate $5\times10^{-5}$, LoRA rank 16, and dropout 0.05.

\paragraph{Detailed Evaluation of CRFT.}
Figure~\ref{fig:crft_pie} evaluates CRFT across all sixteen \textsc{CausalPhys} subcategories using ACC, EF, RA, and DC.
CRFT shows consistent improvements across diverse physical reasoning tasks, especially in anticipation-oriented scenarios.

% Bootstrap sensitivity analysis has been moved to
% Appendix~\ref{ap:bootstrap} (Additional Experiments and Analyses).

% \subsection{Token Consumption Analyses}

% For the evaluatuion, LLM as a judge stages, we estimate the token consumed.

\clearpage
\section{Case Study}

Based on the case studies below, we observe a consistent pattern: when a model’s rationale is well aligned with the ground-truth causal graph, its final answer is usually correct. In contrast, when this alignment breaks down, the model’s prediction becomes unreliable and often appears to rely on guesswork rather than genuine reasoning. The following cases provide detailed examples from three representative models on selected instances.

\begin{figure}[h]
    \centering
    \includegraphics[width=1\linewidth]{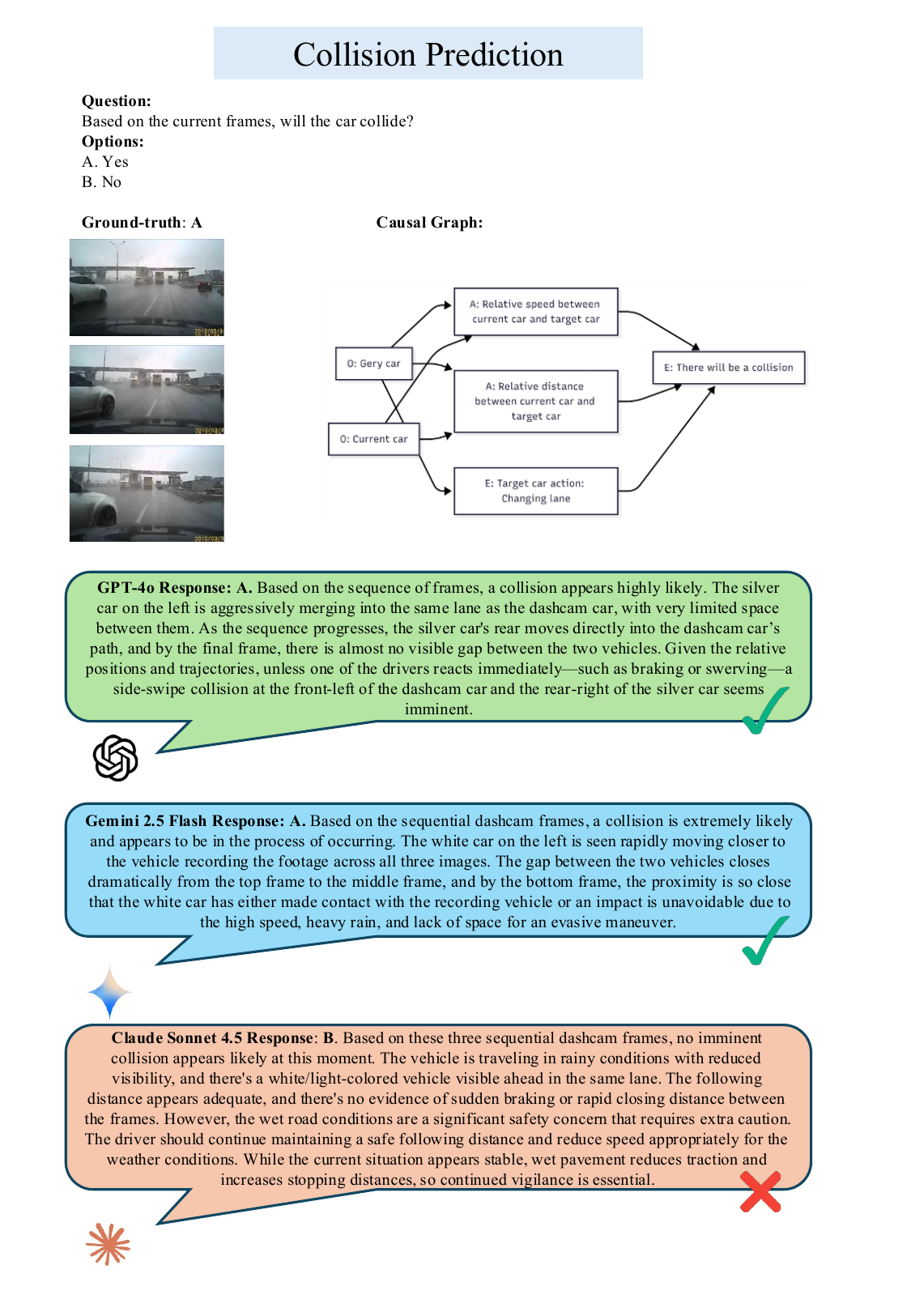}
    \caption{Sample cases of Collision Prediction}
    \label{fig:1}
\end{figure}
\newpage

This highlights the distinctive value of \textbf{CausalPhys}: its fine-grained causal annotations expose reasoning failures that conventional VQA benchmarks cannot detect, enabling more diagnostic evaluation and guiding the development of VLMs with genuinely causal reasoning.
\vspace{1.1cm}

\begin{figure}[h]
    \centering
    \includegraphics[width=1\linewidth]{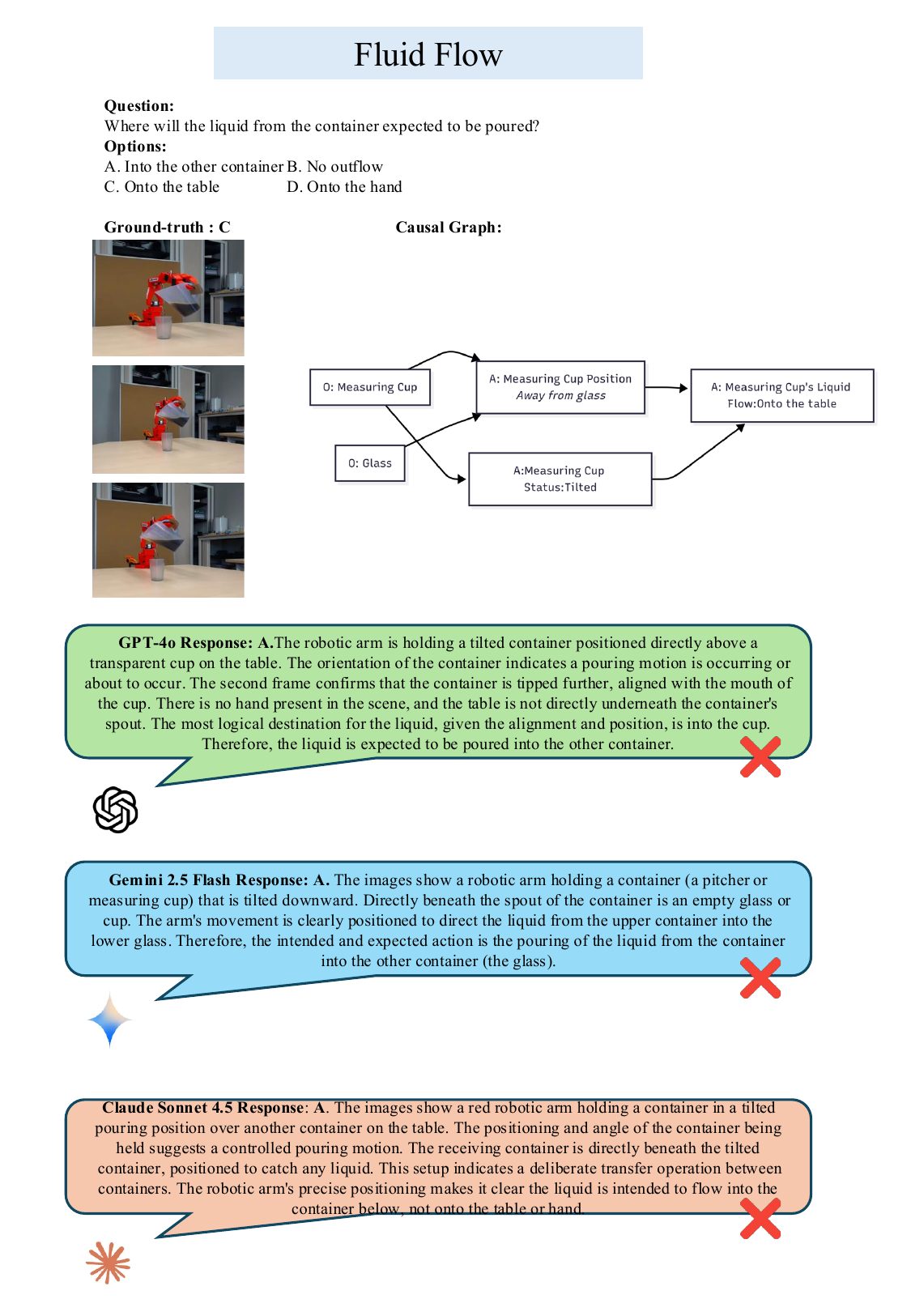}
    \caption{Sample cases of Fluid Flow}
    \label{fig:2}
\end{figure}

\newpage
\begin{figure}
    \centering
    \includegraphics[width=1\linewidth]{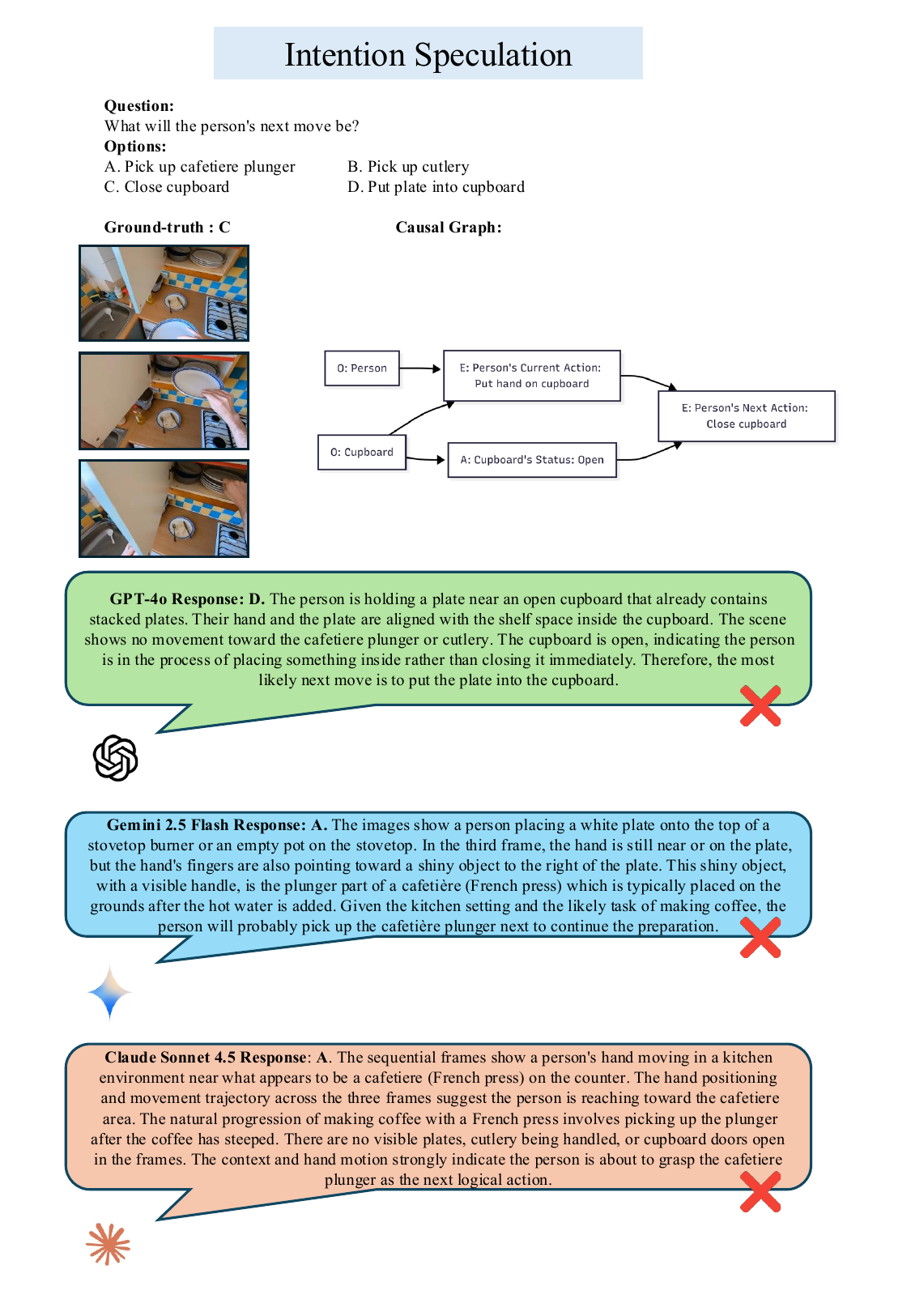}
    \caption{Sample cases of Intention Speculation}
    \label{fig:3}
\end{figure}

\newpage
\begin{figure}
    \centering
    \includegraphics[width=1\linewidth]{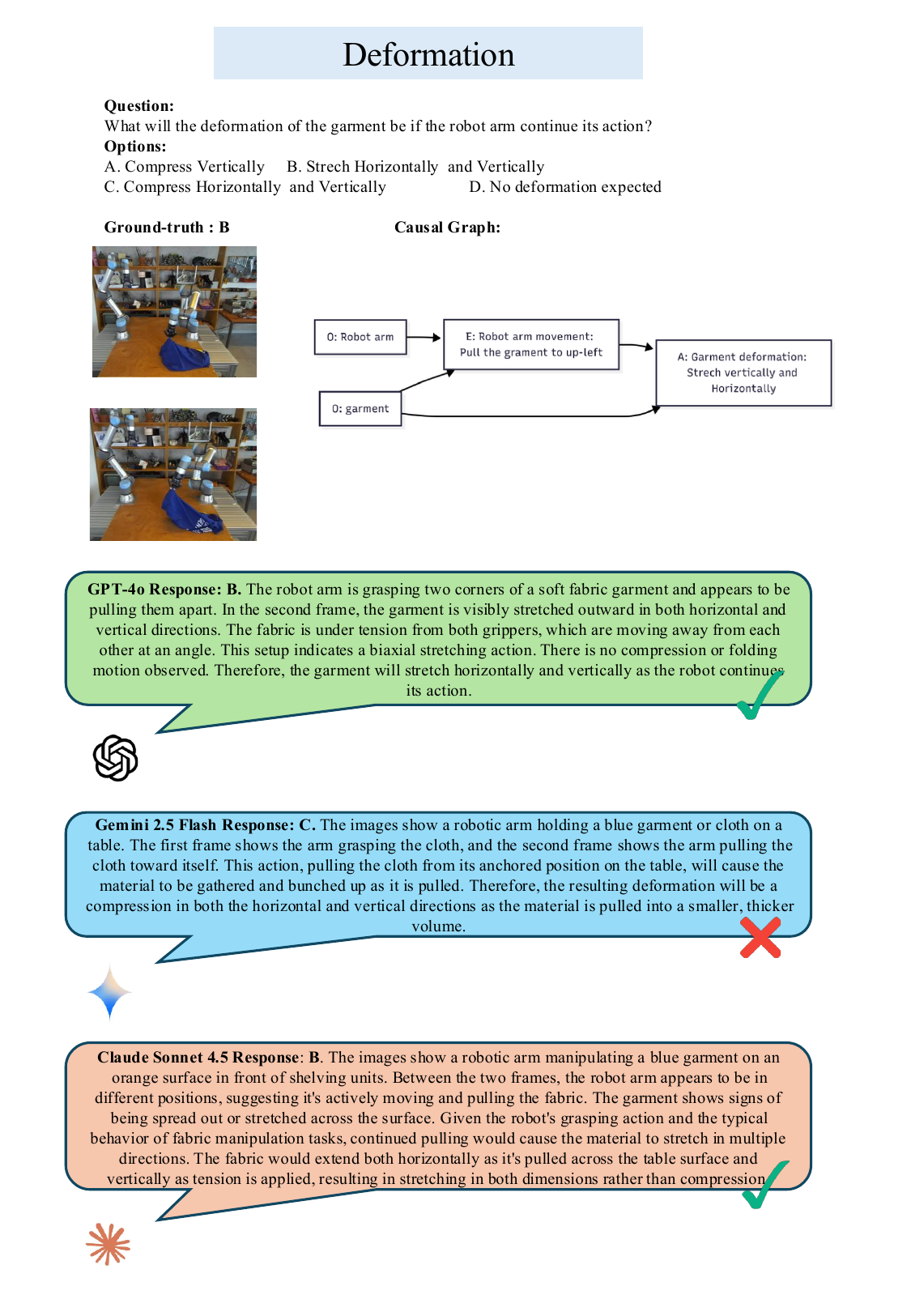}
    \caption{Sample cases of Deformation}
    \label{fig:4}
\end{figure}

\clearpage
\begin{figure}
    \centering
    \includegraphics[width=1\linewidth]{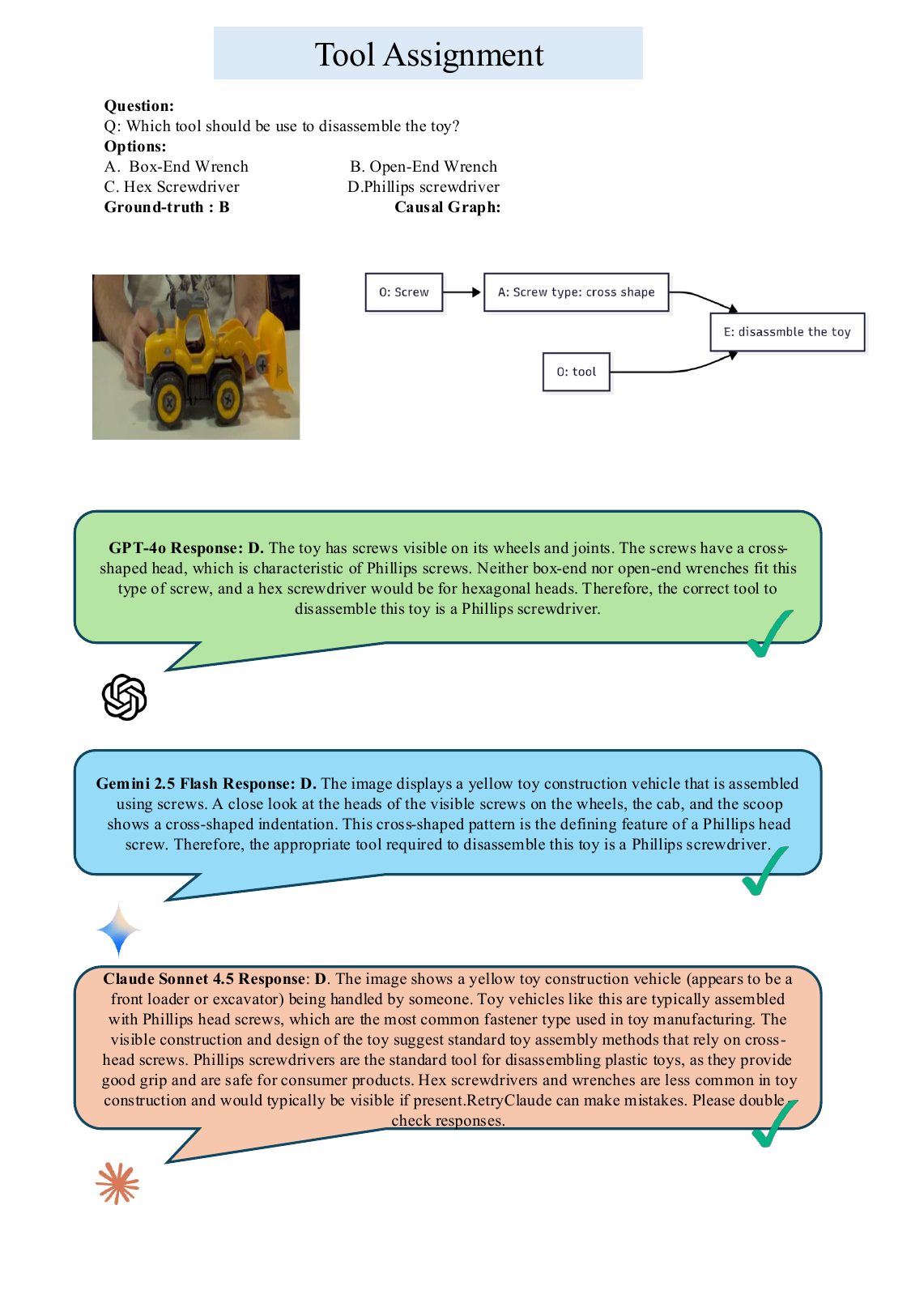}
    \caption{Sample cases of Tool Assignment}
    \label{fig:5}
\end{figure}

\newpage
\begin{figure}
    \centering
    \includegraphics[width=1\linewidth]{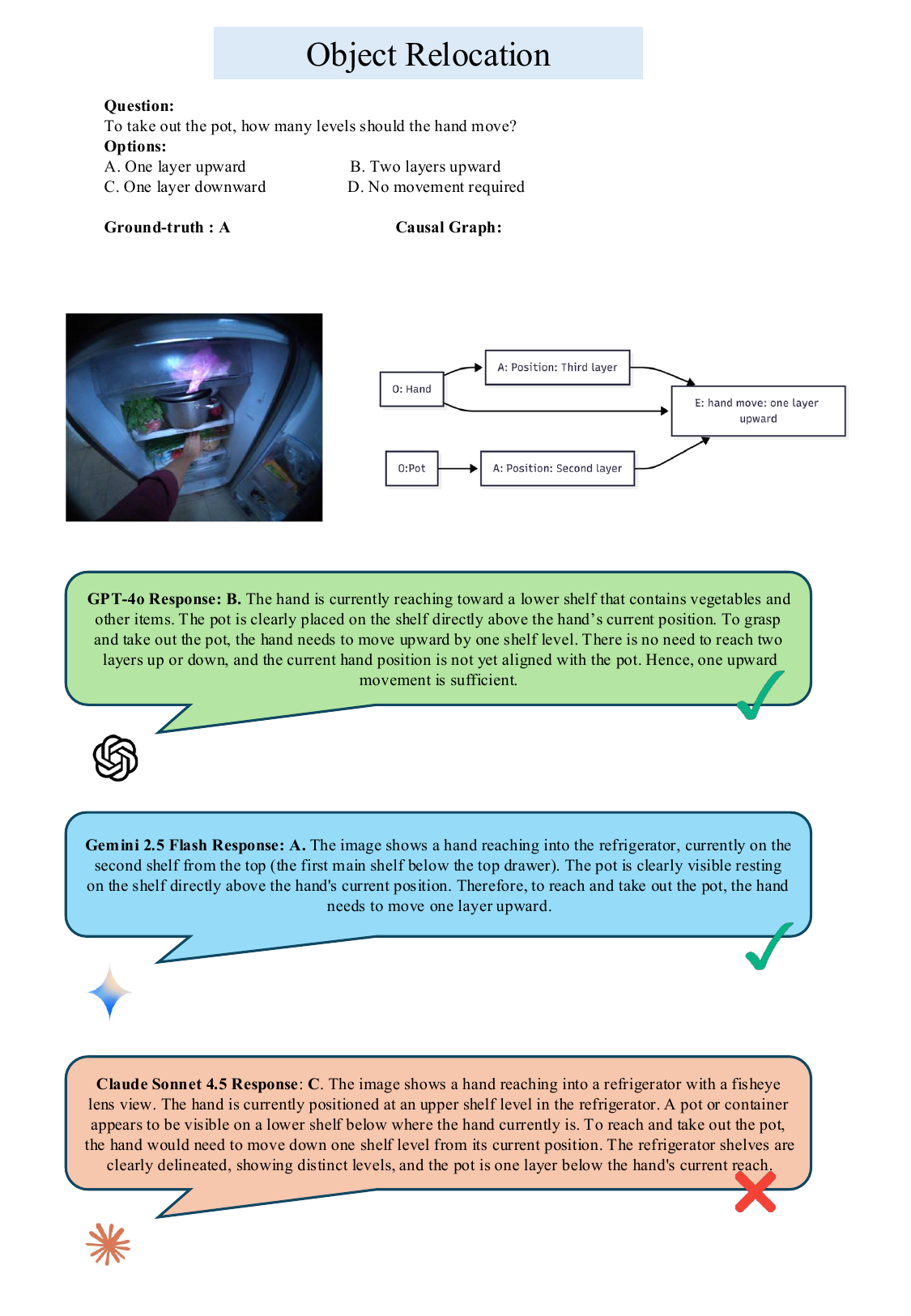}
    \caption{Sample cases of Object Relocation}
    \label{fig:6}
\end{figure}

\clearpage
\begin{figure}
    \centering
    \includegraphics[width=1\linewidth]{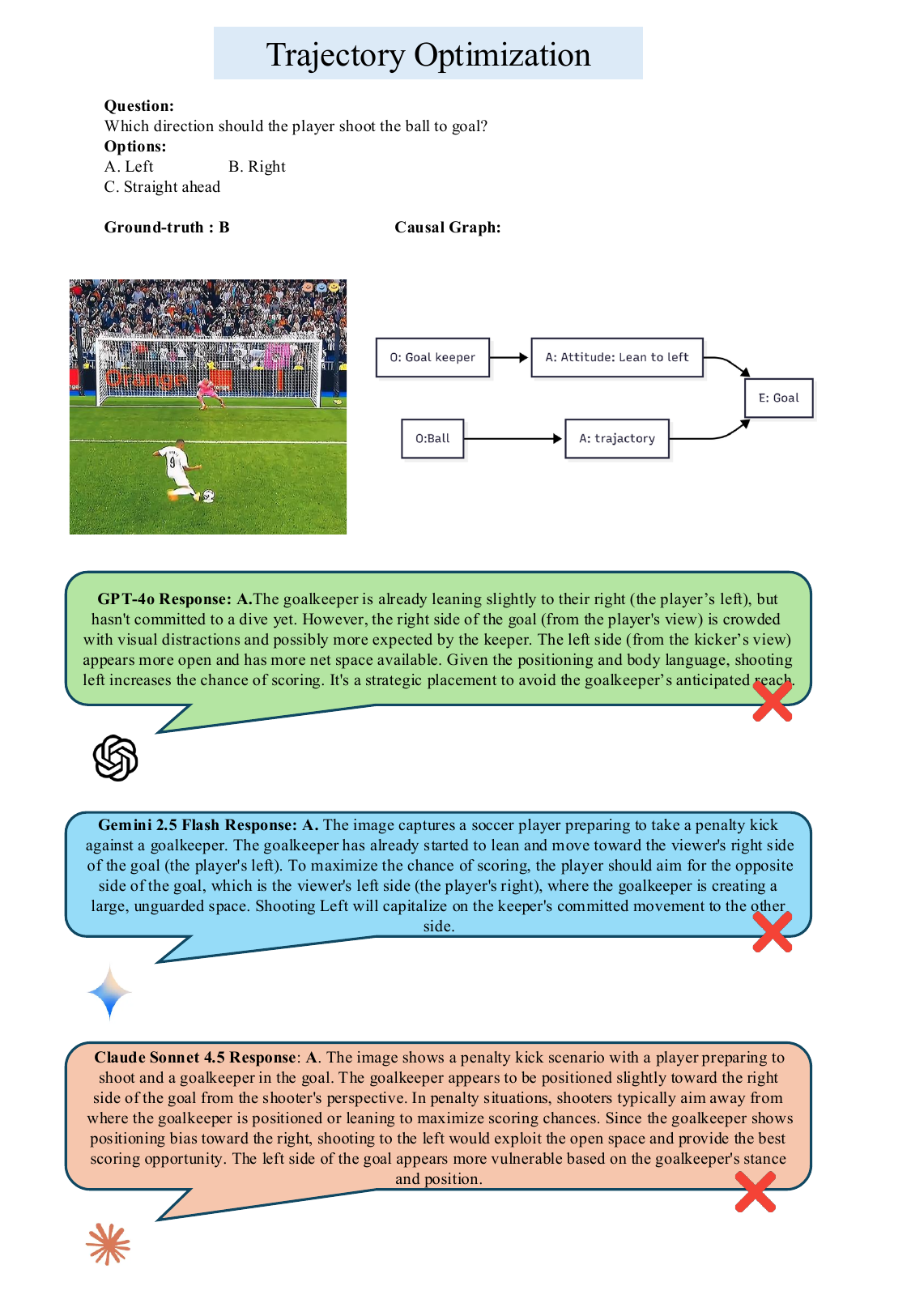}
    \caption{Sample cases of Trajectory Optimization}
    \label{fig:7}
\end{figure}
 
\newpage
\begin{figure}
    \centering
    \includegraphics[width=1\linewidth]{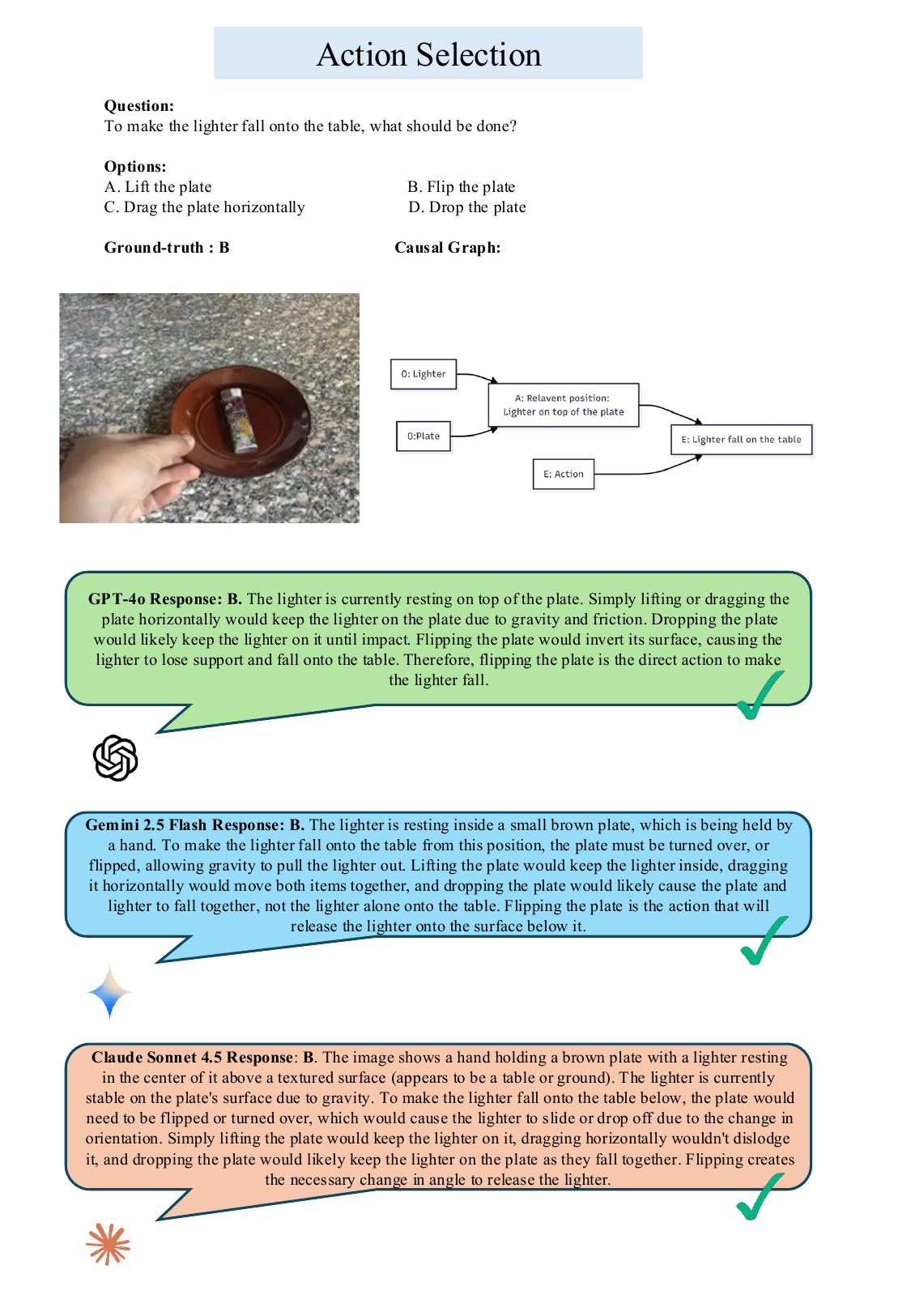}
    \caption{Sample cases of Action Selection}
    \label{fig:8}
\end{figure}

\clearpage
\begin{figure}
    \centering
    \includegraphics[width=1\linewidth]{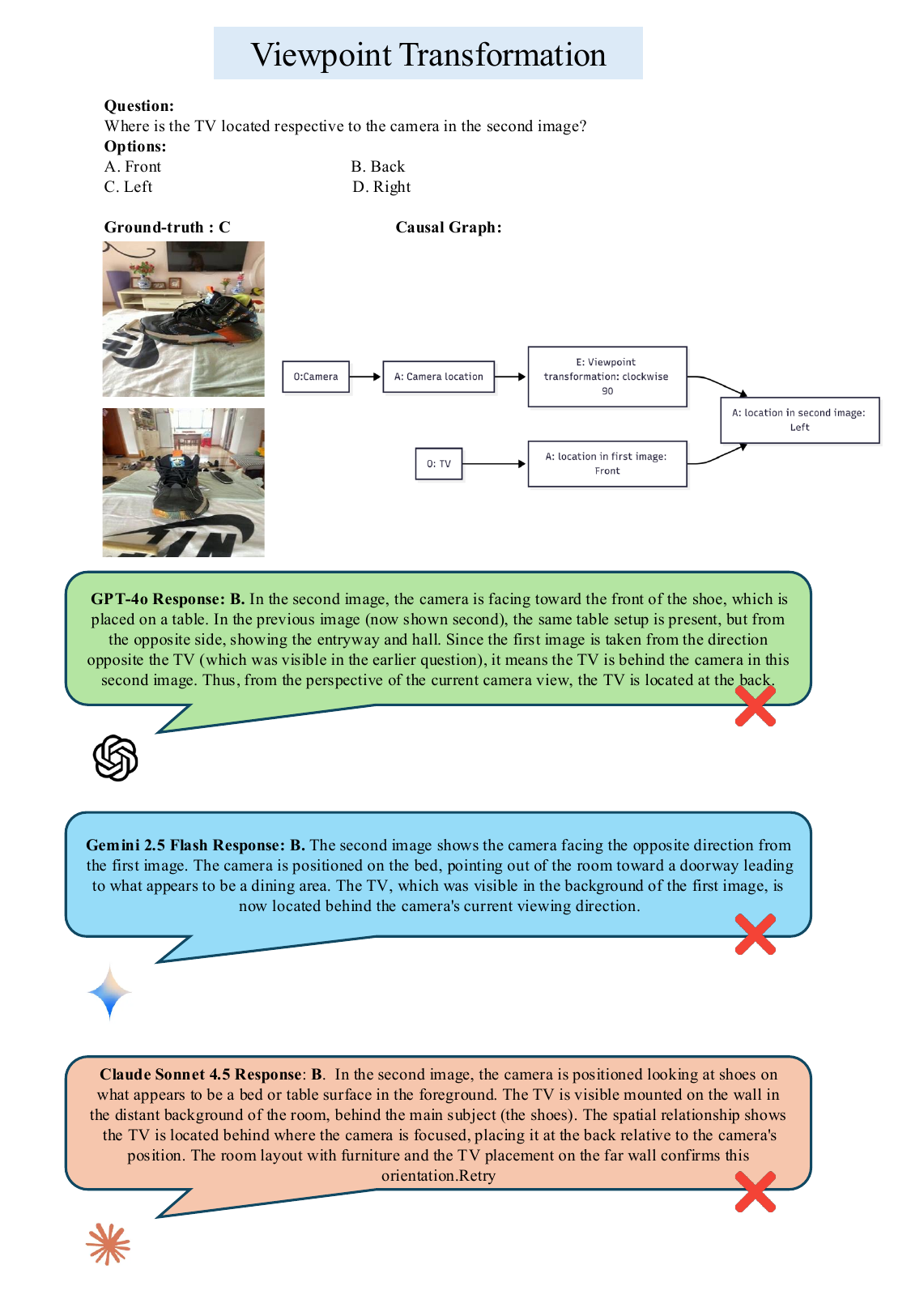}
    \caption{Sample cases of Viewpoint Transformation}
    \label{fig:9}
\end{figure}

\newpage
\begin{figure}
    \centering
    \includegraphics[width=1\linewidth]{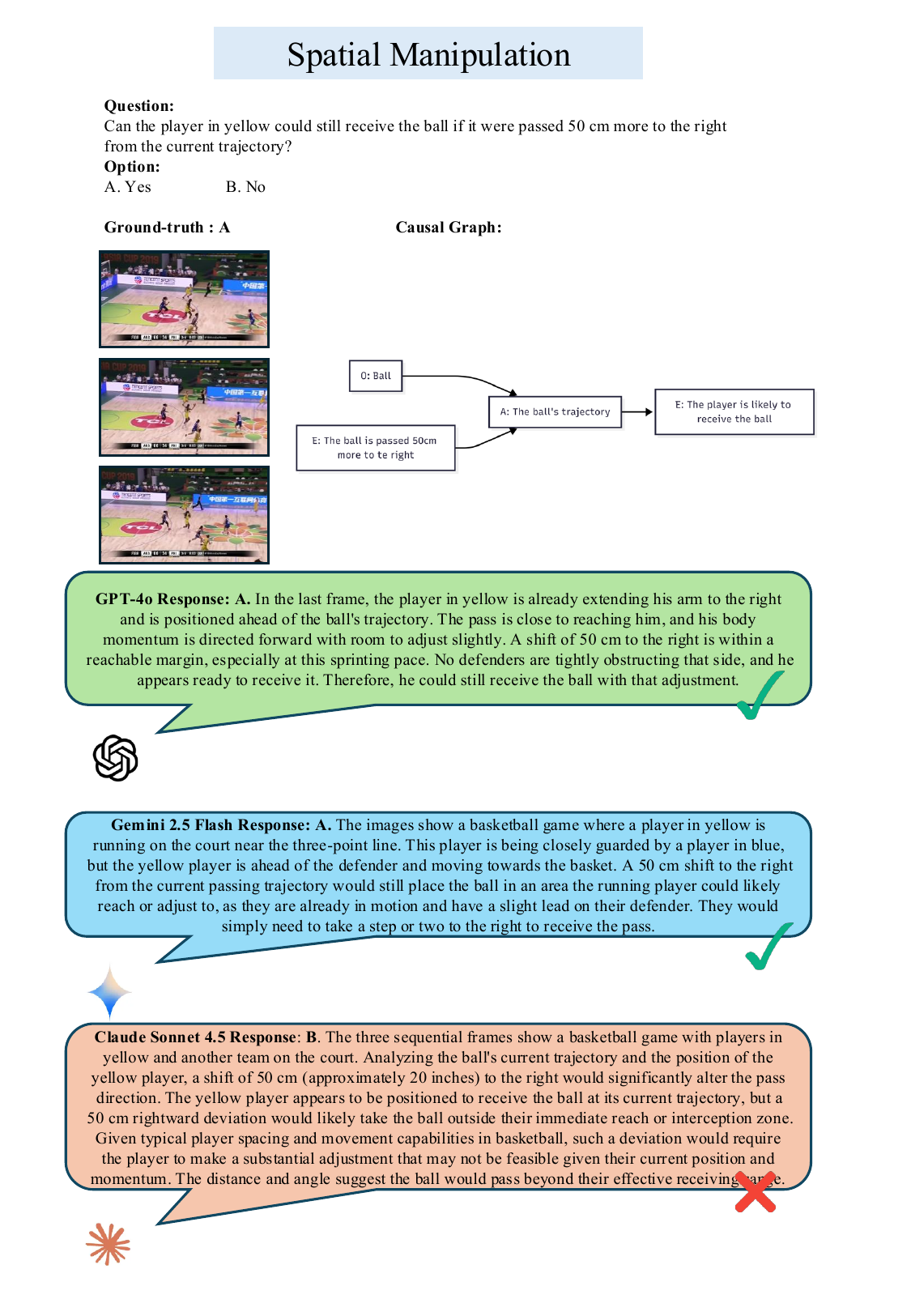}
    \caption{Sample cases of Spatial Manipulation}
    \label{fig:10}
\end{figure}

\clearpage
\begin{figure}
    \centering
    \includegraphics[width=1\linewidth]{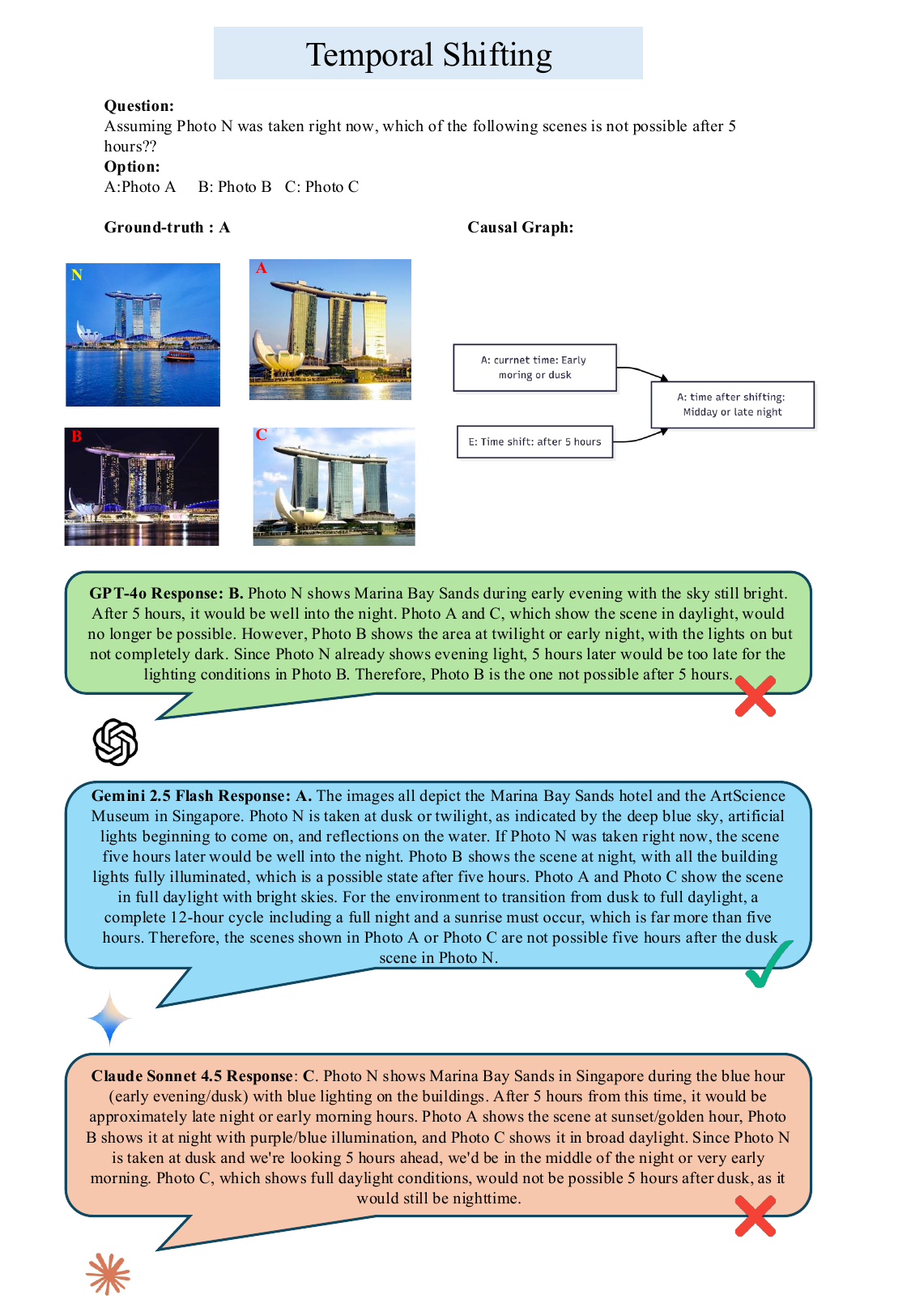}
    \caption{Sample cases of Temporal Shifting}
    \label{fig:11}
\end{figure}

\newpage
\begin{figure}
    \centering
    \includegraphics[width=1\linewidth]{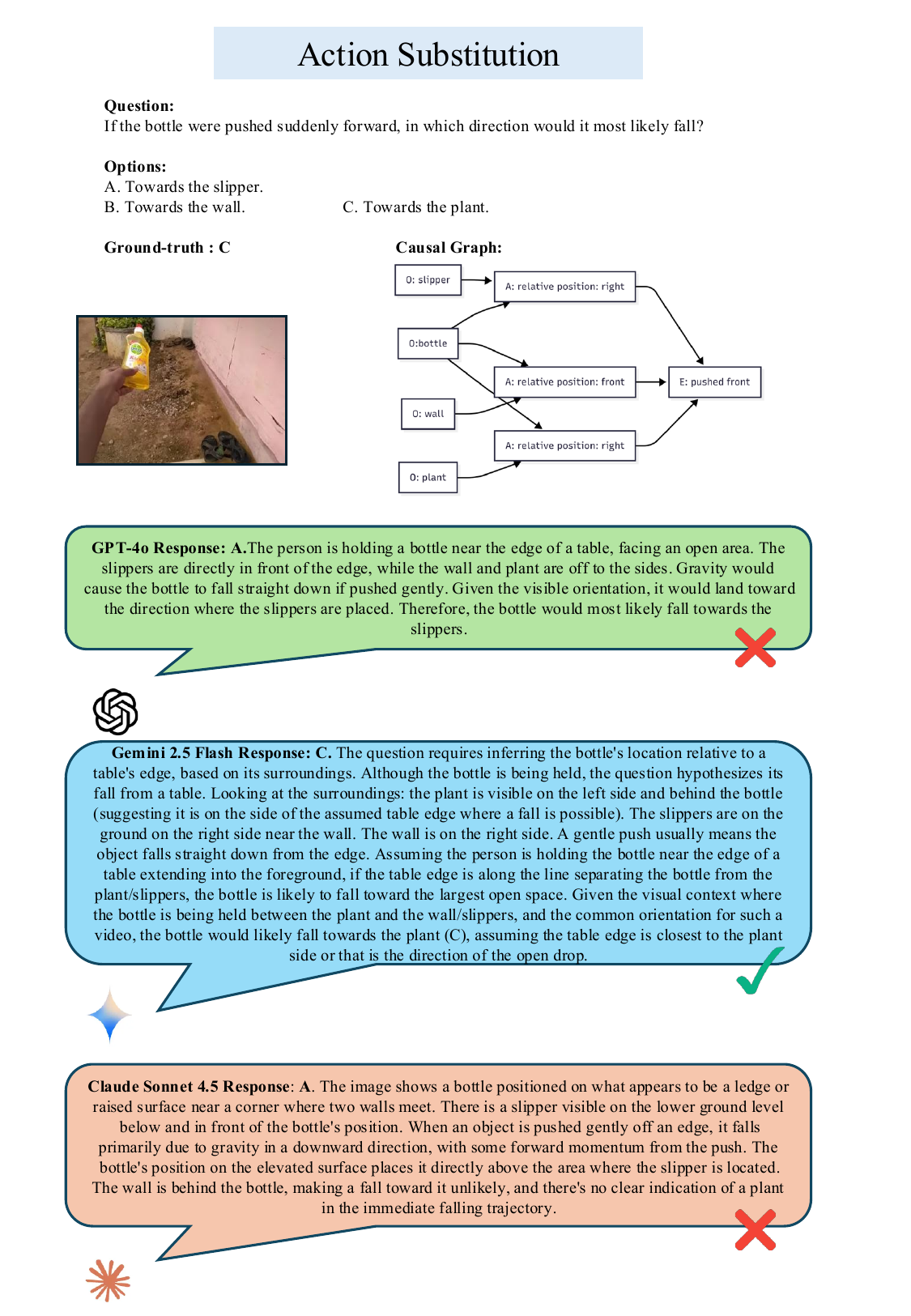}
    \caption{Sample cases of Action Substitution}
    \label{fig:12}
\end{figure}

\clearpage
\begin{figure}
    \centering
    \includegraphics[width=1\linewidth]{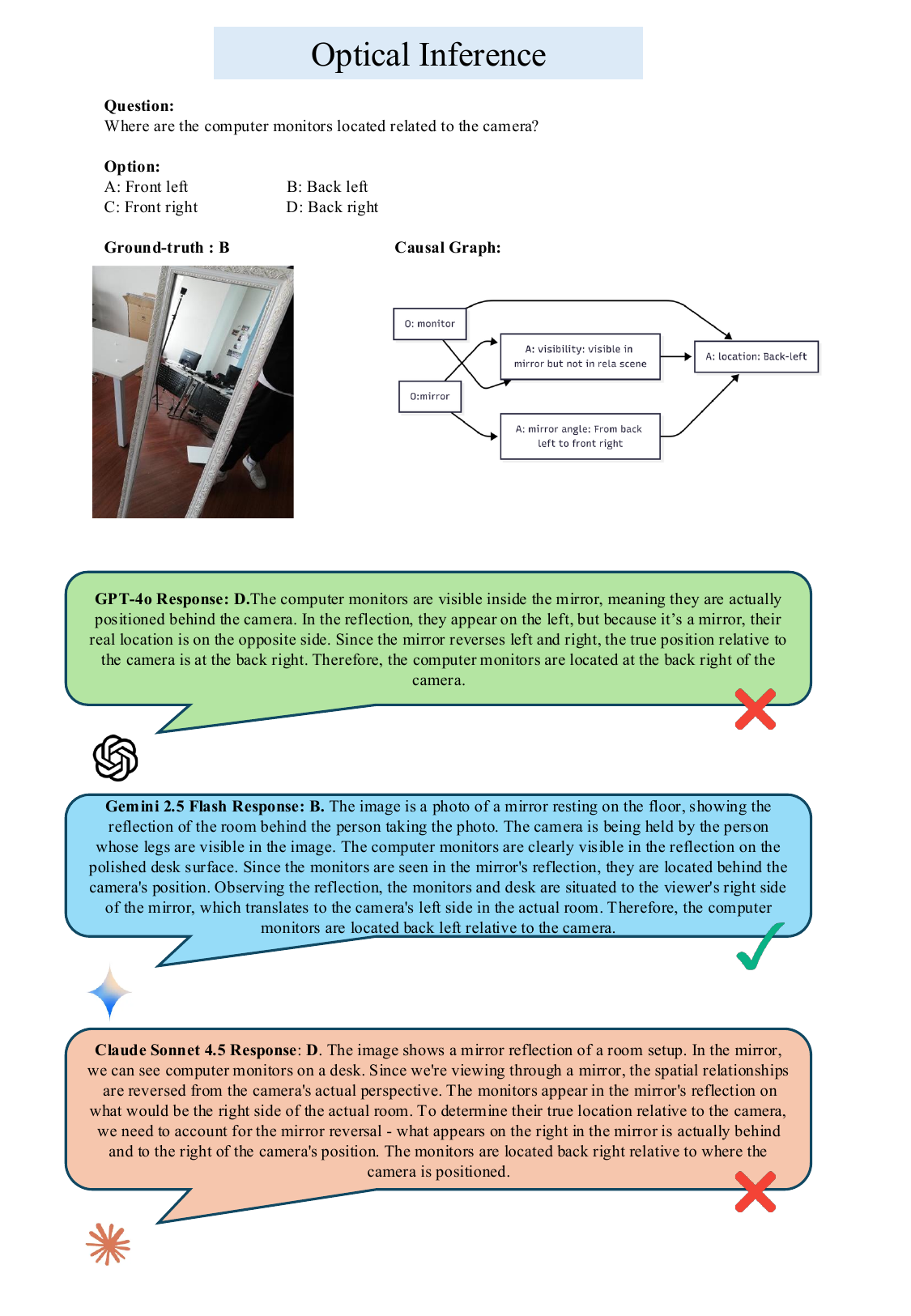}
    \caption{Sample cases of Optical Inference}
    \label{fig:13}
\end{figure}

\newpage
\begin{figure}
    \centering
    \includegraphics[width=1\linewidth]{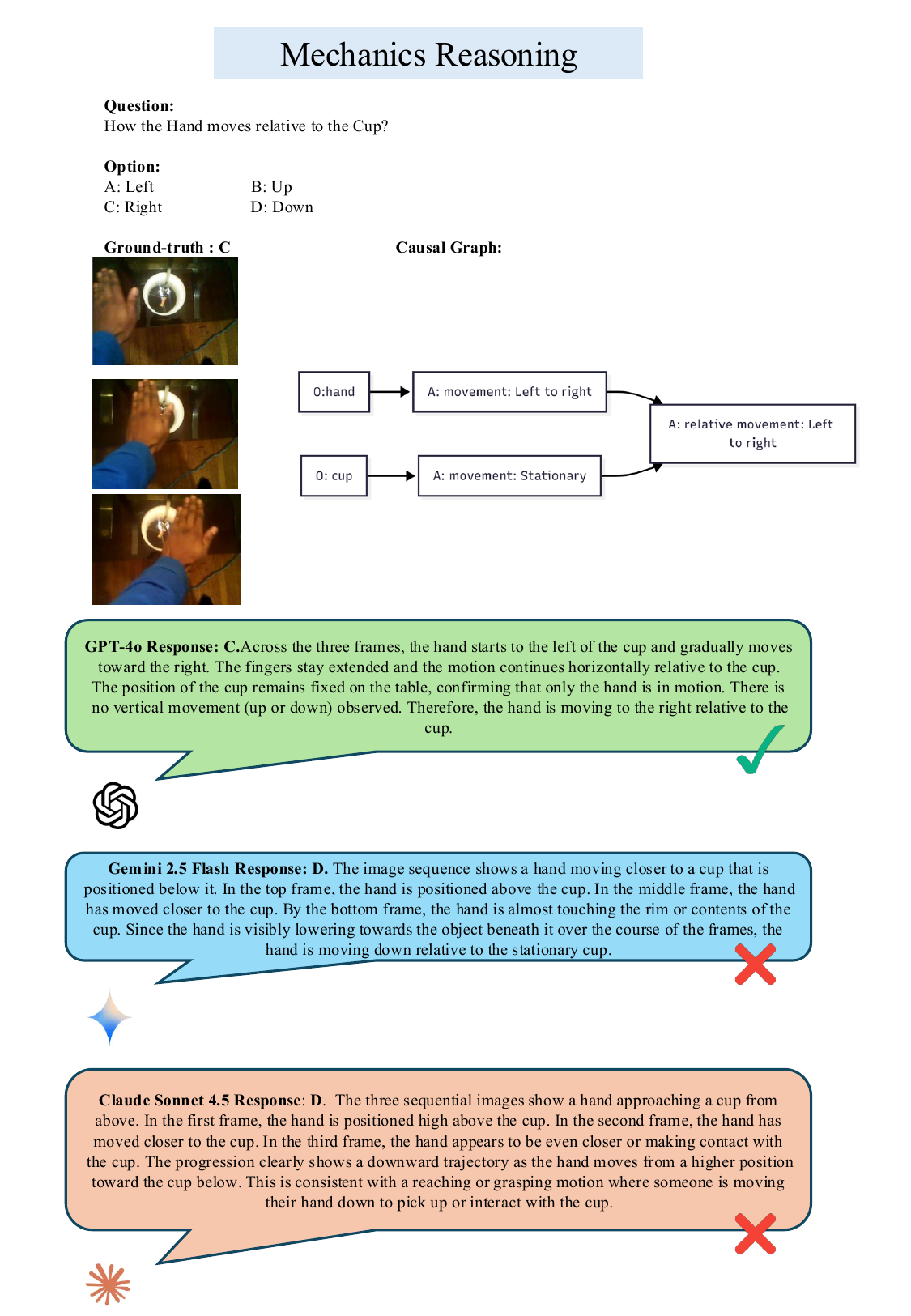}
    \caption{Sample cases of Mechanics Reasoning}
    \label{fig:14}
\end{figure}

\clearpage
\begin{figure}
    \centering
    \includegraphics[width=1\linewidth]{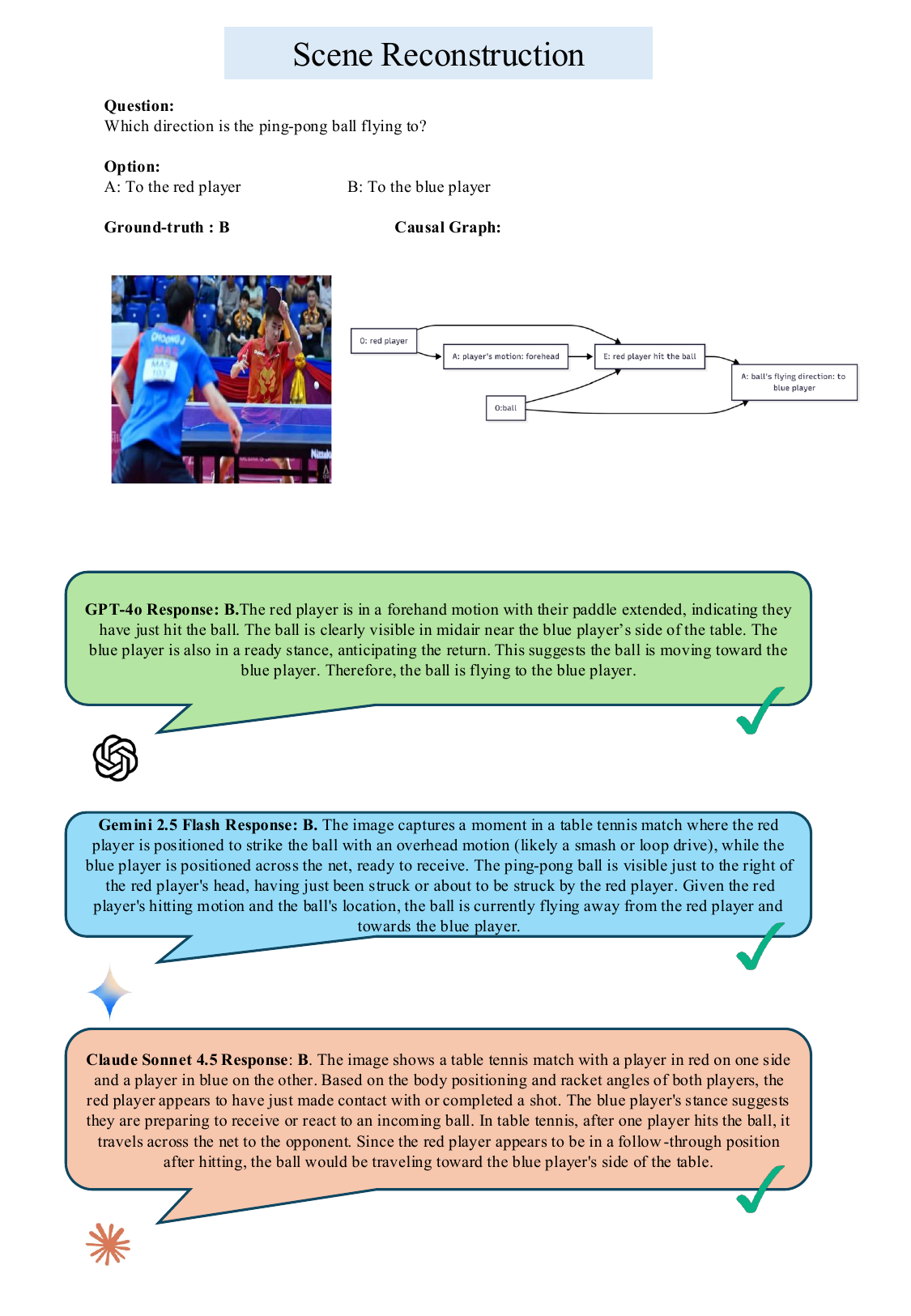}
    \caption{Sample cases of Scene Reconstruction}
    \label{fig:15}
\end{figure}

\begin{figure}
    \centering
    \includegraphics[width=1\linewidth]{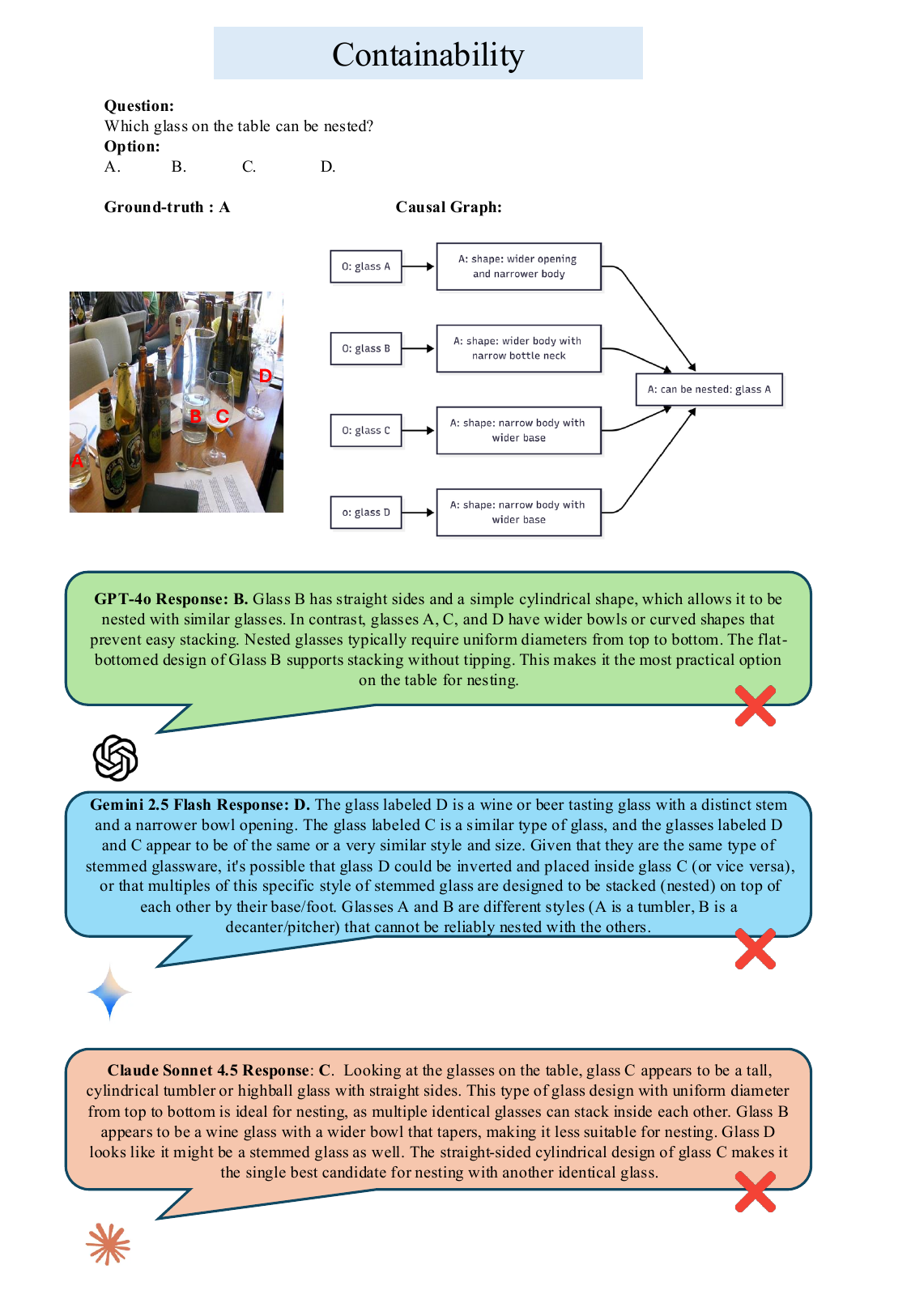}
    \caption{Sample cases of Containability}
    \label{fig:16}
\end{figure}

\clearpage
\end{document}